\documentclass[11pt,final]{article}
\pdfoutput=1
\usepackage{MRnotes}

\newcommand{\ibf}[1]{\operatorname{B}\left(#1\right)}

\newcommand{\tFo}[1]{\,{}_{2}F_{1}\left(#1\right)}
\newcommand{\harm}[1]{\operatorname{Har}\left(#1\right)}
\newcommand{\gE}{\gamma_{E}}

\newcommand{\ctor}{\zeta}  
\newcommand{\br}{\xi}
\newcommand{\ri}{\rho} 	
\newcommand{\rib}{\bar{\ri}} 	
\newcommand{\rit}{\tilde{\ri}} 	
\newcommand{\Dz}{\mathbb{D}}
\newcommand{\Dzeta}{D_\ctor}
\newcommand{\bw}{\mathfrak{w}}
\newcommand{\bq}{\mathfrak{q}}

\newcommand{\od}{{\sf s}}

\newcommand{\ann}{\mathscr{M}}

\newcommand{\TT}{\mathbb{T}}
\newcommand{\VV}{\mathbb{V}}
\newcommand{\ScS}{\mathbb{S}}
\newcommand{\ai}{\alpha}
\newcommand{\bi}{\sigma}

\newcommand{\bk}{\vb{k}}
\newcommand{\bx}{\vb{x}}

\newcommand{\sen}[1]{\varphi_{_{#1}}}
\newcommand{\dils}{\chi_s}
\newcommand{\dilv}{\chi_v}
\newcommand{\cpen}[1]{\pi_{_{#1}}}

\newcommand{\In}{\text{\tiny{in}}}

\newcommand{\Gin}{G_{_\ann}^\In}
\newcommand{\GinN}{G_{_{-\ann}}^\In} 
\newcommand{\GinNc}{\check{G}_{_{-\ann}}^\In} 
\newcommand{\GinNt}{\widetilde{G}_{_{-\ann}}^\In} 

\newcommand{\Kin}{K_{_\ann}^\In}
\newcommand{\KinN}{K_{_{-\ann}}^\In}

\newcommand{\XiNN}{\Xi_{\text{nn}}}
\newcommand{\XiNNR}{\Xi_{\text{nn}}^{\text{R}}}

\newcommand{\ctphi}[1]{c_{_\varphi}^{\scriptscriptstyle{(#1)}}}
\newcommand{\ctpi}[1]{c_{_\pi}^{\scriptscriptstyle{(#1)}}}

\newcommand{\snMar}{ \breve{\Phi}}

\newcommand{\JMar}{J}
\newcommand{\JnMar}{\breve{J}}

\newcommand{\Rev}{\text{\tiny{rev}}}
\newcommand{\Out}{\text{\tiny{out}}}
\newcommand{\SKs}{\text{\tiny{SK}}}
\newcommand{\Grev}{G_{_\ann}^\Rev}
\newcommand{\Gout}{G_{_\ann}^\Out}
\newcommand{\GrevN}{G_{_{-\ann}}^\Rev}

\newcommand{\Krev}{K_{_\ann}^\Rev}
\newcommand{\KrevN}{K_{_{-\ann}}^\Rev}

\newcommand{\incfM}{\mathfrak{c}_{_\ann}^{\In}}
\newcommand{\hawkcfM}{\mathfrak{c}_{_\ann}^{\Rev}}
\newcommand{\nB}{n_{_B}}

\newcommand{\Cm}{\mathscr{C}}
\newcommand{\Vm}{\mathscr{V}}
\newcommand{\vMax}{\bar{\Phi}}
\newcommand{\sMax}{\bar{\Psi}}
\newcommand{\psMax}{\bar{\Pi}}
\newcommand{\pJMax}{\bar{P}}
\newcommand{\sMaxD}{\bar{\Phi}_{\text{\tiny{D}}}}
\newcommand{\EMax}{\mathcal{E}^{\text{\tiny{Max}}}}
\newcommand{\EMaxD}{\mathcal{E}^{\text{\tiny{D}}}}
\newcommand{\muN}{\mu_{_\ann}}
\newcommand{\bmu}{{\bm \mu}_{\text{\tiny{D}}}}
\newcommand{\CmD}{\mathscr{C}^{\text{\tiny{D}}}}
\newcommand{\VmD}{\mathscr{V}^{\text{\tiny{D}}}}

\newcommand{\ctV}[1]{c_{_\Vm}^{\scriptscriptstyle{(#1)}}}
\newcommand{\snMarQ}{\breve{\mathcal{Q}}}
\newcommand{\JMarQ}{\mathcal{A}}
\newcommand{\Jcft}{J^\text{\tiny{CFT}}}

\newcommand{\tGR}{\Phi}
\newcommand{\vGR}{\Psi}
\newcommand{\AGR}{\mathscr{A}}
\newcommand{\FGR}{\mathscr{F}}

\newcommand{\Tcft}{T^{\text{\tiny{CFT}}}}
\newcommand{\TcftP}{P}
\newcommand{\snMarP}{\breve{\mathcal{P}}}
\newcommand{\JMarP}{\bm{\gamma}}
\newcommand{\sEinD}{\bar{\Phi}_{\text{\tiny{D}}}}

\newcommand{\bwt}{\mathfrak{w}}

\newcommand{\skR}{\text{\tiny R}}
\newcommand{\skL}{\text{\tiny L}}
\newcommand{\Kbeta}{\bm{\beta}}

\title{Effective field theory of stochastic diffusion  from  gravity}

\author[a]{Jewel K. Ghosh,}
\author[a]{R. Loganayagam,}
\author[a]{Siddharth G. Prabhu,}
\author[b]{Mukund Rangamani,}
\author[a]{\\ Akhil Sivakumar,}
\author[a]{V. Vishal}

\affiliation[a]{
	International Centre for Theoretical Sciences (ICTS-TIFR), \\ 
	Tata Institute of Fundamental Research, Shivakote, Hesaraghatta, Bangalore 560089, India.}
\affiliation[b]{
	Center for Quantum Mathematics and Physics (QMAP)\\
	Department of Physics \& Astronomy, University of California, Davis, CA 95616 USA}

\emailAdd{jewel.ghosh@icts.res.in}
\emailAdd{nayagam@icts.res.in}
\emailAdd{siddharth.prabhu@icts.res.in}
\emailAdd{mukund@physics.ucdavis.edu}
\emailAdd{akhil.sivakumar@icts.res.in}
\emailAdd{vishal.vijayan@icts.res.in}

\abstract{
Planar  black holes in AdS have long-lived quasinormal modes which capture the physics of charge and momentum diffusion in the dual field theory. How should we characterize  the effective dynamics  of a probe system coupled to the conserved currents of the dual field theory?  Specifically, how would such a probe record the long-lived memory of  the  black hole and its Hawking fluctuations?  We address this question by exhibiting a universal gauge invariant framework which captures the physics of stochastic diffusion in holography: a designer scalar with  a gravitational coupling governed  by a single parameter, the  Markovianity index. We argue that the physics of gauge and gravitational perturbations of a planar Schwarzschild-AdS black hole can be efficiently captured by such designer scalars. We demonstrate that this framework allows one to decouple, at the quadratic order, the  long-lived quasinormal and Hawking modes from the short-lived ones. It furthermore provides a template for analyzing fluctuating open quantum field theories with memory. In particular, we use this set-up  to analyze the diffusive Hawking photons and gravitons about a planar Schwarzschild-AdS black hole and derive the quadratic effective action that governs fluctuating hydrodynamics of the dual CFT. Along the way we also derive results relevant for  probes of  hyperscaling violating backgrounds  at finite temperature.
}

\begin{document}
\maketitle


\section{Introduction: \emph{Remembrance of a black hole's past}}
\label{sec:intro}

Black holes when disturbed settle down  after classically ringing in quasinormal modes \cite{Vishveshwara:1970cc,Press:1971wr}. In addition, they undergo simulated radiation in Hawking modes \cite{Hawking:1974sw} (which may also be spontaneous owing to vacuum fluctuations). As in any quantum statistical system these are two intimately related features of a thermodynamic ensemble of states. The quasinormal modes signify dissipation into a medium and the Hawking modes correspond to the attendant quantum statistical fluctuations. They are related through fluctuation-dissipation relations.  The timescales for the ring-down are typically short for non-extremal black holes with compact horizon topology. Typically, black holes carry no long-term memory. 

However, black holes in negatively curved anti-de Sitter (AdS) spacetimes with non-compact horizons exhibit long-lived quasinormal modes.  Quasinormal modes of AdS black holes correspond to thermalization rates of the dual field theory \cite{Horowitz:1999jd}.  Massless spin-1 and spin-2 fields in planar AdS black holes have long-lived quasinormal modes with dispersions that are characteristic of hydrodynamic fluctuations \cite{Policastro:2001yc,Policastro:2002se,Policastro:2002tn}. These long lived modes correspond to the charge and momentum diffusion, and  (attenuated)  sound waves in the dual field theory plasma. One can trace their origins to the underlying gauge invariance of  massless spin-1 and spin-2 fields manifested as global  conservation laws for charge currents and energy-momentum tensor. For field theories on $\mathbb{R}^{d-1,1}$ these conserved currents decay slowly.\footnote{ While these modes are present for planar AdS black holes with non-compact horizon topology,  as explained in \cite{Bhattacharyya:2007vs} one expects nearly long-lived modes in large AdS black holes.}  These results which were originally derived for linearized perturbations also hold at the non-linear level as manifested by the fluid/gravity correspondence \cite{Bhattacharyya:2008jc,Hubeny:2011hd}. 

While the study of the dissipative dynamics of long-lived quasinormal modes of AdS black holes has a rich and storied history (cf., \cite{Berti:2009kk,Morgan:2009pn} for comprehensive reviews), the corresponding discussion for the fluctuations of the long-lived modes (i.e., long-lived Hawking modes) is less well developed. Our goal in this work is to provide a unified treatment of the diffusive modes of AdS black holes and obtain  an effective action governing their dynamics.

\subsection{Open quantum systems  with memory}
\label{sec:oqmemory}

We find it helpful to phrase the problem in the language of open quantum systems following \cite{Jana:2020vyx}. It was demonstrated in this work that  holography is a natural arena to study open quantum field theories by  allowing one to compute real-time QFT observables without the limitations of  perturbation theory.  Consider a holographic field theory at finite temperature, e.g., $\mathcal{N}=4$ SYM,   coupled to an external probe. The probe which evolves as an open quantum system, carries in its effective dynamics (obtained after integrating out the holographic system), an imprint of the thermal medium's correlations. Thus, the probe's real-time dynamics should encode both the dissipation in the plasma and the fluctuations it experiences.  Consequently, such a probe,  if coupled to the CFT's conserved currents,  should be able to record the long memory of AdS black holes.  We seek to understand how open quantum systems work in this non-Markovian regime where long-time memory is retained.

The simplest examples of such probes are quantum mechanical ($0+1$ dimensional probes). Indeed, it was demonstrated in \cite{deBoer:2008gu,Son:2009vu} that external point particle probes (quarks)  exhibit stochastic Brownian motion in the plasma. More recently, building on the proposal of \cite{Glorioso:2018mmw} for computing real-time (Schwinger-Keldysh) observables in AdS/CFT (see \cite{Son:2002sd,Herzog:2002pc,Skenderis:2008dh,Skenderis:2008dg,vanRees:2009rw} for important earlier works on the subject), \cite{Chakrabarty:2019aeu} were able to derive the non-linear fluctuation-dissipation relations for the Brownian particle. Crucial in this regard was the use of time reversal isometry that gives a clean construction of the Hawking fluctuations from the in-falling dissipative  modes. 

Quantum field theoretic probes however are more interesting  as  the problem of constructing local open effective field theories remains a challenge using standard techniques (see \cite{Jana:2020vyx} for a brief commentary on this issue).  In \cite{Jana:2020vyx} it was shown that holographic engineering of  local open EFTs is quite straightforward. They considered a scalar probe $\Psi$ coupled to a single-trace bosonic operator $\mathcal{O}$ in the holographic field theory.\footnote{ This discussion was recently generalized to fermionic open EFTs in \cite{Loganayagam:2020eue,Loganayagam:2020iol} and to rotating backgrounds in \cite{Chakrabarty:2020ohe}. } The $\Psi$ effective action has  coefficients that are determined by the real-time correlators of $\mathcal{O}$. The latter can be computed using the   real-time geometry dual to the field theory Schwinger-Keldysh contour, dubbed the \emph{gravitational Schwinger-Keldysh saddle} or grSK geometry for short. For operators which have short-lived quasinormal modes, the correlations decay quite fast. The thermal plasma viewed as a bath/environment for the probe field $\Psi$ has a very short memory. We will refer to  such probes as Markovian; they are forgetful and do not keep a detailed record of the black hole state. Indeed one of the reasons for the success in the holographic modeling of Markovian open EFTs in \cite{Jana:2020vyx} may be attributed to the fact that holographic thermal baths scramble information optimally.

But there are other physical questions about open EFTs where there is a bath degree of freedom which exhibits long time correlations and has to be retained in the effective description even as the rest of the bath is integrated out.  The non-Markovian behaviour of this kind often happens for a physical reason: either because of a Goldstone mode emerging out of a spontaneous breaking of a continuous symmetry (e.g., holographic superfluids), or because of the effect of Fermi surface with long-lived quasiparticles,  or because there is a conserved charge/energy/momentum whose relaxation happens at macroscopic time scales as described above.  It is interesting to ask what holography can  teach us about the structure of open  EFTs in such situations. 

Unfortunately, the most straightforward models which incorporate the kind of physics described above are technically involved. They typically involve systems  whose standard description in the holographic geometry exhibits gauge invariance  with its usual companions,   Gauss constraints and Bianchi identities. Thus, understanding the physics of Hawking modes and fluctuations in such systems run into two issues: one conceptual and another technical. The conceptual issue is to come up with a way to think about the long-lived part of the Hawking radiation (dual to hydrodynamic fluctuations) by decoupling it from the short-lived part without spoiling gauge invariance. This should be distinguished from the technical issue of making suitable choice of gauge to solve for the Hawking fluctuations. For instance, much  of the AdS/CFT literature uses a radial gauge, which is a poor  choice in this context \cite{Glorioso:2018mmw}. Physically, one may  understand its inefficacy by the fact that the time-reversal isometry used to directly construct the Hawking modes in  \cite{Chakrabarty:2019aeu,Jana:2020vyx} does not preserve the radial gauge.

We aim to address both the conceptual  and the technical issues, though it will  prove convenient to first decouple the two and address them independently. To this end, it will prove helpful to first build intuition regarding the key physical aspects of non-Markovian probes of black holes. We shall do so by analyzing a probe scalar field with suitably dressed gravitational interactions. This allows us  to decouple the questions of long-time correlations from those of gauge invariance. 

Inspired by the above, we introduce a class of models of scalar probes coupled to a thermal holographic bath. These probes will be characterized by a single parameter $\ann$ which we will refer to as the  \emph{Markovianity index}. Probes with $\ann >-1$ will have short-lived memory and behave analogous to the massive scalar probes studied in \cite{Jana:2020vyx}. Probes with $\ann <-1$, however, retain long-term memory.\footnote{ The zero-point here has been chosen so that a  minimally coupled scalar in \AdS{d+1}  has $\ann = d-1$. Our choice is engineered for a simple analytic continuation rule from Markovian to non-Markovian fields.}  These capture the essence  of the non-Markovian physics we wish to analyze. In the gravity dual we will model such probes using a dilatonic coupling. The heuristic intuition is that one wants to amplify the coupling of the field near the horizon where the thermal atmosphere of the black hole is the strongest, thus amplifying the low-lying IR modes. Relatedly, we want to suppress the coupling to the UV modes since the dynamics of the long-lived modes such as those that appear in hydrodynamics are largely universal, and insensitive to the detailed microscopic description.  We will study thus a class of designer scalar probes where the dilaton, which governs the coupling of the probe to the geometry, is a simple function of the radial holographic coordinate and is characterized by $\ann$. The class of models we study has been previously examined in the holographic literature in \cite{Chamblin:1999ya,Charmousis:2010zz,Iizuka:2011hg} in the context of holographic RG flows and applications to AdS/CMT.\footnote{ We will also demonstrate that the models we consider may equivalently be  viewed in terms of the  dynamics of scalar probes in thermal hyperscaling violating backgrounds, cf., \cref{sec:desis}.}

While the designer scalar parameterized by the Markovianity index  is a convenient proxy for analyzing the dynamics of non-Markovian probes, it also  allows us to capture the physics  of gauge invariant degrees of freedom. Specifically,  we will  demonstrate that the dynamics of holographic diffusion, be it charge or momentum, can be mapped to the study of such scalars. This can be engineered by organizing the perturbation of spin-1 gauge fields and spin-2 gravitons in gauge invariant combinations, e.g., as  carried out in \cite{Kodama:2003jz,Kodama:2003kk} in the study of black hole perturbations. The designer scalar actions (along with variational boundary terms and boundary counterterms) naturally descend from the underlying Maxwell or Einstein-Hilbert dynamics for spin-1 and spin-2 fields respectively. In other words, the scalar probes we study are effectively coupling to particular polarizations of the boundary charge current or  energy-momentum tensor. The Markovianity index for charge diffusion is $\ann = -(d-3)$ while that for momentum diffusion is $\ann = -(d-1)$. 

This gauge agnostic treatment has several advantages compared to earlier studies. In the fluid gravity correspondence literature one treats the radial Gauss constraint(s) distinctly from the radial evolution equations. Solving the radial evolution equations usually takes into account the effects of the fast Markovian modes whereas the hydrodynamics of the long-lived non-Markovian modes is encoded within the conservation laws inherent in radial Gauss constraints. For example, in the gravitational problem the diffeomorphism Gauss constraints on the radial hyper-surface become the Navier-Stokes equations for the dual CFT plasma. In contrast, we will solve all the bulk equations of motion including the radial Gauss constraints in our discussion below. We will still keep the non-Markovian fluid modes off-shell  by turning on appropriate non-normalizable modes.

There is another way to understand the inefficacy of the radial gauge for gauge and gravitational dynamics. The grSK geometry has two timelike boundaries which correspond to the L and R (or bra and ket) components of the boundary Schwinger-Keldysh contour. However, these are connected through the AdS bulk  -- one therefore finds only a single radial gauge constraint. This automatically puts the difference Schwinger-Keldysh fields on-shell and thus imposes the conservation of the difference currents. This is a problem: if we attempt to derive a generating function for current correlators we end up missing a degree of freedom from the difference current. One way to proceed is to attempt  to take this difference current off-shell. This, as far we understand, is the proposal of \cite{Glorioso:2018mmw}, who do so by imposing an explicit source for the difference fields at the horizon. However, it makes more physical sense to let the gravitational dynamics lead the way and impose all the bulk equations of motion.  Then one has a more natural parameterization in terms of the boundary expectation values.

Having motivated the study of the designer scalar system, and its connection to the dynamics of conserved currents of a holographic thermal baths, let us now describe what should one compute, given the dynamics. For the Markovian sector with $\ann > -1$ it is clear that one can follow as in \cite{Jana:2020vyx} the usual rules of the AdS/CFT correspondence, now uplifted to the grSK geometry, and compute the generating function of real-time correlation functions (with Schwinger-Keldysh time ordering). For non-Markovian probes such a generating function is guaranteed to be non-local -- one is integrating out long-lived modes.  While this would be a fine approach to take, it is more conducive in the spirit of EFT to ask if there is an alternative that allows us to derive a local effective action for non-Markovian probes. We will argue that there indeed is one, which is obtained by a Legendre transformation of the generating function of connected correlators into a Wilsonian SK action parameterized  by the non-Markovian operators (or their expectation values in AdS/CFT parlance). We will demonstrate that such a Wilsonian influence functional for real-time non-Markovian observables is well defined and obtain an expression for the same at the quadratic order (in amplitudes). From this functional one can of course obtain the real-time correlators for the non-Markovian fields. We will indeed recover the known physics of diffusion as well as the stochastic noise associated thereto.

\subsection{Synopsis of salient results}
\label{sec:summary}

To aid the reader in navigating the paper let us quickly record some of the salient  results we  derive in the discussion below:  

\begin{itemize}[wide,left=0pt]
\item We obtain the expected diffusive dynamics of charge and momentum  recovering the expected form of  the retarded Green's functions of charge  and energy-momentum  current. Specifically, for momentum diffusion  we find the dispersion:
\begin{equation}\label{eq:Tdisperse}
\begin{split}
0 = 
	i\, \omega- \frac{\beta}{4\pi} \,k^2 + \frac{\beta}{4\pi} \, \harm{\frac{2}{d}-2} \omega^2  + \cdots 
\end{split}	
\end{equation}	
which gives the famous shear diffusion pole. The coefficient of the $\omega^2$ contains contributions from two hydrodynamic transport  coefficients $k_R$ and $k_\sigma$ in the notation of \cite{Haehl:2015pja}. As generalizations, we compute the next order  (cubic in gradients) term in the retarded  Green's function and also obtain results that pertain to probes of hyperscaling violating backgrounds at finite temperature, cf., \eqref{eq:disp3} for the general result parameterized by the Markovianity index $\ann$. 
\item One may  interpret our results for the retarded Green's function in terms of the  greybody factors  of the planar \SAdS{d+1} black holes for photons and gravitons, accurate to cubic order in a (boundary) gradient expansion. The relevant expressions are given in 
\eqref{eq:WIFMax} and \eqref{eq:WIFEin}, for photons and gravitons respectively. Our results match with earlier derivations from holography \cite{Policastro:2002tn,Baier:2007ix,Arnold:2011ja} for energy momentum and charge correlators. Explicit expressions for the stress tensor correlators of  thermal $\mathcal{N}=4$ SYM plasma are given in \eqref{eq:N4TenRet} and \eqref{eq:N4VecRet}, respectively.
\item In addition we also derive the fluctuations associated with the dissipative modes using the Schwinger-Keldysh formalism. We obtain the leading contribution  (at quadratic order) to the Hawking noise correlations associated with  thermal photons and gravitons. We package this information in the Wilsonian influence functional as mentioned above. Explicit expressions for stress tensor correlators for $\mathcal{N}=4$ SYM plasma can be found in \eqref{eq:N4TenKel} and \eqref{eq:N4VecKel}, respectively.
\item One of the  key technical insights  of the  discussion is a  clean  separation in systems with gauge invariance of the Markovian  and non-Markovian degrees of freedom.  For instance, for stress tensor dynamics we show that the non-Markovian transverse momentum diffusion modes can be cleanly separated from the Markovian transverse  tensor polarizations. This separation uses a classic gauge invariant decomposition of linearized perturbations (cf., \cite{Kodama:2003jz}) which has been employed in various studies of gravitational problems over the years.
\item Finally, a corollary  of our  work is a derivation of an effective action for diffusive modes. In particular, we verify  the structural form of the class L action for non-dissipative coefficients conjectured from fluid/gravity data in \cite{Haehl:2015pja}. 
\end{itemize}

While we are not the first to ask these questions our treatment of  the problem will be quite different from the discussion in the literature hitherto for the most part. For  instance, the semi-holographic models of \cite{Faulkner:2010tq}  provided a template for analyzing the effective dynamics of holographic systems with long-lived excitations. This motivated \cite{Nickel:2010pr} to provide a framework for  understanding holographic liquids. Furthermore,  \cite{Crossley:2015tka,deBoer:2015ija} attempted to derive effective actions for hydrodynamics using holography (primarily in the non-dissipative sector at  the ideal fluid level). Closely related to our present discussion is the work of \cite{Glorioso:2018mmw,deBoer:2018qqm} who used the grSK geometry to study effective dynamics of charge diffusion. Our treatment here while having some elements of commonality,  differs substantially in that our primary focus is on the physics of outgoing Hawking modes.\footnote{On the field theory side there have been many constructions of the SK effective action for the hydrodynamic modes, see eg.,  \cite{Kovtun:2012rj,Grozdanov:2013dba,Kovtun:2014hpa,Haehl:2014zda,Crossley:2015evo,Haehl:2015uoc,Jensen:2017kzi,Haehl:2018lcu,Jensen:2018hse,Chen-Lin:2018kfl}. At the quadratic order, our effective action agrees trivially with all these constructions. Since the non-trivial aspects of these constructions appear only at the interacting level, we will postpone a detailed comparison to these formalisms to future.\label{fn:hydroactions}
}  However, for the most part we will carefully analyze the physics of infalling quanta. Once  we understand this  in a gauge agnostic manner, by exploiting the time-reversal isometry mentioned above, we  can extract the outgoing modes efficiently. 

In other words, we seek a treatment of the problem, that  respects  time-reversal properties, has a sensible local gradient expansion amenable to effective field theory analysis, and captures  the physics of long-term black hole memory. The fact that  all  of these  can  be done self-consistently is a central  thesis of  our work.

\subsection{Outline of the paper}
\label{sec:}

The outline  of  the  paper  is as follows: in  \cref{sec:grsk} we review some of  the salient aspects of  the grSK geometries summarizing the salient features described in  \cite{Jana:2020vyx}. We introduce then in \cref{sec:designer} the class of designer fields which we use to  characterize Markovian and non-Markovian dynamics. The reader will find a succinct summary here of  the connection to probes coupled to conserved  currents. The central thesis of our  discussion of memory is given in \cref{sec:trscalar}. Here using the auxiliary  scalar system we explain the salient differences between Markovian and non-Markovian probes, and explain how we should encode effectively  the physics of probes which  are sensitive to the long-term memory  of the black hole. \cref{sec:scalarM} re-purposes the discussion of \cite{Jana:2020vyx} to describe Markovian probes, while \cref{sec:scalarnM} deals with non-Markovian probes. As we describe in detail there, we do not resolve the dynamics of the non-Markovian probes, but rather, exploit an analytic continuation in the Markovianity  index to extract the necessary physical information. In \cref{sec:grSKsol} we put  together the intuition gleaned from the designer scalar analysis and compute the two-point real-time correlation functions for both Markovian and non-Markovian probes. In \cref{sec:trgauge} we finally turn to probe gauge fields and demonstrate that a suitable gauge invariant decomposition allows us to  map their dynamics to the designer scalars, and use this intuition to extract the physics  of current correlators in the holographic field theory. In \cref{sec:gravity} we turn to the dynamics of transverse tensor and vector gravitons and show that  they  too can be mapped  to our designer scalars. One of  our central results which  we describe  in this section  is the quadratic fluctuations (including stochastic noise) of gravitons in the black brane background. We end with a brief discussion  in \cref{sec:discuss}.

There  are several supplemental appendices which contain some  of the technical details. \cref{sec:descanal} contains various details related to our designer scalar system which we draw upon in \cref{sec:scalarM} and \cref{sec:scalarnM}. Some of the details relating to gauge field probes can be found in \cref{sec:gaugeapp}, where we also describe the physics in the (suboptimal) radial gauge. The reduction of gravitational dynamics onto our designer systems is explained in \cref{sec:appgravity}. In \cref{sec:normalizable} we give the expressions for the conserved currents in terms of the designer scalar fields, which may be employed to directly compute the current correlation functions.

The final two appendices \cref{sec:ibf} and \cref{sec:harmonics} contain salient details of incomplete beta functions and plane wave harmonics, that we employ in our analysis throughout.

\section{Review of the grSK geometry}
\label{sec:grsk}

\begin{figure}[h!]
\begin{center}
\begin{tikzpicture}[scale=0.6]
\draw[thick,color=rust,fill=rust] (-5,0) circle (0.45ex);
\draw[thick,color=black,fill=black] (5,1) circle (0.45ex);
\draw[thick,color=black,fill=black] (5,-1) circle (0.45ex);
\draw[very thick,snake it, color=orange] (-5,0) node [below] {$\scriptstyle{r_h}$} -- (5,0) node [right] {$\scriptstyle{r_c}$};
\draw[thick,color=black, ->-] (5,1)  node [right] {$\scriptstyle{r_c+i\varepsilon}$} -- (0,1) node [above] {$\scriptstyle{\Re(\ctor) =0}$} -- (-4,1);
\draw[thick,color=black,->-] (-4,-1) -- (0,-1) node [below] {$\scriptstyle{\Re(\ctor) =1}$} -- (5,-1) node [right] {$\scriptstyle{r_c-i\varepsilon}$};
\draw[thick,color=black,->-] (-4,1) arc (45:315:1.414);
\draw[thin, color=black,  ->] (9,-0.5) -- (9,0.5) node [above] {$\scriptstyle{\Im(r)}$};
\draw[thin, color=black,  ->] (9,-0.5) -- (10,-0.5) node [right] {$\scriptstyle{\Re(r)}$};  
\end{tikzpicture}
\caption{ The complex $r$ plane with the locations of the two boundaries and the horizon marked. The grSK contour is a codimension-1 surface in this plane (drawn at fixed $v$). As indicated the direction of the contour is counter-clockwise and it encircles the branch point at the horizon.}
\label{fig:mockt}
\end{center}
\end{figure}
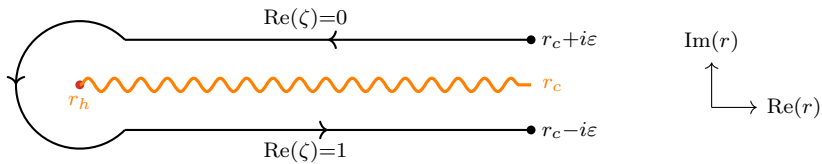

The grSK geometry dual to the Schwinger-Keldysh contour of a $d$ dimensional field theory studied in \cite{Glorioso:2018mmw,Chakrabarty:2019aeu,Jana:2020vyx} is characterized by the metric (we will mostly follow the conventions introduced in \cite{Jana:2020vyx})\footnote{A useful identity which helps in various simplifications is $x \, \dv{x}f(x) = d \, (1-f(x))$. This also explains why we choose to parameterize the horizon radius $r_h = \frac{1}{b}$ as in \cite{Bhattacharyya:2008jc}.}
\begin{equation}\label{eq:sadsct}
ds^2 = -r^2\, f(br) \, dv^2 +  i\, \beta\, r^2 \, f(br)\,  dv\, d\ctor + r^2\, d\vb{x}^2 \,, \qquad f(\br) = 1 - \frac{1}{\br^d} \,.
\end{equation}	
Here $\ctor$ is the \emph{mock tortoise coordinate}, which parameterizes a curve in the complexified radial plane as we illustrate in 
\cref{fig:mockt}. The coordinate $\ctor$ is defined by the  differential relation ($\beta$ is the inverse temperature)
\begin{equation}\label{eq:ctordef}
\frac{dr}{d\ctor} = \frac{i\,\beta}{2} \, r^2\, f(br) \,, \qquad \beta = \frac{4\pi b}{d} \equiv \frac{4\pi}{d \,r_h}\,,
\end{equation}	
subject to the following boundary conditions at the cut-off surface $r=r_c$
\begin{equation}\label{eq:ctorbc}
\ctor(r_c+i\,\varepsilon) = 0 \,, \qquad \ctor(r_c-i\,\varepsilon) = 1\,.
\end{equation}	

To understand this geometry, consider the familiar planar-\SAdS{d+1} geometry in ingoing Eddington-Finkelstein coordinates:
\begin{equation}\label{eq:efads}
ds^2 = -r^2\, f(br) \, dv^2 + 2\, dv\, dr + r^2\, d\vb{x}^2 \,.
\end{equation}	
The asymptotic timelike boundary has a time coordinate $v$, which for the Schwinger-Keldysh path integral contour should be viewed as a curve in the complex time domain. The grSK spacetime fills in this Schwinger-Keldysh contour with a metric of constant negative curvature and provides a saddle point for the real-time gravitational path integral.\footnote{To be clear, here we only refer to the fact that given the asymptotic boundary conditions the grSK geometry is a smooth spacetime satisfying the Einstein's equations with the prescribed boundary conditions. It is an open question whether the geometry is the unique or dominant saddle, but the fact that it gives results consistent with field theory expectations \cite{Glorioso:2018mmw,Chakrabarty:2019aeu,Jana:2020vyx} strongly argues in its favour. For further comments about real-time gravitational saddles see \cite{Colin-Ellerin:2020mva}.} The spacetime as such comprises of two copies of the planar black hole: one at $\Re (\ctor )=0$ and another at $\Re(\ctor) =1$ glued smoothly across a horizon-cap. For earlier discussions of this construction see \cite{vanRees:2009rw}. In what follows we will find the following non-dimensional radial coordinates helpful to simplify expressions:\footnote{ Our parameterization here differs from the choice made in \cite{Jana:2020vyx} which used $\varrho  = \frac{1}{\ri}$. The advantage of $\ri$ is that we will be able to express various functions in their principal branch since $\ri \in (0,1]$ in the domain of outer communication.}
\begin{equation}\label{eq:rhodef}
\br \equiv  b\,r \,, \qquad 
\ri \equiv \left(\frac{1}{b\,r} \right)^d =\br^{-d}
 \;\; \Longrightarrow \;\; \dv{}{\ctor} = -2\pi i\, \ri^{1-\frac{1}{d}} \,  (1-\-\ri) \, \dv{}{\ri}\,.
\end{equation}	

As noted in \cite{Jana:2020vyx} one can obtain a closed form expression for $\ctor(r)$ by solving \eqref{eq:ctordef} in terms of a hypergeometric function. We will find it convenient to switch to a parameterization in terms of the incomplete Beta function 
$\ibf{p,q;z}$, see \cite[section 8.17]{NIST:DLMF}.  We have the formal solution
\begin{equation}\label{eq:ctorbeta}
\ctor(r) = \frac{d}{2\pi i}\, \int_{\infty+ i0}^{\br} \;  \frac{y^{d-2}\, dy}{y^d-1}  = -\frac{1}{2\pi i} \, \ibf{\frac{1}{d}, 0; \left(\frac{1}{\br}\right)^d} \equiv 
-\frac{1}{2\pi i}\, \ibf{\od, 0; \ri} \,,
\end{equation}	
where we have introduced a useful shorthand 
\begin{equation}\label{eq:oddef}
\od \equiv \frac{1}{d} \,.
\end{equation}	

It will be convenient to introduce certain derivative operators that will appear extensively in our analysis. We define:
\begin{equation}\label{eq:Dz}
\Dz_\pm = r^2\, f(r) \, \pdv{}{r} \pm \, \pdv{}{v} \,, \qquad \Dz_\pm = r^2\, f(r) \, \pdv{}{r} \mp \, i \, \omega \,, 
\end{equation}	
in the time and frequency domain, respectively.  The above operator is related to the derivation $\Dzeta^\pm$ introduced in \cite{Jana:2020vyx} by a constant rescaling $\Dzeta^\pm = \frac{i\beta}{2}\, \Dz_\pm$ and avoids the occurrence of  factors of $i$ in various expressions.
 
While the choice of ingoing coordinates breaks the explicit time-reversal invariance, the geometry retains a time-reversal $\mathbb{Z}_2$ isometry: the transformation $v\mapsto i\beta\ctor-v$ preserves the form of the metric. The  operator $ \Dz_+$ is  naturally covariant under this isometry as can be seen  from the identity
\begin{equation}\label{eq:trops}
\begin{split}
dr\, \partial_r+dv\,\partial_v+dx^i\,\partial_i
&= \frac{dr}{r^2f} \ \Dz_++\left(dv-\frac{dr}{r^2f}\right)\partial_v +dx^i\,\partial_i \\
&=\frac{i\beta}{2}d\ctor\ \Dz_+ +\left(dv-\frac{i\beta}{2}d\ctor\right)\partial_v +dx^i\partial_i\ ,
\end{split}
\end{equation}
where we have used the real domain expression $\Dz_+=r^2 f\partial_r+  \partial_v$. The 1-forms that appear in r.h.s.,
$\{\frac{dr}{r^2 f} , dv - \frac{dr}{r^2 f}, dx^i\}$ furnish a basis of cotangent space that is covariant under the time-reversal $\mathbb{Z}_2$. It follows that the dual derivative operators $\{\Dz_+,\partial_v, \partial_i\}$ furnish a natural basis of the bulk tangent space covariant under time-reversal.

\paragraph{A word on our conventions:} Uppercase Latin indices are used for the bulk AdS spacetime, with lowercase Greek indices reserved for the timelike boundary. Spatial directions along the boundary are further indexed by lowercase mid-alphabet Latin indices. $g_{AB}$ is the bulk metric, $\gamma_{\mu\nu}$ the induced boundary metric, and $n^A$ the unit normal to the boundary.

\section{Designing gravitational probes with memory}
\label{sec:designer}

The \SAdS{d+1} provides for us a holographic thermal bath which we wish to probe. Our primary interest is in understanding probes that couple to the conserved currents, global charge current or energy-momentum current, of the dual field theory which exhibit long-lived  diffusive behaviour. It will transpire that one can give a unified presentation in terms of a scalar probe of the grSK background, albeit one that not only couples to the gravitational background, but also to an auxiliary dilaton. We will demonstrate below that the conserved current dynamics can be mapped directly on this scalar system for a specific choice of the dilaton profile which depends on the nature of the current.

\subsection{Designer scalar and gauge probes}
\label{sec:desis}
Our prototypical holographic probe field is  what we call a designer scalar. It is a massless Klein-Gordon field $\sen{\ann}$ coupled minimally to gravity and to a dilaton $\dils$. Specifically, consider an action of the form
\begin{equation}\label{eq:KGDil}
\begin{split}
S_{ds}  = -\frac{1}{2} \, \int \, d^{d+1}x \sqrt{-g}\ e^{\dils}\,  \nabla^A\sen{\ann} \nabla_A\sen{\ann}+ S_{bdy}\,, \qquad e^{\chi_s} \equiv r^{\ann+1-d}  
\end{split}
\end{equation}
The auxiliary dilaton which depends on a designer parameter $\ann$ serves the purpose of modulating the coupling  of our probe as a function of the energy scale.\footnote{Choosing $\ann = d-1$ gives us a massless minimally coupled scalar, which was studied in \cite{Jana:2020vyx}.}  Depending on the sign of $\ann$ the coupling of the scalar to the geometry can be modulated across the energy scale. For $\ann$ sufficiently negative the scale is weakly coupled deep in the bulk relative to its dynamics in the asymptotic region. This mimics the modes of the dual field theory which, say like the hydrodynamic modes, represent a small number of UV degrees of freedom, but a large number of IR degrees of freedom. This heuristic is based on the UV/IR dictionary of the AdS/CFT duality \cite{Susskind:1998dq} along with the idea that bulk couplings   roughly correspond inversely to the number of CFT degrees of freedom.  

In Fourier domain, the action \eqref{eq:KGDil} gives rise to an equation of motion of the form\footnote{We have  yet  to formulate the variational principle, which involves specifying the boundary terms for \eqref{eq:KGDil}. In addition we will also have to introduce boundary counterterms. We will address these terms in due course.}
\begin{equation}\label{eq:gseom1}
\begin{split}
\frac{1}{r^\ann} \Dz_+  \left( r^\ann\, \mathbb{D}_+\sen{\ann}\right)+\left(\omega^2-k^2f\right) \sen{\ann}=0\ 
\end{split}
\end{equation}
We will argue that for sufficiently negative $\ann$, this system does indeed exhibit long-lived correlations and hence gives a simple toy model for a bath with a memory. While this might seem like an oversimplified model with no gauge invariance etc., we will see that this system indeed captures the essence of many realistic gauge systems. 

A second probe which we will examine in detail is a bulk Maxwell system, also dilatonically coupled, designed with an action of  the form
\begin{equation}\label{eq:MaxDil}
\begin{split}
S_{dv} =  -\frac{1}{4} \, \int \, d^{d+1} x\, \sqrt{-g}\ e^{\dilv} \,\Cm^{AB}\Cm_{AB}\,, \qquad e^{\dilv} \equiv r^2\, e^{\dils} = r^{\ann+3-d} 
\end{split}
\end{equation}
where $\Cm_{AB} = 2\,\nabla_{[A} \Vm_{B]}$ denotes the Maxwell field strength and $\Vm$ the 1-form potential. We will demonstrate how this system  with gauge invariance also exhibits long-lived correlations associated with the physics of charge diffusion in the dual CFT. In particular, we note that setting $\ann=d-3$ corresponds to the standard Maxwell system. Furthermore, in a precise sense, we will exhibit how this gauge system can be reduced to the scalar problem \eqref{eq:KGDil}. This explains why we continue to employ the same symbol $\ann$ to characterize the Markovianity index for the designer vector.

Both the scalar and the gauge field will be taken to probe the grSK saddle geometry \eqref{eq:sadsct}. On this two sheeted geometry the Lagrangian density for the systems is integrated over the grSK contour depicted in \cref{fig:mockt}. Effectively, this means that the radial integral  is morphed into a contour integral $\int dr \mapsto \oint d\ctor$ for the real-time part of the evolution. We won't write this out at the moment, since for the most part our analysis can be done on a single sheet and thence upgraded onto the two-sheeted grSK contour.

\paragraph{Designer probes and hyperscaling violation:}
There is an useful metaphor that helps understand our designer scalar system and sharpens the intuition that the parameter $\ann$ allows for tuning the relative coupling of low and high  energy modes of the field to the geometry. Our designer scalar can be mapped onto the dynamics of a massless scalar probing a thermal state in a Lorentzian hyperscaling violating background with $z=1$. More specifically, 
consider the following  Einstein-dilaton action in $\bar{d}+1$ dimensions which has been analyzed extensively in 
\cite{Charmousis:2010zz} (see also \cite{Chamblin:1999ya}):
\begin{equation}\label{eq:EDaux}
\begin{split}
S_{ED} &= \frac{1}{16\pi G_N} \, \int d^{\bar{d}+1} x \; \sqrt{-\bar{g}}  \left(R+ V_0\, e^{\bar{b}\,\chi} -\frac{1}{2} \, \bar{g}^{AB}\, 
\partial_A\chi\, \partial_B\chi\right)\\
\text{with}&  \qquad
V_0 = (\bar{d}+1-\theta)(\bar{d}-\theta) \,, \qquad \bar{b}^2 = \frac{2\theta}{\bar{d}(\theta-\bar{d})} \,.
\end{split}
\end{equation}
which has a black hole solution 
\begin{equation}\label{eq:z1hypscale}
\begin{split}
d\bar{s}^2 &=  r^{-2\frac{\theta}{\bar{d}}} \left( -f(br)\, r^2 \, dv^2 + 2 \, dv\, dr + r^2 \,d\vb{x}^2_{\bar{d}}\right) \\
e^{\bar{b}\,\chi} &= r^{-2\frac{\theta}{\bar{d}-1}} \,, \qquad f(b r) = 1-\left(\frac{1}{br}\right)^{\bar{d}+1-\theta}\,.
\end{split}
\end{equation}	
A massless scalar $\sen{\ann}$ probing the geometry \eqref{eq:z1hypscale} with dynamics
\begin{equation}\label{eq:}
	\int d^{\bar{d}+1} x\, \sqrt{-\bar{g}} \, e^{\bar{\alpha}  \,\bar{b}\, \chi}\, \bar{g}^{AB}\, \partial_A \sen{\ann} \, \partial_B \sen{\ann}\,,
\end{equation}	
would map to the action \eqref{eq:KGDil}  with the identifications:
\begin{equation}\label{eq:EDKGmap}
\bar{d} -\theta = d \,, \qquad \ann =  \bar{d} -1-\theta  \left( 1+ \frac{2\,\bar{\alpha}}{\bar{d}-1}\right)  = d-1-\frac{2\,\bar{\alpha}\,\theta}{\theta+d-1}\,.
 \end{equation}	
As discussed extensively in the AdS/CMT literature (cf., \cite{Hartnoll:2016apf} for an excellent review) the Lorentzian hyperscaling violating geometries can be interpreted as RG flows arising when we turn on dangerously irrelevant  primaries. The IR dynamics in these examples strongly deviates from the UV leading to emergent scaling dynamics at low energies. The construction of designer probes is also reminiscent of attempts to understand holographic duals of non-conformal branes \cite{Kanitscheider:2008kd} and Ricci-flat holography \cite{Caldarelli:2013aaa} using generalized dimensional reduction \cite{Gouteraux:2011qh}.

\subsection{Origins of the designer fields}
\label{sec:desorigin}

Before we proceed, let us explain the rationale behind our choice of the designer fields. The designer scalar and gauge field arise quite naturally when we consider the dynamics of Maxwell fields and gravitons in the \SAdS{d+1} background. We give  here a quick synopsis of the salient statements deferring the details to subsequent sections.\footnote{ Our statement that $\ann=-(d-3)$ in the scalar sector of gravity requires careful interpretation, as  it primarily refers to the relative fall-off between modes with $k\neq  0$. We will explore this  sector which has qualitatively different physics in  a  later  publication.}

\begin{table}[h!]
\centering
\begin{tabular}{|l|c|c|c|}
\hline \hline
 \shadeR{massless field} & \shadeR{Scalar sector} & \shadeR{Vector sector} & \shadeR{Tensor sector} \\
 \hline \hline
Scalar & $d-1$ & $-$ & $-$ \\  
 \hline
Gauge field & $-(d-3)$ & $d-3$ & $-$ \\   
 \hline
Metric & $-(d-3)$ & $-(d-1)$ & $d-1$ \\  
  \hline
Nambu-Goto (linearized)& \multicolumn{3}{|c|}{2}\\  
\hline
\end{tabular}
\caption{Values of $\ann$ obtained for linearized perturbations of  various fields of interest in \SAdS{d+1} background. Note that both positive and negative values of $\ann$ are thus obtained.}
\label{tab:Nval}
\end{table}

We give a quick reference summary of the values of $\ann$ obtained from the study of various massless fields in the \SAdS{d+1} background in \cref{tab:Nval}. More specifically, the following statements hold, as we will demonstrate in due course:
\begin{enumerate}[wide,left=0pt]
\item The field $ \int_k \, \sen{d-1} \, \ScS$ solves the massless Klein-Gordon equation. Here $\ScS = e^{-i\omega v + i\, \bk\cdot \bx}$ is the plane wave on $\mathbb{R}^{d-1,1}$.  In particular, its definition includes the $e^{-i\omega v}$ frequency dependence.\footnote{Our conventions for the planar harmonics are described in \cref{sec:harmonics}.}  We also introduce a short-hand notation for momentum integrals: 
\begin{equation}
\int_{k}\equiv \int \frac{d\omega}{2\pi}\int \frac{d^{d-1}k}{(2\pi)^{d-1}}\ .
\end{equation}

\item\label{it:max} The 1-form $\Vm$ parameterized as 
\begin{equation}\label{eq:maxVpar}
\Vm_A\,dx^A = \int_k 
	\left[\frac{1}{r^{d-3}}\left( dv\ \Dz_+-dr\, \dv{r} \right) \sen{3-d}\,\ScS+ \sum_{\ai=1}^{N_V} \sen{d-3}^{\ai}\,  \VV_i^\ai dx^i \right]
\end{equation}
solves the Maxwell equations \cref{sec:trgauge}.   

Here  $\VV_i^\ai$ denote the $N_V=(d-2)$ transverse planar vector waves  
with $k_i \VV_i=0$ on $\mathbb{R}^{d-1,1}$. They transform in  the spin-1 representation of $\mathrm{SO}(d-2)$ transverse to $\bk$ and represent electromagnetic waves polarized along  black brane which quickly fall into the brane.  
On the other hand,  $\sen{3-d}$ corresponds to the radially polarized electromagnetic wave grazing the black brane. The latter  is  a long-lived (non-Markovian) mode  dual to the charge diffusion mode in the CFT.

\item\label{it:grav} Finally, the spin-2 symmetric tensor combination  
\begin{equation}\label{eq:grvspar}
\delta g _{AB}\, dx^A \, dx^B = 
	r^2\, \int_k 
		\left( 2\ \frac{ dx^i}{r^{d-1}}\left(  dv\ \Dz_+-dr \, \dv{r} \right)\ 
		\sum_{\ai=1}^{N_V} \, \sen{1-d}^{\ai} \,\VV^\ai_i+\sum_{\bi=1}^{N_T} \, \sen{d-1}^\bi\, \TT_{ij}^\bi \,dx^i\, dx^j + \ldots\right) 
 \end{equation}
 solves the linearized Einstein equations \cref{sec:gravity}.  The ellipses denotes the scalar polarizations, dual to the CFT  sound mode,  which we ignore in our discussion.

 We have introduced  $\TT_{ij}^\bi$  which  are the  transverse, symmetric, trace-free, tensor plane waves  on $\mathbb{R}^{d-1,1}$ with $k_i \,\TT_{ij}^\bi=0$ and $\TT_{ii}^\bi=0$.  They form a spin-two representation of   $\mathrm{SO}(d-2)$  transverse to $\bk$. There are  $N_T=\frac{1}{2}d(d-3)$ such linearly independent tensor waves and they represent  the in-falling graviton modes which decay very quickly. In contrast, $\sen{1-d}^\ai$ corresponds to the  graviton modes grazing the black brane. They are long-lived (non-Markovian) and are dual to the momentum diffusion/shear modes in the CFT.
\item  A corollary of points \ref{it:max} and \ref{it:grav} above is that the gravitational perturbations in the tensor and vector sector can be expressed in terms of a Klein-Gordon scalar and  designer vector both with $e^{\chi_s} = 1$ and $e^{\chi_v} =r^2$, respectively.
\end{enumerate}

The proof of these statements will be given in \cref{sec:trgauge,sec:gravity} and \cref{sec:gaugeapp,sec:appgravity}. The reader familiar with the study of linearized gravity in terms of the diffeomorphism invariant combinations starting from the early work of Regge-Wheeler-Vishweshwara-Zerelli \cite{Regge:1957td,Vishveshwara:1970cc,Zerilli:1970se,Zerilli:1971wd} and the more recent analysis of Kodama-Ishibashi \cite{Kodama:2003jz,Kodama:2003kk} will find our parameterization natural.  These have been adopted in the AdS/CFT literature in the course of the study of quasinormal modes and linearized hydrodynamics \cite{Kovtun:2005ev,Michalogiorgakis:2006jc,Gubser:2007nd,Morgan:2009pn,Diles:2019uft}. We will outline the connection with these works when we explain the derivation of the designer system. While the discussion in the main  text will  be in terms of gauge  invariant variables,  we  have chosen to write the gauge field and metric in  \eqref{eq:maxVpar} and \eqref{eq:grvspar}  in a particularly convenient  gauge  fixed  form (in a Debye gauge).\footnote{ Similar statements hold in global \AdS{d+1} where  the different gauge and gravitational perturbations can be reduced to  scalar fields with different Ricci-shifted mass terms \cite{Ishibashi:2004wx} (cf., \cite{Kodama:2003jz,Kodama:2003kk}). Note that therefore our decomposition is background dependent; the dynamical content captured in our designer scalars should be viewed as being tailored to the specific background of interest.}

We note that the perturbations \eqref{eq:maxVpar} and \eqref{eq:grvspar} are written in terms of the time-reversal covariant derivations introduced in \cref{sec:grsk}, see \eqref{eq:trops}.  In particular, the derivative operator  $  dv\, \Dz_+-dr\, \partial_r$ is the natural  operator built of $\mathbb{Z}_2$ covariant 1-forms and $\mathbb{Z}_2$ covariant derivative operators. This operator  is odd under the  time-reversal $\mathbb{Z}_2$ isometry $v\mapsto i\beta\ctor-v$. This fact will play a crucial role in our analysis.

The field equations for the designer fields under discussion are also invariant under this isometry. 
More precisely, one gets an action of this $\mathbb{Z}_2$ on the space of solutions as follows: if 
$\sen{\ann}(\ctor,\omega,\vb{k})$ is a solution of \eqref{eq:gseom1}, then it can be shown that $
e^{-\beta\omega\ctor} \,\sen{\ann}(\ctor,-\omega,\vb{k})$  is also a solution \cite{Jana:2020vyx}. 
Depending on the physical problem under study, the field $\sen{\ann}$ can also have an \emph{intrinsic time-reversal parity} which 
means that we will sometimes define the time-reversed solution with an extra minus sign, i.e., we take  
$-e^{-\beta\omega\ctor}\,  \sen{\ann}(\ctor,-\omega,\vb{k})$
to be the time reversed solution. In the example above, given that $  dv\, \Dz_+-dr \,\partial_r$ is odd under time-reversal, the intrinsic time-reversal parity of  $ \sen{3-d}$ and $ \sen{d-3}$ are opposite to each other. Similarly   $\sen{1-d}$ and $ \sen{d-1}$ fields above should also have opposite  intrinsic time-reversal parity.

\section{Markovianity and lack thereof: memories lost and regained}
\label{sec:trscalar}

Having introduced our two designer systems, we will now examine the central features of each of them. It will suffice for us to explore the designer scalar dynamics in some detail, for it will transpire that the designer gauge fields can be mapped with a suitable parameterization onto designer scalars. In this section we will begin exploring some general features of the designer scalar, emphasizing the key differences between models with $\ann >-1$ and those with $\ann < -1$. We will argue that the former correspond to \emph{Markovian probes} while the latter are \emph{Non-Markovian probes} which carry non-trivial memory far into the future. 

Since our aim here is to delineate the salient, universal features of the model, we will analyze the wave equation \eqref{eq:gseom1} in the \SAdS{d+1} background directly in the standard radial coordinate parameterization. Translation to the grSK saddle will be straightforward once we  understand the central features of the solutions. To this end, we work with the wave equation written out more explicitly in the form:
\begin{equation}\label{eq:gsrad}
\begin{split}
\dv{r}( r^{\ann+2}\, f\, \dv{\sen{\ann}}{r} )-i\omega \left[\dv{r}\left(r^{\ann}\sen{\ann}\right)+r^{\ann}\dv{\sen{\ann}}{r} \right]
	-k^2\,r^{\ann-2} \sen{\ann} =0\ .
\end{split}
\end{equation}
%

\subsection{Analytic versus Monodromy modes and their interpretation} 
\label{sec:ammodes}

Let us first characterize the solutions of the differential equation \eqref{eq:gsrad}.  To begin with,  set $\omega, \vb{k}$ to zero and look for 
solutions that are constant along the boundary. We have then a homogeneous second order ordinary differential equation 
\begin{equation}\label{eq:gszeroeq}
\begin{split}
\dv{r} (r^{\ann+2}f\dv{r}\sen{\ann} )=0\,,
\end{split}
\end{equation}
 whose general solution is of the form  
 \begin{equation}\label{eq:gszeromode}
\begin{split}
\sen{\ann}^{(0)} 
&= 
	c_a + c_m \int_{\br_c+i0}^{\br} \frac{y^{d-2}\;dy}{y^\ann (y^d-1)}\,, 	\\
&=
	c_a + \tilde{c}_m \int^{\ri}_{\ri_c+i0} \, \frac{ \rib^{\,\od(\ann+1)} \, d\rib}{1-\rib}
\end{split}
\end{equation}
where $r_c \gg 1$ is the radial cutoff chosen to regulate the UV region ($\br_c$ and $\ri_c$ similarly defined).

Let us note  some crucial features of this solution:
\begin{itemize}[wide,left=0pt]
\item When $c_m=0$, the solution is analytic and ingoing at the future horizon. When uplifted to the grSK geometry by replacing $r(\ctor)$  this solution is smooth. In particular, it does not have a branch cut -- it is identical on both sheets of the grSK saddle.
\item When $c_m\neq 0$, the solution has a logarithmic branch cut at the horizon $r = \frac{1}{b}$. This is manifest from the change of variables to the dimensionless $\ri$ coordinate.  If we start off with a value of $(c_a,c_m)$ on one sheet of the grSK saddle,  then we pick up a monodromy  from the logarithmic branch-cut as we cross over to the other sheet. Specifically, $c_a$ gets shifted by an amount proportional to $c_m$, while $c_m$ is unchanged, under this crossing, i.e., 
\begin{equation}\label{eq:monodromy}
\JMar_\skR = \JMar_\skL + \frac{2\pi i}{d}\, c_m \,, \qquad c_a(\text{R}) = \JMar_\skR \,, \;\; c_a(\text{L}) = \JMar_\skL \,.
\end{equation}	
\end{itemize}
Hence we will refer to the two modes multiplying $c_a$ and $c_m$ as the analytic and monodromy modes, respectively.

Based on the above, we can equivalently parameterize the general solution of the homogeneous equation \eqref{eq:gszeroeq}  in terms of the boundary values $\JMar_\skL$ and $\JMar_\skR$ as
 \begin{equation}\label{eq:zeroRL}
\begin{split}
\sen{\ann}^{(0)} 
= \JMar_\skL +  \frac{d}{2\pi i} \, (\JMar_\skR-\JMar_\skL)\,   \int_{\br_c+i0}^{\br}  \frac{y^{d-2-\ann}\, dy}{y^d-1}\ .
\end{split}
\end{equation}
We note that the function multiplying $\JMar_\skR-\JMar_\skL$ behaves as $r^{-\ann-1}$ near conformal boundary of AdS. Therefore, turning on 
$\JMar_\skR-\JMar_\skL$ is equivalent to `dressing' the original  analytic solution by adding a solution which behaves as $r^{-\ann-1}$ near infinity. Clearly, the nature of the solution depends on whether the dressing decays or grows at large $r$.  This leads us to the two advertised cases of Markovian and non-Markovian dynamics, respectively.

\subsubsection{The Markovian case: \texorpdfstring{$\ann+1 >0$}{positive}}
\label{sec:markoff}

Let us first examine the case where $\ann +1>0$, which we term to be a Markovian probe, for reasons that will become clear soon.  For such a designer scalar turning on $\JMar_\skR-\JMar_\skL$  is equivalent to turning on a normalizable mode. The monodromy mode  sourced by  $\JMar_\skR-\JMar_\skL$   is sub-dominant at large $r$ relative to the analytic mode. We furthermore recall from \cite{Jana:2020vyx} which analyzed a massless Klein-Gordon field, a particular exemplar with $\ann = d-1 >0$, that turning on $\JMar_\skR-\JMar_\skL$   turns on Hawking radiation in the bulk which is ultimately a normalizable mode in the black brane background.

Let us understand the implications of this statement for the grSK saddle. An immediate consequence of the normalizability is that on the grSK geometry, a double Dirichlet boundary condition problem, which specifies  the coefficient of  non-normalizable mode near both the left and right boundaries, is  thus well-posed.  Specification of $\JMar_\skR$ and $\JMar_\skL$ uniquely determines the bulk solution. This should be viewed as the SK version of the standard boundary condition in AdS/CFT.  Moreover, inspection of the wave equation \eqref{eq:gsrad} makes clear that this conclusion holds even when $\omega, k \neq 0$ and thus the solutions to the differential equation can be obtained order by order in a long wavelength, low frequency, expansion (as illustrated in \cite{Jana:2020vyx} for massive Klein-Gordon fields).

The classical picture from the study of the wave equation can easily be upgraded to the quantum realm.  The designer system can be quantized by performing a path integral over the normalizable modes, taking them off-shell in the process, whilst leaving the left and right 
non-normalizable modes (the sources) frozen at the  respective boundaries. Since the normalizable modes, by definition, have a finite action (with the addition of suitable counterterms as usual), they contribute to the semiclassical path integral with an amplitude determined by the said action. Summing over all the off-shell modes leads to an answer which is a functional purely of the non-normalizable sources. This is the standard GKPW \cite{Gubser:1998bc,Witten:1998qj} dictionary of AdS/CFT applied to the grSK saddle, as indeed argued for in some detail in \cite{Jana:2020vyx} for some specific values of $\ann$. 

\subsubsection{The non-Markovian case:  \texorpdfstring{$\ann+1 \leq0$}{negative}}
\label{sec:nonmarkoff}

Let us turn to the case $\ann +1\leq 0$. Realize that turning on $\JMar_\skR-\JMar_\skL$ is now tantamount to turning on a 
 non-normalizable mode that \emph{grows} at infinity as $r^{\abs{\ann}-1}$. In the most general solution at zero derivative order,
 the coefficient of this growing solution, $c_m$ in \eqref{eq:gszeromode}, is the same in both left and the right branches
 of the grSK geometry. Thus, \emph{imposing double boundary conditions on non-normalizable modes has no solution at the zero 
 derivative order}! Again for reasons that will become clear shortly, we will term such a designer field with $\ann +1\leq 0$ where this
 phenomena  happens  as a non-Markovian probe.

How should one proceed in this case? One may wish to change the boundary condition to freeze the growing mode. This would clearly work and would involve changing the variational principle to keep the mode parameterized by $c_m$ to be frozen. Operationally this is a Legendre transformation on the space of boundary conditions, akin to the choice made for highly relevant operators \cite{Klebanov:1999tb} or multi-trace deformations in AdS/CFT \cite{Witten:2001ua}.  Indeed, explicit computations reveal that the non-normalizable modes start differing on the two branches as we go to higher orders. This would however have to be done at the expense of abandoning the gradient expansion and letting $c_m$ have non-analytic dependence on $\omega$ and $k$. With these changes we would be able to write down a solution with double boundary conditions on the non-normalizable modes. With these modifications one could come up with an `alternate' GKPW dictionary for such probes, enabling us thereby compute their Schwinger-Keldysh correlation functions.
 
 It is however worth reflecting on the physics prior to abandoning the gradient expansion altogether. The failure of  the long wavelength, low frequency expansion in the bulk, is indicative of the fact that there are slow propagating degrees of freedom in the probe. These prevent us from  approximating the correlators by contact terms  at long distances and times. In other words the system retains memory of its origins, justifying the characterization of such designer probes as \emph{non-Markovian}.\footnote{Analogously one justifies the characterization of the Markovian probes discussed in \cref{sec:markoff} -- such probes retain no memory and have no lasting effect on the system.}  Attempting to solve the bulk problem with double boundary conditions on non-normalizable modes is tantamount to integrating out  these slow degrees of freedom by going beyond the  derivative expansion.
 
But what if we did not want to integrate out these modes? We might want to freeze the slow modes and integrate out only the fast modes (as in the derivation of  the Born-Oppenheimer effective potential for nuclei). The latter can be tackled in a gradient expansion. With this motivation,  we ask what is the class of solutions in the bulk SK geometry where a derivative expansion is
 possible? As we will explain now, there is indeed such a class of solutions.

\subsection{The well of memory:  hydrodynamic moduli space}
\label{sec:hydromod}

As we have seen the non-Markovian probes are characterized by the presence of long-lived, low lying, Goldstone type modes which if integrated out lead to a non-local functional of the boundary sources (the generator of connected correlators). We will now give a precise characterization of how to deal with such modes and use them to define a low energy moduli space. Since such modes appear in the study of designer probes arising from gauge or gravitational perturbations, which in the dual field theory, corresponds to the dynamics of conserved currents, we will refer to the low energy space as the \emph{hydrodynamic moduli space}. 

\subsubsection{Analytic continuation into the hydrodynamic moduli space}
\label{sec:hmodc}

To begin with, a mathematical characterization of solution to the non-Markovian probe's equation of motion, may be obtained via an 
analytic continuation in the designer exponent $\ann$ from positive to negative values.  When $\ann$ is in the Markovian regime as argued in \cref{sec:markoff} we can set-up double Dirichlet boundary conditions on the (analytic) non-normalizable modes and a solution may be obtained order by order in the gradient expansion.   Furthermore, as in the fluid/gravity correspondence \cite{Bhattacharyya:2008jc,Hubeny:2011hd}, given slowly varying non-normalizable data on the two boundaries, call them $\JMar_\skR$ and $\JMar_\skL$, respectively, as in \eqref{eq:zeroRL}, one can write down a solution of the grSK geometry which admits a local series expansion in derivatives along the boundary. The main point to note here is that the normalizable modes are fixed completely at each order in this gradient expansion in terms of the pair of non-normalizable data ${\JMar_\skR, \JMar_\skL}$. For the non-Markovian probe we propose to  take the  Markovian solution for a given  $\ann > 1$ and analytically continue  it to a non-Markovian solution  with $\ann < -1$.\footnote{For the most part of our discussion we will focus on $\ann < -1$ non-Markovian fields, leaving a detailed analysis of the marginal case, $\ann = -1$, for the future. The issue in this case is that the asymptotic behaviour is analogous to minimally coupled massive scalars at the Breitenlohner-Freedman bound, viz., solutions with an admixture of a logarithmic mode. Note $\ann =-1$ is physically realized for a probe Maxwell field in \SAdS{5} background. }
  
The analytic continuation in the exponent $\ann$ from the Markovian to non-Markovian regime, roughly speaking,  flips the normalizable
and the non-normalizable pieces. More precisely, the normalizable modes on the Markovian side analytically continue to non-normalizable modes on the non-Markovian side. On the other hand, as we will argue now, the non-normalizable data, $\JMar_\skR$ and $\JMar_\skL$   on the Markovian side analytically continue into normalizable \emph{Wilsonian SK effective fields} $\snMar_\skR$ and $\snMar_\skL$, respectively, on the non-Markovian side.

To see this, note that by analytic continuation we do  indeed obtain a space of solutions for the non-Markovian system 
parameterized by a pair of boundary  data, $\{\snMar_\skR ,\snMar_\skL\}$. Moreover, the solutions are, by construction, obtained in a gradient expansion in these boundary  parameters.  This space of solutions constitutes  the hydrodynamic moduli space parameterized by the effective fields $\{\snMar_\skR ,\snMar_\skL\}$. In any such solution we can determine  the doubled non-normalizable data in terms of these hydrodynamic moduli.  This is equivalent to determining the hydrodynamic equations of motion for $\{\snMar_\skR ,\snMar_\skL\}$.  

 In fact, the non-normalizable mode is determined as a derivative operator acting on $\snMar_{\skR,\skL}$. This, in turn, means that if we want to set the non-normalizable modes to  zero (or any fixed value for that matter),  these fields should satisfy appropriate differential equations.  At a linearized level, in the Fourier domain, setting the derivative operator to zero results in a \emph{dispersion relation} for these fields parameterizing the boundary degrees of freedom of the non-Markovian field. We will see that the dispersion relation for the designer scalar takes the form of a diffusive mode,  of the form:
\begin{equation}\label{eq:diffdisp}
\omega = -i \, \frac{d}{4\pi \, (\abs{\ann}+1)\,T}  k^2 + \cdots 
\end{equation}	
leading to a diffusion constant $\mathcal{D} = \frac{d}{4\pi\, (\abs{\ann}+1)\, T}$. For gravitational perturbations in the vector sector which give rise to the momentum diffusion mode, using $\abs{\ann} = d-1$, from \cref{tab:Nval} we find $\mathcal{D} = \frac{1}{4\pi \,T}$ which is a rewriting  of the famous relation for the shear viscosity since $\mathcal{D}= \frac{\eta}{\epsilon+ P} = \frac{\eta}{T\,s}$ 
 \cite{Policastro:2001yc}. We will actually demonstrate a more remarkable fact: the non-Markovian dispersion relation can be obtained by analytically continuing the retarded Green's function of the corresponding Markovian field, cf., \cref{sec:twoobs}.

The SK Green's function for the non-Markovian probe can be then obtained by  solving the hydrodynamic equations for $\snMar_{\skR,\skL}$ and inverting  these moduli in terms of the doubled non-normalizable data. In this approach, at the first step one obtains the bulk solution as a  local expression in terms of the fields $\snMar_{\skR,\skL}$. The non-locality of the SK correlators appears only in the second step where we solve for $\snMar_{\skR,\skL}$ in terms of the appropriate non-normalizable sources. These two steps then have a very natural Wilsonian interpretation: the first step involves integrating out the fast modes and  parameterizing the solution in terms of the state (because these are normalizable data) of the slow modes. In the second step, we then solve for the slow modes in terms of their sources which give rise to correlations over long distances and times. 

These facts justify why the intermediate objects $\snMar_{\skR,\skL}$  can be interpreted as the effective fields on the SK contour. Our strategy may be summarized as starting in the forgetful sector and analytically continuing into the hydrodynamic moduli space, regaining memory when we finally solve for the dispersion of the long-lived, low momentum, hydrodynamic modes. This picture characterizes, quite universally, the essential physics of diffusive dynamics.
 
Equivalently, one might be interested in the local Wilsonian SK effective action which yields the above hydrodynamic equations for non-Markovian probes. This can  be done by deriving  the Legendre transform of the on-shell action parameterized by non-normalizable modes. As we will argue later, freezing non-normalizable modes at the AdS boundary requires one to quantize the non-Markovian probes with Neumann boundary conditions.  In turn, one requires a variational boundary term to the free designer scalar action \eqref{eq:KGDil} to impose such a Neumann boundary condition. Fortunately, the Legendre transform we are after cancels the aforementioned variational boundary term. Thus, a direct on-shell evaluation of the free designer  scalar action \eqref{eq:KGDil} in terms  of normalizable modes  $\{\snMar_\skR ,\snMar_\skL\}$ will give us the required local Wilsonian SK effective action.\footnote{ For a free massless scalar one sees that the change from the Dirichlet to Neumann boundary conditions involves a variational boundary term $\phi \partial \phi$, which is the product of the field and its conjugate momentum. The Legendre transformation switching between the generating function of correlators and the Wilsonian effective action parameterized  by the normalizable modes requires an addition of  $-\phi \partial \phi$, which exhibits the fortunate cancellation alluded to above.  } We will refer to this Schwinger-Keldysh effective action as the  \emph{Wilsonian influence functional} for reasons that will be elucidated below.

  \subsubsection{Observables on hydrodynamic moduli space}
\label{sec:obshmod}

A general system would have both Markovian and non-Markovian modes interacting with each other.   As a prototypical example, one can consider dynamics of energy-momentum tensor in a  CFT, which is dual to gravitational  Einstein-Hilbert dynamics. The tensor sector of the energy-momentum  dynamics is Markovian, but the vector and scalar sectors are  non-Markovian (see \cref{tab:Nval}). These sectors interact  with each other (beyond the quadratic order) due to the non-linearities of gravity. We would like to characterize the observables in such a  system.

We can parameterize the Markovian field data with non-normalizable sources $\{\JMar_\skR, \JMar_\skL\}$.  The non-Markovian sector  we continue to parameterize with the right/left (normalizable) hydrodynamic moduli $\{\snMar_\skR, \snMar_\skL\}$ introduced above. 
The full bulk solution can be parameterized in  terms of  this data in a boundary gradient expansion. 

Evaluating the on-shell action of this bulk solution we will get a local action,  viz., something which takes the form:
 \begin{equation}\label{eq:LWIF} 
\mathcal{S}_\text{WIF}\left[\JMar_\skR, \JMar_\skL, \snMar_\skR, \snMar_\skL\right]  = \int d^dx\ \mathcal{L}_\text{WIF}\left[\JMar_\skR, \JMar_\skL, \snMar_\skR, \snMar_\skL\right] ,
 \end{equation}
 where $ \mathcal{S}_\text{WIF}$ is the Wilsonian influence functional for the hydrodynamic moduli. We can add non-Markovian sources, $\{ \JnMar_\skR, \JnMar_\skL \}$, to this action and obtain the hydrodynamic equations  for   $\{\snMar_\skR, \snMar_\skL\}$  in the presence of these sources. Alternately, as explained  in \cref{sec:hmodc}, we may  obtain these hydrodynamic equations by looking at the non-normalizable (source) data for the non-Markovian sector in the full bulk solution. Furthermore, the normalizable data  for the non-Markovian sector yield the expectation values of the hydrodynamic moduli, i.e., for the dual gauge invariant boundary operators (single trace primaries). 

Our procedure is similar in spirit to the holographic Wilsonian renormalization group \cite{Heemskerk:2010hk,Faulkner:2010jy} and semi-holographic models \cite{Faulkner:2010tq}. In that case one solves for the bulk dynamics with Dirichlet boundary conditions at some intermediate radial position (corresponding to mixed boundary conditions at the AdS boundary) and thence integrating over the chosen Dirichlet data on this fiducial surface.  Closer in spirit is the seminal discussion of \cite{Nickel:2010pr} who were interested in deriving the universal low energy dynamics of holographic liquids.  

Our main goal in the rest of the paper is to explicitly construct the space of solutions advertised above, order by order in a gradient expansion, for the designer scalars and gauge fields, and establish a clear link to the advertised physics of charge and momentum diffusion. We will obtain from our analysis the Wilsonian influence functional \eqref{eq:LWIF} in these models, which serves as our input to the integral over the hydrodynamic moduli space. In the following sections we will implement this exercise at the quadratic order for probe scalars, gauge fields, and gravitons.

\section{Time-reversal invariant scalar system 1: Markovian dynamics}
\label{sec:scalarM}

We now turn to a detailed analysis of the designer scalar system introduced in \cref{sec:designer}. We first elaborate on the construction of the ingoing solution which is analytic at the horizon. We have already explained that there is a unique analytic solution once we fix the overall normalization, albeit one that has a very different interpretation depending on whether we are dealing with a Markovian ($\ann+1 >0$) or a non-Markovian ($\ann+1 <0$) probe. 

For the Markovian fields as explained in \cref{sec:markoff} an asymptotic Dirichlet boundary condition serves to uniquely pick out the  analytic Green's function with fixed normalization. Once we have the ingoing Green's function we can determine the outgoing Green's function by exploiting the time-reversal isometry \eqref{eq:trops} of the geometry. 

With this understanding we will now describe the explicit solutions for  the ingoing Green's functions in a gradient expansion along the boundary for Markovian scalars $\ann >-1$. We work in the Fourier domain and denote by $\Gin(\omega, r, \bk)$ the solution that is analytic at the horizon. In the Markovian sector, this is the retarded boundary to bulk Green's function which encodes the infalling quasinormal modes of  $\sen{\ann}$. 

\subsection{Ingoing solution in derivative expansion} 
\label{sec:inG}

We parameterize $\Gin$  in a derivative expansion as follows:\footnote{The expressions can be made much more compact by passing to the dimensionless $\ri$ coordinate \eqref{eq:rhodef}. We however stick to the radial variable as it is more familiar in the black hole context. }
\begin{equation}\label{eq:Ginpar3}
\begin{split}
\Gin &= 
	e^{-i b \omega F(\ann,\br) }\bigg[1- \bq^2 \, H_k(\ann,\br)-\bw^2 \, H_\omega(\ann,\br) +i \, \bw \,\bq^2\,  I_k(\ann,\br)+i \bw^3 \, I_\omega(\ann,\br)  
 	+\cdots\bigg] \,,
\end{split}
\end{equation}
where we have introduced dimensionless frequencies and momenta, 
\begin{equation}\label{eq:bwbqdef}
\bw = b\omega = \frac{d}{4\pi} \, \beta \omega \,, \qquad \bq = bk = \frac{d}{4\pi} \, \beta k\,.
\end{equation}	
In writing the expansion we have exploited the fact that the spatial reflection symmetry of the background guarantees the absence of terms odd in the momenta $\bk$ and have restricted attention to the first three orders in the gradient expansion. The choice of parameterization above, where we separate out the factor $e^{-i\, b \omega \, F(\ann,\br) }$ is made to simplify the structure of the gradient expansion. Our parameterization is largely inspired by the fluid/gravity literature especially \cite{Bhattacharyya:2008mz}. 

We  employ the following normalization convention for our ingoing Green's function, demanding,
\begin{equation}\label{eq:bcsGin}
\lim_{\br\to \infty} \, \Gin(\omega, r, \bk)  = 1 \,, 
\end{equation}	
and that $ \Gin(\omega, r, \bk)$ be analytic everywhere. This implies that in the Markovian case of $\ann > -1$ we will simply require the functions $F$, $H_k$, $H_\omega$ and other higher order gradient terms to vanish asymptotically. Thus we seek purely normalizable solutions to the wave equation \eqref{eq:gsrad} at higher orders in the derivative expansion. The non-Markovian results will be obtained by analytically continuation in $\ann$. 

Plugging in our  ansatz \eqref{eq:Ginpar3} into \eqref{eq:gsrad}, we get a hierarchy of radial second order ODEs of the form:
\begin{equation}\label{eq:radderinv}
\begin{split}
 \dv{r}(r^{\ann+2}\,f\, \dv{\mathfrak{F}(r)}{r} ) = \mathfrak{J}(r).
\end{split}
\end{equation}
where $\mathfrak{F}$ is a placeholder for a function that appears at some order in the gradient expansion and $\mathfrak{J}$ refers to the `source' constructed out of lower order terms in the gradient expansion. We therefore need to invert the differential operator 
$ \dv{r}(r^{\ann+2}\,f\, \dv{r} ) $ which  bears a close similarity to the fluid/gravity discussion of \cite{Bhattacharyya:2008jc}. 

Moreover, we already know the two homogeneous solutions of this differential operator from \eqref{eq:gszeromode}. The constant mode is analytic and is killed by the boundary conditions, while the monodromy mode is singular at the horizon. Hence the solution to the \eqref{eq:radderinv} is unique at any order in the derivative expansion.

The strategy to find this unique solution is then straightforward: we start with a particular solution to the equation and subtract from its derivative a piece that is a multiple of $\frac{1}{r^{\ann+2}}$, so as to make it analytic, and then integrate it back up. In the final integral we impose the vanishing of the functions at infinity. Schematically, we have (we use the dimensionless variable $y$ in the expressions below):
\begin{equation}\label{eq:gsolbr}
\mathfrak{F}(\br) = \int_\infty^{\br}\,  \frac{dy}{y^{\ann+2}\, f(y)} \; \int_1^{y} \mathfrak{J}(y') \, dy'
\end{equation}	
Since the inner integral usually involves the difference of function evaluated at some radial position from its value at the horizon, we will introduce a shorthand notation for this combination by decorating the symbol with a hat, viz., for any $\mathfrak{F}(\ann,\xi)$ we define
\begin{equation}\label{eq:hatfns}
\widehat{\mathfrak{F}}(\ann,\xi)  \equiv \mathfrak{F}(\ann,\xi) -\mathfrak{F}(\ann,1)\,.
\end{equation}	

It is amusing to write the iterated integral expression in \eqref{eq:gsolbr} as follows in the $\ri$ coordinate of \eqref{eq:rhodef}:
\begin{equation}\label{eq:gsolrhoBE}
\mathfrak{F}(\ri) = \int_{0}^{\ri}\, d\rit\, \frac{\rit^{\,\od(1+\ann)-1}}{1-\rit} \,  \int_1^{\rit} \mathfrak{J}(\ri') \, d\ri'\,,  \qquad \od = \frac{1}{d}
\end{equation}	
This is a Bose-Einstein integral. (For more direct verification introduce a further change of variables $\ri = e^{-z}$).  This suggests a formal analogy between the solutions to the wave equation in the black hole background to the Chapman-Enskog and Grad moment methods for solving the Boltzmann equation by perturbing around the equilibrium Maxwell-Boltzmann distribution (cf., \cite{Uhlenbeck:1963lec}). The really curious fact, though one we are quite used to from AdS black holes, is that the wave equation immediately is aware of the statistical distribution functions. In a certain sense the above form suggests that the black hole solution itself should be viewed as a coherent state of gravitons distributed according to bosonic statistics with the radial direction playing the role of an energy scale.

\subsection{Explicit parameterization of ingoing Green's function}
\label{sec:expingoing}

Proceeding in a manner described above, we reduce the problem to the following set of equations for the  functions $\{ F, H_k, H_\omega, I_k,I_\omega\}$ that appear at the first three orders in the gradient expansion. The first two of these satisfy a simple first order equation after integrating up once, see \cref{sec:appphigradexp}.

The differential equations above can be solved using incomplete Beta functions $\ibf{a,0;\ri}$ with one of its arguments being zero.\footnote{Some basic facts about this subclass of incomplete Beta functions are collected in \cref{sec:ibf}.} To keep our expressions compact we will write the solutions using the shorthand $\od = \frac{1}{d}$ and use the dimensionless $\ri$ coordinate \eqref{eq:rhodef}. For the functions $F$ and $H_k$ we immediately find:
\begin{equation}\label{eq:FHkSol}
\begin{split}
F(\ann,\br)&= \od  \, \ibf{\od ,0; \ri} - \od \, \ibf{ \od(\ann+1),0; \ri} \ ,\\
H_k(\ann,\br)&= \frac{\od}{\ann-1} \bigg[\ibf{2\od,0; \ri} -\ibf{ \od(\ann+1),0; \ri} \bigg]\ .\\
 \end{split}
\end{equation}
They satisfy the defining ODE obtained from \eqref{eq:gsrad} at $\order{\omega}$ and $\order{k^2}$, respectively. The differential equations themselves can be found in \eqref{eq:FHkeq}. 

The solution for $H_\omega$ can be also written down directly, but it is helpful to first define a new function $\Delta(\ann,\ri)$ via
\begin{equation}\label{eq:Deltadef}
\Delta(\ann,\br)  \equiv \od \, \ibf{\od(1-\ann) ,0; \ri} - \od \, \ibf{\od(1+\ann),0; \ri} ,
\end{equation}	
This combination is antisymmetric in $\ann \to -\ann$ and is introduced to simplify aspects of the analytic continuation from the Markovian to the non-Markovian case. It too satisfies a simple ODE \eqref{eq:DeltaEq}.  We parameterize the solution for the  function $H_\omega$  in terms of $\Delta$ as
\begin{equation}\label{eq:HwSol}
\begin{split}
H_\omega(\ann,\br)
&=
	 \od\, \bigg[\frac{1}{2} \, \Delta(\ann,\ri)- \Delta(\ann,1)\bigg] \, \ibf{ \od(1+\ann),0; \ri}  \\
&
	\qquad +
	 \frac{\od^2}{2}\, \sum_{n=0}^\infty \, \left(\frac{1}{n+ \od\, (1-\ann)}-\frac{1}{n +\od(1+\ann)}\right)  \ibf{n+2\, \od,0; \ri}\,.
\end{split}
\end{equation}
One can check that this function satisfies \eqref{eq:Hweq2} and the boundary conditions \eqref{eq:bcsGin}. 

A similar analysis applies for the third order functions $I_k$ and $I_\omega$ which are sourced by $\widehat{H}_k$ and $\widehat{H}_\omega$, respectively. The reader can find the details in \cref{sec:appphigradexp}, where we solve the equations using similar tricks; for instance we introduce new functions $\Delta_k$ and $\Delta_\omega$ in parallel with $\Delta$ to simplify the analytic continuation and extraction of asymptotics.

All of the functions $\{F,H_\omega, H_k, I_\omega, I_k\}$  vanish as a power law for Markovian probes ($\ann>-1$) as we approach the asymptotic boundary. In particular, we have up to the second order: 
\begin{equation}\label{eq:FHsasym}
F(\ann,\br) \sim \order{\br^{-1}} \,, \quad H_\omega(\ann,\br) \sim \order{\br^{-2}} \,, \quad H_k(\ann,\br) \sim \order{\br^{-2}}\,, \qquad \text{as} \; \; \br \to \infty\,.
\end{equation}	
The complete series solution is in fact easily read off using the defining series representation of the incomplete beta function \cref{eq:ibfser}.  The non-trivial behaviour is that of the auxiliary function $\Delta(\ann, \br)$ which one can check asymptotes as (for $\ann > -1$)
\begin{equation}\label{eq:Delasym}
\Delta(\ann,\br) \sim \order{\br^{\ann-1}} \,, \qquad \text{as}\;\; \br\to \infty \,.
\end{equation}	
This is in fact one reason for introducing the function (and others of its kind at higher orders) -- one can use it to isolate the modes that grow rapidly as we approach the boundary.\footnote{ There is another more important reason: the analytic continuation to $\ann <-1$, which we will describe in \cref{sec:scalarnM}.}

With this data the ingoing Green's function for the Markovian problem can be written down given the parameterization in \eqref{eq:Ginpar3}.  As noted above there are no subtleties with the asymptotic behaviour since these modes satisfy the standard Dirichlet boundary conditions at the AdS boundary (and the source  has been fixed to unity). To extract the boundary correlators we need to supply some counterterms  as  function of the boundary sources  as usual. We turn to these issues next.

\subsection{Counterterms and boundary correlators: Markovian scalar}
\label{sec:ccbdyM}

To wrap up the discussion, and to obtain the boundary observables, we will give a quick summary of the canonical conjugate momentum of the designer scalar system \eqref{eq:KGDil}.  For the Markovian probe the canonical conjugate is simply the  normalizable mode, whose large $r$ expansion yields the expectation value of the dual single trace primary of the boundary CFT.\footnote{ In the non-Markovian case one will instead be dealing with the source of the dual field theory (at a given point in the hydrodynamic moduli space). }

For the action \eqref{eq:KGDil} the momentum conjugate to radial evolution is given by the normal derivative. Letting $n^A$ be the unit spacelike normal to the fixed $r$ hypersurface, and $\gamma_{\mu\nu}$ the induced timelike metric on the hypersurface (see \cref{sec:grsk}), we have the projected derivative
\begin{equation}\label{eq:phinorm}
\partial_n \, \sen{\ann} \equiv n^A \nabla_A \sen{\ann} = \frac{1}{r\,\sqrt{f}}\, \Dz_+\sen{\ann}  \,,
\end{equation}	
in terms of which the canonical momentum density is given by 
\begin{equation}\label{eq:pin}
\cpen{\ann}= -\sqrt{-\gamma} \, e^{\chi_s} \,  \partial_n \, \sen{\ann} = -r^\ann\, \Dz_+ \sen{\ann} \,,
\end{equation}	
where we used  $e^{\chi_s} = r^{\ann+1-d}$.
 
As is often the case in AdS/CFT we are interested in the renormalized value of this canonical conjugate density 
evaluated on the ingoing solution, given in  a derivative expansion. An explicit computation gives up to the third order in gradient expansion the following:
\begin{equation}\label{eq:BareCanN}
\begin{split}
r^\ann \, \Dz_+ \Gin 
&= 
	\frac{r^{\ann-1}}{\ann-1} \, (k^2-\omega^2)\, \Gin+\cdots\\
&
	\quad 
	+ \frac{1}{b^{\ann+1}} 
		\Bigg\{-i\, \bw-\frac{\bq^2}{\ann-1} +\bw^2
		\left(\Delta(\ann,br) +\frac{(b r)^{\ann-1}}{\ann-1}-\Delta(\ann,1)\right) \\
&	\qquad \qquad \quad 
	-	2i\, \bw \left[ \bq^2\,  H_k(\ann,y)+ \bw^2\,  H_\omega(\ann,y)\right]_{y=1}^{y=b r} 
	+\cdots \Bigg\} \,\Gin \,.
\end{split}
\end{equation}
We have organized the result to isolate the terms that are pure UV effects from the point of view of the boundary CFT and those that have non-trivial knowledge of the black hole (and thus IR physics in the CFT). In the first line we have collected a set of terms where a temperature-independent operator acts on $\Gin$. In the Markovian case with $-1<\ann<d$, these vacuum  UV contributions, which  diverge as $r\to\infty$, whereas  the last two lines have a finite limit. In making this statement, we have used the asymptotic expansions given in \eqref{eq:FHsasym} and \eqref{eq:Delasym}.

To remove the vacuum contribution, we add, as usual, counterterms to our original action. The correct counterterm we need to add can  be inferred from the above expansion 
\begin{equation}\label{eq:KGDilCt}
\begin{split}
S_{ds}[\sen{\ann}]  &= -\frac{1}{2} \, \int \, d^{d+1}x \sqrt{-g}\ e^{\dils}\,  \nabla^A\sen{\ann} \nabla_A\sen{\ann}  
 -\frac{\ctphi{2}}{2} \, \int\, d^dx\, \sqrt{-\gamma}\  e^{\dils}\,   \gamma^{\mu\nu}\partial_\mu \sen{\ann}\partial_\nu \sen{\ann}\,.
\end{split}
\end{equation}
One can check that the desired counterterm coefficient is fixed by the asymptotic behaviour of the solution to be 
\begin{equation}\label{eq:ctcfM}
\ctphi{2}  = -\frac{1}{\ann-1}\,.
\end{equation}	
Including this boundary counterterm,  the renormalized canonical conjugate density to $\sen{\ann}$ is 
\begin{equation}\label{eq:pinrenorm}
\begin{split}
\cpen{\ann}\big|_\text{ren} =  -r^\ann \Dz_+ \sen{\ann}  -\frac{1}{\ann-1} \, \sqrt{f}\ r^{\ann-1} 
	\left(\partial_i\partial_i-\frac{1}{f}\partial_v^2\right)\sen{\ann}  .
\end{split}
\end{equation} 
Note that we have only retained the counterterms till quadratic order in the gradient expansion. Higher derivative counterterms may be necessary depending on $\ann$ and the spacetime dimension.

To cubic order, we can finally give the expression for the renormalized retarded boundary two-point function, $\Kin(\omega,k)$, of the operator dual to the field $\sen{\ann}$. We find taking the asymptotic limit:
\begin{equation}\label{eq:KinMark}
\begin{split}
\Kin(\omega,k)
	&\equiv  - \lim_{r\to\infty} \pi_{_\ann}\big|_\text{ren} \\
&=\frac{1}{b^{\ann+1}}	\left\{-i\,\bw-\frac{\bq^2}{\ann-1}
-\bw^2\Delta(\ann,1) +2i \, \bw \left[\bq^2 H_k(\ann,1)
	+\bw^2 \, H_\omega(\ann,1)\right]+\cdots \right\} 
\end{split}
\end{equation}
The explicit values of the functions appearing in the gradient expansion at the horizon are given in \eqref{eq:Deltahor} and \eqref{eq:Fhkomhor} using which we can  write down an explicit formula for the retarded boundary correlator for a Markovian scalar with Markovianity index $\ann$ dual to a field theory operator in $d$ spacetime dimensions. This is the central result for the Markovian sector. We will use it later to compute the full set of thermal Schwinger-Keldysh correlators from the grSK geometry in \cref{sec:grSKsol}.

\section{Time reversal invariant scalar system 2: non-Markovian dynamics}
\label{sec:scalarnM}
We have understood the Markovian designer scalar as a generalization of the massive scalar probes studied in \cite{Jana:2020vyx} and now are in a position to tackle the non-Markovian probe.  As noted in \cref{sec:nonmarkoff} we are not allowed to simply impose Dirichlet boundary conditions in the non-Markovian case. The analytic solution is normalizable at leading order in the gradient expansion to begin with. As we proceed in the gradient expansion we would need to additionally turn on appropriate non-normalizable modes to support the normalizable mode. This is a tedious way to proceed. Fortunately, as explained in \cref{sec:nonmarkoff} we can sidestep the issue by demanding that the non-Markovian solution for a given $\ann<-1$ be related to a Markovian solution for a corresponding value $-\ann > 1$.\footnote{For technical reasons we will refrain from considering the special values $\ann = \pm1$ in our discussion. This is an interesting degenerate class that should be dealt with separately.} Per se, this is merely a convention for parameterizing the solutions: any other parameterization can be related to our  prescription by a field redefinition of the non-Markovian modes, which is always permitted.

Before we embark on our analysis let us pause to make a note regarding our terminology.  Our use of the adjectives  `normalizable' and `non-normalizable' refers a-priori to the particulars of the fall-off behaviour of the mode functions. A mode that grows as $r\to \infty$ will be classified as non-normalizable if  its near-boundary expansion starts off with a  non-negative  exponent. These are modes we imagine having to freeze at the AdS boundary and taking them to specify the boundary conditions for radial evolution. How we treat such modes and construct  a classical phase space and thence the Hilbert space  by  imposing canonical commutation relations depends on boundary conditions, counterterms etc., which dictate what the inner product ought to be. We will of course see in a while that our adjectives are indeed appropriate, as we will exhibit a well defined variational problem for the non-Markovian fields with Neumann boundary conditions.

\subsection{Parameterization of the ingoing solution}
\label{sec:nMGin}

We will start with the explicit solutions for  the ingoing solution in a gradient expansion along the boundary for our non-Markovian scalars $\ann <-1$. We will continue to work in the Fourier domain and denote the inverse Green's function which is analytic at the horizon by $\GinN(\omega, r, \bk)$. This function is no longer the retarded boundary to bulk Green's function describing the infalling quasinormal modes of $\sen{-\ann}$. Rather it corresponds to the inverse Green's function, which gives the sources in terms of field expectation values. In other words this is the derivative operator that defines the hydrodynamic equations.

We parameterize $\GinN(\omega,r,\bk)$ in analogy with \eqref{eq:Ginpar3}
\begin{equation}\label{eq:Ginpar3nM}
\begin{split}
\GinN &= 
	e^{-i \,\bw\, F(-\ann,\br) }\bigg[1- \bq^2 \, H_k(-\ann,\br)-\bw^2 \, H_\omega(-\ann,\br) \\
&\qquad \qquad \qquad \qquad 
	+i \,\bw\bq^2\,  I_k(-\ann,\br)+i \,\bw^3 \, I_\omega(-\ann,\br)  
 	+\cdots\bigg]  .
\end{split}
\end{equation}
We may once again follow the steps outlined in \cref{sec:expingoing} and solve for the functions order by order in a gradient expansion.

Fortunately,  we have already done the heavy lifting. We can now reveal the rationale behind the function $\Delta(\ann,\br)$ which we had introduced in \eqref{eq:Deltadef}. We recall that it was antisymmetric in $\ann \to -\ann$. As a result it allows us to find the analytic continuation of  $\{F, H_k, H_\omega\}$ from positive to negative values of $\ann$. It is straightforward to verify that:
\begin{equation}\label{eq:NMtoMfns}
\begin{split}
F(-\ann,\br)
&=
	F(\ann,\br)-\Delta(\ann,\br)\,,\\
H_k(-\ann,\br)&=
	-\frac{\ann-1}{\ann+1}\, H_k(\ann,\br) +\frac{1}{\ann+1}\,\Delta(\ann,\br)\,,\\
H_\omega(-\ann,\br)
&=
	-H_\omega(\ann,\br)+\Delta(\ann,1)\, \Delta(\ann,\br)-\frac{1}{2}\,\Delta(\ann,\br)^2\,.	
	\end{split}
\end{equation}
For $\ann<-1$ (the non-Markovian case) we use these definitions for the analytic continuations. Expressions at higher orders may similarly be derived, cf., \eqref{eq:Delkomdef} which are relevant at the third order. We will need these for the construction of the boundary observables. This completes for us the specification of the ingoing inverse Green's function at the first few orders in the gradient expansion. 

\subsection{The non-Markovian inverse Green's function and dispersion relations}
\label{sec:nMdisperse}

Armed with $\GinN$ we can proceed as suggested in \cref{sec:hydromod}. The primary novelty in our discussion lies for the non-Markovian fields, the Markovian sector being a simple extension of the analysis in \cite{Jana:2020vyx} as discussed hitherto in \cref{sec:ccbdyM}. We will proceed to delineate non-Markovian  observables,  elaborate on some of the points made in  \cref{sec:hydromod} and extract the dispersion relation for these long-lived modes.

Let us examine more closely the derivative expansion for the non-Markovian ingoing inverse Green's function $\GinN(\omega,r,\bk)$. We use the parameterization \eqref{eq:Ginpar3nM} which we write out explicitly using \eqref{eq:NMtoMfns}, as
\begin{equation}\label{eq:GinNmain}
\begin{split}
\GinN
&= 
	1-i\, \bw\, F(-\ann,\br) -\frac{\bw^2}{2} \, F(-\ann,\br)^2-\bq^2\, H_k(-\ann,\br)- \bw^2 \, H_\omega(-\ann,\br) 
	+\cdots\\
&= 
	1-i\, \bw \, F(\ann,\br) -\frac{\bw^2}{2} \, F(\ann,\br)^2
		+\frac{\ann-1}{\ann+1}\, \bq^2\, H_k(\ann,\br)+ \bw^2\, H_\omega(\ann,\br) \\
&
	\qquad\quad
		-\left[-i \, \bw + \frac{\bq^2}{\ann+1}+\bw^2\,  \Delta(\ann,1)- \bw^2\, F(\ann,\br) \right]\Delta(\ann,\br) +\cdots\,.
\end{split}
\end{equation}
As discussed in \eqref{eq:FHsasym} for positive $\ann$, the functions $\{F,H_k,H_\omega\}$ vanish at infinity whereas \eqref{eq:Delasym} shows that  $\Delta(\ann,\br)\sim \br^{\ann-1} $ at large $r$. It follows that  the asymptotic behaviour  of $\GinN$ is given up to second  order in boundary  gradients  by\footnote{ Explicit results, accurate to third order in the gradients, are given in \cref{sec:nMGin3}.}
\begin{equation}\label{eq:GinNasym}
\begin{split}
\GinN(\omega,r,\bk) 
	\sim  
	\frac{(b r)^{\ann-1}}{\ann-1} \left[-i\, \bw +\frac{\bq^2}{\ann+1}+ \bw^2 \, \Delta(\ann,1) +\cdots\right] \,.
\end{split}
\end{equation}
This demonstrates explicitly  our earlier claim: in the non-Markovian case turning on a normalizable ingoing solution inevitably turns on the non-normalizable mode at higher orders in derivative expansion (at generic $\{\omega,\bk\}$). There is however a subset of  $\{\omega, \bk\}$ defined by a dispersion relation where this non-normalizable mode vanishes. At these loci alone can one have a purely normalizable ingoing solution. 

To extract this more efficiently, let us parameterize the non-Markovian inverse  Green's function in a particularly convenient form: 
We write:
\begin{equation}\label{eq:GinnMpar}
\GinN(\omega,r,\bk) = \GinNt(\omega,r,\bk) \left[ 1- \frac{1}{b^{\ann-1}} \, \KinN(\omega,\bk) \, \XiNN(\omega,r,\bk)\right]
\end{equation}	
The three pieces introduced above in the parameterization are the following:
\begin{itemize}[wide,left=0pt]

\item $\KinN(\omega,\bk)$ is a dispersion function whose vanishing locus parameterizes a hyperspace of the $(\omega,\bk)$ space where normalizable modes exist. Explicitly, we find:
\begin{equation}\label{eq:Kdispdef}
\KinN(\omega,\bk) = b^{\ann-1}	 \left[-i\, \bw+\frac{\bq^2}{\ann+1}  + \bw^2\, \Delta(\ann,1) + \cdots \right]
\end{equation}	
up to the second order in the gradient expansion. At third order the relevant expression can be found in \eqref{eq:Kmn3rd}. 

\item $\XiNN(\omega, \bk)$  is the non-normalizable mode function that is generically turned on at higher orders in the gradient expansion, as anticipated previously. An  explicit expression accurate  to third order can be found in \eqref{eq:XiNN3rd}. 

\item Finally, $\GinNt(\omega,r,\bk)$ is a purely normalizable component of the Green's function given in  \eqref{eq:GinNnorm3rd} up to third order in gradients. We note that $\GinNt(\omega,r,\bk)$ is very closely related to $\Gin(\omega,r,\bk)$, the ingoing boundary to bulk Green's function of the Markovian case obtained earlier in \eqref{eq:Ginpar3} (there  are small differences in the scaling of the various functions, see \cref{sec:nMGin3}).
\end{itemize}

Let us examine the locus where the ingoing non-Markovian Green's function is purely normalizable. The hypersurface is parameterized in the form of a dispersion relation which from \eqref{eq:Kdispdef} takes the form:
\begin{equation}\label{eq:disp2}
\begin{split}
0&=i\, \bw -\frac{\bq^2}{\ann+1}-\bw^2 \Delta(\ann,1)+\cdots
\end{split}
\end{equation}
We recognize this as the dispersion relation of a diffusion mode with the diffusion constant $\frac{b}{\ann+1}$ as advertised in \eqref{eq:diffdisp}, along with some higher order corrections. Thus the moduli space of purely normalizable ingoing solutions in non-Markovian case is identical to the moduli space of solutions of a generalized diffusion equation. This should be contrasted against the Markovian case 
where there is no purely normalizable ingoing solution and the corresponding moduli space is an empty set. 

Using the results of \cref{sec:nMGin3,sec:horvalues} we can write down the dispersion relation to the third order in the gradient expansion for a $d$ dimensional boundary field theory as:
\begin{equation}\label{eq:disp3}
\begin{split}
0
& = 
	i\, \bw- \frac{ \bq^2}{\ann+1}   	-\bw^2 \Delta(\ann,1) 
 		+  2i \, \bw\, \bq^2 \frac{\ann-1}{\ann+1} \, H_k(\ann,1)  +2 i\,  \bw^3 \,H_\omega(\ann,1) + \cdots\,, \\
&=
	i\, \bw- \frac{ \bq^2}{\ann+1} - \bw^2 \frac{\psi\left( \frac{\ann+1}{d}\right) - \psi\left(\frac{1-\ann}{d} \right)}{d}
	+  2i \, \bw\, \bq^2 \frac{\psi\left(\frac{\ann+1}{d} \right) - \psi\left(\frac{2}{d}\right)}{d(\ann+1)}  \\
&	\qquad 
	+ i \, \bw^3
	\left\{
			\frac{\psi\left(\frac{\ann+1}{d}\right) 
			\left[ \psi\left(\frac{\ann+1}{d}\right)  - \psi\left( \frac{1-\ann}{d}\right) \right]}{d^2}  
	 +  \sum_{n=0}^\infty 	\left[ \frac{1}{n+ \frac{1+\ann}{d}} - \frac{1}{n + \frac{1-\ann}{d}}\right] \psi\left(n+\frac{2}{d}\right) 	 
	\right\}	+ \cdots\,,
\end{split}	
\end{equation}	
where $\psi(x)$ is the  digamma function.

\subsection{Two observations about non-Markovian scalars}
\label{sec:twoobs}

Before we proceed to discuss further details about the Markovian scalars, let us make note of two observations that while empirical at this point, suggest a deeper principle in operation. 

\paragraph{The dispersion function \& the Markovian Green's function:} We had deliberately denoted  the  renormalized Markovian retarded Green function as $\Kin(\omega,\bk)$ in \eqref{eq:KinMark}.  As can be readily checked, this function is the analytic continuation of the non-Markovian dispersion function $\KinN(\omega,\bk)$  obtained in \eqref{eq:Kdispdef} (for the third order expression see \eqref{eq:Kmn3rd}). We thus come to a remarkable conclusion: not only are the solutions in the Markovian and non-Markovian bulk-boundary ingoing Green's functions related by analytic continuation, but more interestingly, 
\emph{the dispersion function of the non-Markovian field can also be obtained by analytically continuing the retarded boundary Green's function of the Markovian field}. While we have not given  a general argument why this should be the case, the fact that this holds till third order in derivative expansion is  a strong evidence for this statement. If this statement is assumed to be true, it gives the simplest way to compute  the dispersion function on the non-Markovian side without ever having to solve the non-Markovian problem.

\paragraph{A field redefinition to diffusion:} 
We had mentioned in \cref{sec:hydromod}, with a suitable field redefinition, the dispersion relation obtained above can be made into a explicit diffusion dispersion. Let us describe now in detail how this could be achieved. Consider a rescaling of the non-Markovian inverse Green's function by a factor, viz.,
\begin{equation}
\begin{split}
\GinNc 
&\equiv 
	\left(1-i\, \bw  \, \Delta(\ann,1)+\frac{\bq^2}{\ann+1} \Delta(\ann,1)+\cdots \right) \GinN \\
&= e^{-i\,\bw\, \Delta(\ann,1)} \times e^{-i\, \bw  F(-\ann,\br) } \, 
	\left[1-\bq^2\,  \left(H_k(-\ann,\br)-\frac{\Delta(\ann,1)}{\ann+1}\right) \right. \\
&
\hspace{5.2cm}
	\left.	- \bw^2 \,\left(H_\omega(-\ann,\br) -\frac{1}{2}\, \Delta(\ann,1)^2\right) +\cdots \right]
\end{split}
\end{equation}
where we have distributed the constant shift by $\Delta(\ann,1)$ into the various functions appearing in the parameterization. By construction, $\GinNc$ also solves the generalized KG equation with the exponent $\ann$. It has an asymptotic behaviour
\begin{equation}
\begin{split}
\GinNc
\sim  \frac{(b r)^{\ann-1}}{\ann-1} \left(-i\, \bw +\frac{\bq^2}{\ann+1}\right) +\cdots\,.
\end{split}
\end{equation}
which as presaged gives us the correct diffusive dispersion relation, accurate to second order.

\subsection{Counterterms and boundary correlators: non-Markovian scalar}
\label{sec:ccbdynM}

By direct computation we obtain the canonical conjugate momentum of the non-Markovian scalar to be 
\begin{equation}\label{eq:pinM}
\begin{split}
\cpen{-\ann} &= 
- r^{-\ann}\,\Dz_+ \GinN(\omega, r,\bk) \\
&= 
	-r^{-\ann}  \, \Dz_+ \GinNt(\omega, r,\bk) + \frac{r^{-\ann}}{b^{\ann-1}}\,
	\KinN(\omega,\bk) \, \Dz_+ \left[\GinNt(\omega, r,\bk) \, \XiNN(\omega,r,\bk)\right]\\
&=
	-\KinN(\omega,\bk) +\text{subleading}
\end{split}
\end{equation}
In  deriving the above we have used the asymptotic behaviour obtained in  \eqref{eq:Xiasym} and \eqref{eq:GinNhasym}.  This  can also be directly obtained from direct differentiation of \eqref{eq:GinNasym}. 

What we see here is that the  canonical momentum conjugate  to the non-Markovian designer scalar picks out the non-normalizable mode of the field.  Moreover, since the asymptotic behaviour of $\cpen{-\ann}$  is finite, one needs no counterterms to regulate it.\footnote{ In  making this statement we  are restricting attention to  the Gaussian (free)  non-Markovian designer scalar. If there are bulk self-interactions then one would  perhaps need additional counterterms to account for renormalization of bulk Witten diagrams (the analog of source renormalization in the grSK contour discovered  in \cite{Jana:2020vyx}). We  will  leave it for future work to deduce such effects, if present.} 
These facts are a-priori quite counter-intuitive. The standard  AdS/CFT dictionary relates  the non-normalizable mode to the asymptotic field value, and not to the conjugate momentum  (this  is indeed the case  for the Markovian scalar, see \cref{sec:ccbdyM}).  The non-Markovian field is very unconventional in this regard. One can discern that the dilatonic coupling in \eqref{eq:KGDil} is highly damped in the near AdS-boundary region and is clearly responsible for this unconventional behaviour.  

Armed with the observation above, we can ask how does one quantize the non-Markovian designer field with AdS asymptotic boundary conditions. The usual rules of AdS/CFT tell us that the modes that grow asymptotically are to be frozen. We see here that $\cpen{-\ann}$ is 
the non-normalizable mode which  we need to freeze. The standard AdS boundary conditions freeze the field, but freezing the conjugate momentum is easy to do. Instead of  quantizing $\sen{-\ann}$ with Dirichlet boundary conditions we impose Neumann boundary conditions on the non-Markovian field. 

In practice  implementing Neumann boundary conditions is simple: one simply starts with the usual  Dirichlet boundary conditions where the asymptotic field value is frozen and then Legendre transforms to the Neumann boundary  condition where the momentum  is frozen instead.  This is the  usual story  for the alternate and multi-trace boundary conditions in AdS/CFT  \cite{Klebanov:1999tb,Witten:2001ua}. To wit, the variational problem of the non-Markovian field, which defines the classical phase space, and subsequently is used to compute  the generating function of connected correlators, requires one Legendre transform to impose  Neumann boundary conditions. One would define the action for the non-Markovian scalar, completing the discussion around \eqref{eq:KGDil}, as
\begin{equation}\label{eq:nMNeumann}
\begin{split}
S_{ds}[\sen{-\ann}] 
&= 
	-\frac{1}{2} \, \int \, d^{d+1}x \sqrt{-g}\ e^{\chi_s}\,  \nabla^A\sen{-\ann} \nabla_A\sen{-\ann} + 
		\int d^dx\,  \sqrt{-\gamma}\, e^{\chi_s}\, \sen{-\ann}\, n^A \nabla_A \sen{-\ann}+S_\text{ct}\,,\\
&=
	-\frac{1}{2} \, \int \, d^{d+1}x \sqrt{-g}\ e^{\chi_s}\,  \nabla^A\sen{-\ann} \nabla_A\sen{-\ann} -  
	\int d^dx\,  \cpen{-\ann}\, \sen{-\ann} +S_\text{ct}\,.
\end{split}
\end{equation}	
Here we note that  $e^{\chi_s} = r^{-\ann+1-d}$ since we are discussing the non-Markovian scalar and used \eqref{eq:pin}. In writing the above we have also acknowledged that there might yet be additional boundary counterterms necessary to compute the boundary generating function
\begin{equation}\label{eq:}
\mathcal{Z}_{SK}[\hat{J}_\skR, \hat{J}_\skL]  =\int [D\sen{-\ann}] \; e^{iS_{ds}[\sen{-\ann}]}\,,
\end{equation}	
with the bulk functional integral being assumed to be carried out on the grSK saddle geometry with non-Markovian sources $\hat{J}_{\skR, \skL}$.

However, as we discussed hitherto in \cref{sec:trscalar}, we compute first the Wilsonian Influence Functional which is parameterized, in the current  parlance, in terms of normalizable modes  (the hydrodynamic moduli). From the Wilsonian Influence Functional we can obtain the generating function of connected correlators by Legendre transforming  the  hydrodynamic moduli onto the  non-Markovian sources.

By a happy circumstance (or clever design depending on one's perspective), these two Legendre transformations however cancel  each other out!  We conclude that \emph{for the non-Markovian scalars, the standard Klein-Gordon action with no additional
variational boundary terms automatically computes the effective action for the non-Markovian fields}.  Thus we may write:
\begin{equation}\label{eq:}
\begin{split}
\mathcal{L}_\text{WIF} [\snMar_\skR, \snMar_\skL]  
&=
	\int [D\sen{-\ann}] \; e^{i \breve{S}_{ds}[\sen{-\ann}] }\,, \\ 
\breve{S}_{ds}[\sen{-\ann}] 
&=  -\frac{1}{2} \, \int \, d^{d+1}x \sqrt{-g}\ e^{\chi_s}\,  \nabla^A\sen{-\ann} \nabla_A\sen{-\ann}  +S_\text{ct}\,. 
\end{split}
\end{equation}	
Consequently, we are in a fortuitous circumstance where the computation of the on-shell action involves no variational boundary terms.  This story strongly parallels the  Markovian case, which makes the analysis quite straightforward, despite the seeming complications. 
 
 There will however be boundary counterterms necessary as indicated in $S_\text{ct}$. While the non-normalizable modes do not need any counterterms for their definition, the normalizable mode and the on-shell action require appropriate counterterms made of non-normalizable modes to cancel the large $r$ divergences. Since the canonical momentum $\cpen{-\ann}$ corresponds to the non-normalizable mode, its canonical conjugate i.e., the field $\sen{-\ann}$ itself with appropriate counterterms should give the normalizable mode. 

The allowed counterterms should be built out of the non-normalizable modes or what we would want to call the `CFT sources' themselves viz., $\partial_n \sen{-\ann}$ following \eqref{eq:phinorm}.  To make this explicit, we will first rewrite $\GinN$ in terms of 
$\Dz_+ \GinN$ to extract  the non-normalizable pieces explicitly. With a bit of algebra, we find  
\begin{equation}\label{eq:GmNminimSub}
\begin{split}
\GinN 
&= 
	\frac{1}{r(\ann-1)}\,
	\left(1-\frac{k^2-\omega^2}{r^2\, (\ann-1)\, (\ann-3)} \right) \Dz_+  \GinN \\
&
	\quad 
		+ e^{-i \,\bw\, F(\ann,\br) } \times\left[1-\frac{\KinN}{b^{\ann-1}} \; \XiNNR(\ann,\br) 
		\right] \times 
		\left[1+\frac{k^2}{r^2(\ann^2-1)}  +\frac{\ann^2-1}{\ann+1} \bq^2\, H_k(\ann,\br) 
		\right.\\
&\left. \qquad \qquad
	 	+ \;	\bw^2 H_\omega(\ann,\br)-i \,\bw\,\bq^2 \, \frac{\ann-1}{\ann+1}\, I_k(\ann,\br)-i \,\bw^3 \, 
		I_\omega(\ann,\br)  +\cdots\right]\ .
\end{split}
\end{equation}

We have isolated the large $r$ divergences in the first line. We see that this is a vacuum contribution from the $b$-independence of the prefactor. We also introduced $\XiNNR(\ann,\br)$ -- this is  the `renormalized' non-normalizable mode function which is engineered to vanish as $r\to \infty$. It is obtained from $\XiNN(\ann,\br)$ by a minimal subtraction scheme to remove the divergent pieces.  An explicit parameterization of this function can be given in terms of data described in \cref{sec:appphigradexp}, especially the functions $\{\Delta(\ann,\br),\Delta_k(\ann,\br), \Delta_\omega(\ann,\br)\} $ introduced there.  We find: 
\begin{equation}
\begin{split}
\XiNNR(\ann,\br)
&\equiv  
	\Delta(\ann,\br)+\frac{\br^{\ann-1}}{\ann-1}\\
&
	-2\left[\frac{\ann-1}{\ann+1}\, \bq^2\, H_k(\ann,\br)+\bw^2\, H_\omega(\ann,\br)\right]
	\left(\Delta(\ann,\br)+\frac{\br^{\ann-1}}{\ann-1}-\Delta(\ann,1)\right)\\
&
	+\left(\frac{\bq^2}{\ann+1}-\frac{\bw^2}{2(\ann-1)}\right)\left(\Delta_k(\ann,\br)- 2\frac{\br^{\ann-3}}{(\ann-1)(\ann-3)}\right)+\bw^2\, \Delta_\omega(\ann,\br)+\cdots
\end{split}
\end{equation}
Using \eqref{eq:Delasym} and  \eqref{eq:Delasym3rd}  we can see that $\XiNNR$ actually vanishes as we take $\br\to\infty$ limit.

Having extracted out the vacuum divergences, we can cancel them by adding appropriate  counterterms made of $\partial_n\sen{-\ann}$ and its derivatives to our original action. Adding the most general counterterms  admissible  up to the third order in boundary derivatives, we obtain
\begin{equation}\label{eq:KGDilctnM}
\begin{split}
\breve{S}_{ds}[\sen{-\ann}] 
&= 
	S_\text{bulk} +S_\text{ct}\\
S_\text{bulk} 
&= 
		-\frac{1}{2}\, \int\, d^{d+1}x\, \sqrt{-g}\, r^{-\ann+1-d}\; \nabla^A \sen{-\ann} \,\nabla_A\sen{-\ann}\\
S_\text{ct}
&=
	\frac{1}{2} \, \int\, d^dx\, \sqrt{-\gamma}\, r^{-\ann+1-d}\,  
 	\left[ \ctpi{0}\, (\partial_n\sen{-\ann})^2 - \ctpi{2}   \,\gamma^{\mu\nu} (\partial_\mu \partial_n\sen{-\ann}) 
 	 (\partial_\nu \partial_n\sen{-\ann}) \right] 
\end{split}
\end{equation}
From the asymptotic behaviour of the solution we can fix the coefficients of the counterterms. We find:\footnote{ We will see in subsequent sections that these coefficients coincide with those for the  Einstein-Hilbert action (with standard Dirichlet boundary conditions) in the special case $\ann = d-1$. }
\begin{equation}\label{eq:ctcfnM}
\ctpi{0} = 	- \frac{1}{\ann-1} \,, \qquad   \ctpi{2} = -\frac{1}{(\ann-1)^2\,(\ann-3)} \,.
\end{equation}	

 Let us check how these counterterms work to give us a finite result for the boundary observables. Let us first check the canonical pairs in the classical phase space prior to addition of counterterms.  The variation of the bulk action $\delta S_\text{bulk}$ in \eqref{eq:KGDilctnM} gives a boundary term $ - \sqrt{-\gamma}\ r^{-\ann+1-d}\ \partial_n\sen{-\ann}\ \delta \sen{-\ann}$. The variational boundary  term required to impose Neumann boundary conditions, further shifts the boundary variation to   
\begin{equation}\label{eq:varnM}
\begin{split}
\delta S_\text{bulk} \bigg|_\text{Neumann}
& \propto \int\, d^dx\,  \sen{-\ann}\ \delta\left( \sqrt{-\gamma}\ r^{-\ann+1-d}\ \partial_n\sen{-\ann}\right)\\
& = \int d^dx\, \sen{-\ann}\, \delta \left( r^{-\ann}\, \Dz_+\sen{-\ann}\right) = -\int d^dx \, \sen{-\ann} \delta\cpen{-\ann}
\end{split}
\end{equation}	
This shows that the canonical conjugate of $-\cpen{-\ann}$ is the field $\sen{-\ann}$ as we had intuitively anticipated earlier and justifies our choices made hitherto.

We can account for the addition of counterterms, and learn that the regulated statement is that $-\cpen{-\ann}$ is conjugate to:
\begin{equation}\label{eq:pinMcts}
\begin{split}
-\cpen{-\ann} &\longleftrightarrow \sen{-\ann}+ \ctpi{0}\,  \partial_n\sen{-\ann} + \ctpi{2}\, r^{-2}\left(\partial_i\partial_i-\frac{1}{f}\partial_v^2\right)\partial_n\sen{-\ann} \,,\\
-\cpen{-\ann} & \longleftrightarrow \sen{-\ann}- \frac{1}{\sqrt{f}}\left[-\frac{\ctpi{0}}{r}+\frac{ \ctpi{2}}{r^{3}} \left(k^2-\frac{1}{f}\omega^2\right)\right]\Dz_+\sen{-\ann}.
\end{split}
\end{equation} 
In the second line we have written out the normal derivative in terms of the derivation $\Dz_+$ introduced in \eqref{eq:Dz} and passed into the Fourier domain.   In the $r\to\infty$ limit, we then find the following:
\begin{equation}
\begin{split}
 \lim_{r\to\infty}\left\{
\GinN - \frac{1}{\sqrt{f}}\left[\frac{1}{r(\ann-1)}-\frac{1}{r^3}\, \frac{1}{(\ann-1)^2(\ann-3)}\left(k^2-\frac{1}{f}\omega^2\right)\right]\mathbb{D}_+\GinN +\cdots\right\}=1\ .
\end{split}
\end{equation}
This shows that, with the choices described above, $\GinN$ describes the state of the non-Markovian scalar 
which is dual to a CFT state with a unit renormalized vacuum expectation value for the CFT single trace primary (along with a CFT source that is needed to maintain this expectation value).

\section{Solution and on-shell action on grSK geometry}
\label{sec:grSKsol}

Having constructed the retarded boundary to bulk Green's functions, we will now begin by constructing the solution on grSK
geometry satisfying the appropriate SK boundary conditions. Once the solution is constructed, we can evaluate the
on-shell bulk action on the solution to obtain the Wilsonian influence phase in the dual CFT defined in \eqref{eq:LWIF}.

As described in \cref{sec:trscalar}, our goal will be to integrate out the fast Markovian degrees of freedom while freezing the slow non-Markovian modes and getting a Wilsonian influence phase in terms of the doubled  Markovian sources (denoted by $\{\JMar_\skR, \JMar_\skL \}$) as well as non-Markovian effective fields (denoted by $\{\snMar_\skR, \snMar_\skL\}$). Furthermore, as we explained in \cref{sec:ccbdynM} such a Wilsonian influence phase is computed by evaluating the  on-shell action on the gravity side without including the variational counterterm that implements Neumann boundary-condition in the non-Markovian sector. 

We will begin our discussion by generating the solutions that describe outgoing Hawking modes. This is most efficiently done by exploiting the $\mathbb{Z}_2$ time-reversal isometry $v\mapsto i\beta\ctor-v$ of the black brane background \cite{Chakrabarty:2019aeu,Jana:2020vyx} which was described in \cref{sec:grsk}. For functions in Fourier domain, this amounts to the map $\omega\mapsto -\omega$ followed by a multiplication with a factor of $e^{-\beta\omega\ctor}$ to go from the ingoing Green's functions to the outgoing Green's functions. We  define the time-reversed Green's function 
\begin{equation}\label{eq:Grevdef}
\Grev(\omega,\bk) \equiv \Gin(-\omega,\bk) \ ,
\end{equation}
so that the outgoing Green's function has the form 
\begin{equation}\label{eq:Goutdef}
\Gout(\omega,\bk) \equiv \Grev(\omega,\bk) \;e^{-\beta \omega \ctor}\equiv \Gin(-\omega,\bk) \; e^{-\beta \omega \ctor} \,.
\end{equation}

Given that all our Green's functions come with a phase factor $e^{-i \bw F(\ann, \br)}$, we conclude that 
the outgoing Green's functions are obtained by reversing the $\omega$'s and performing a shift
\begin{equation}
\begin{split}
F(\ann, \br) \mapsto F(\ann, \br)+i\frac{\beta}{b}\ctor = F(\ann, \br)+\frac{4\pi i}{d}\ctor \ ,
\end{split}
\end{equation}
where we have used the relation \eqref{eq:ctordef} to  relate the inverse Hawking temperature and the inverse horizon radius of the 
\SAdS{d+1} black hole.

We have hitherto observed  in \cref{sec:scalarM} and \cref{sec:scalarnM} that counterterms are needed for the definition of normalizable modes within the ingoing solution. We will now see that the \emph{same} counterterms render finite  the on-shell action evaluated on grSK geometry which, in addition, incorporates the effects of outgoing Hawking modes. In particular, while the effective theory we derive would be non-unitary, only the unitary counterterms of the microscopic theory are needed to get finite answers.

\subsection{Markovian probes}
\label{sec:Markprobe}

Let us begin with the Markovian sector where the analysis parallels that of \cite{Jana:2020vyx}. The most general solution on grSK geometry \eqref{eq:sadsct} is given by 
\begin{equation}\label{eqn:gsSKsol}
\sen{\ann}^{\SKs}(\omega,\ctor,\bk) =  \incfM \: \Gin(\omega,\ctor,\bk) + \hawkcfM\: \Grev(\omega,\ctor,\bk)\, e^{-\beta \omega\ctor}\,.
\end{equation}
We relate the non-normalizable modes of the field $\sen{\ann}$ to the CFT sources at  the left and the right boundaries, i.e., at
 $r = \infty \pm i 0$ where $\ctor$ takes the values $0$ and $1$ respectively:
\begin{equation}
\lim_{r \to \infty + i 0} \, \sen{\ann}^{\SKs} = \JMar_\skL \,, \qquad  \lim_{r \to \infty -i 0} \, \sen{\ann}^{\SKs} = \JMar_\skR \,.
\end{equation}
Our normalization of the ingoing Green's function \eqref{eq:bcsGin} and the action of time-reversal \eqref{eq:Grevdef} together imply that the coefficients $\incfM$ and $\hawkcfM$ are given by 
\begin{equation}
  \incfM  + \hawkcfM= \JMar_L\ ,\qquad  \incfM  +e^{-\beta\omega} \hawkcfM= \JMar_R\,.
\end{equation}

Fixing the constants with the above boundary conditions we see that the solution of the designer scalar on the grSK geometry is then given by the following  \cite{Chakrabarty:2019aeu,Jana:2020vyx}\footnote{ An earlier incarnation of this expression was obtained in \cite{Son:2009vu}  on a single copy of the bulk by analogy with thermal Green's functions in field theory.}
\begin{equation}\label{eq:SonTeaney}
\begin{split}
\sen{\ann}^{\SKs}(\omega,\ctor, \bk) 
&=
	\Gin \left[(\nB+1)\: \JMar_\skR - \nB \: \JMar_\skL \right] - 
	\Grev\, \nB \left(\JMar_\skR-\JMar_\skL\right)  e^{\beta \omega(1-\ctor)}\ \\
&=
	\Gin \, \JMar_a+
	\left[\left(\nB+\frac{1}{2}\right) \Gin- \nB\,  e^{\beta \omega(1-\ctor)} \Grev\right] \JMar_d\,.
\end{split}
\end{equation}
where 
\begin{equation}\label{eq:nBdef}
\nB \equiv \frac{1}{e^{\beta \omega} -1}\,,
\end{equation}	
is the Bose-Einstein factor. In the last line of \eqref{eq:SonTeaney}, we have defined the average-difference or Keldysh basis sources 
\begin{equation}\label{eq:keldJ}
\JMar_a\equiv  \frac{1}{2} \left( \JMar_\skR+ \JMar_\skL\right)\ ,\qquad
\JMar_d\equiv \JMar_\skR-\JMar_\skL\,.
\end{equation}

It can be explicitly checked that if $ \Gin$ has a derivative expansion such that all odd powers at least have 
one factor of $\omega$, then $\sen{\ann}^{\SKs}$ can also be written in a derivative expansion \cite{Jana:2020vyx}. More 
precisely, if $ \Gin$ is known till $n^\text{th}$ order in derivative expansion, $\sen{\ann}^{\SKs}$ can be determined
to $(n-1)^\text{th}$ order in derivative expansion. As indicated at several points in our discussion these statements should not be surprising in the Markovian designer scalar context. The general structure largely parallels the massive scalar probes studied in the aforementioned reference.

We are now ready to compute the one-point function in the presence of sources, by examining the right/left normalizable 
modes. This, as presaged, requires the inclusion of the appropriate counterterms determined hitherto. Putting all the pieces computed in \cref{sec:ccbdyM} we  obtain
\begin{equation}
\expval{\mathcal{O}_{\skL/\skR} } = \lim_{r\to\infty\pm i0}\left[ -r^\ann \Dz_+ \sen{\ann} + \ctphi{2} \sqrt{f}\ r^{\ann-1}\left(\partial_i\partial_i-\frac{1}{f}\partial_v^2\right)\sen{\ann} +\ldots\right] .
\end{equation}
This yields the expressions for the right and left one-point functions to be 
\begin{equation}\label{eq:MoneptJ}
\begin{split}
\expval{\mathcal{O}_\skR(\omega,\bk)}
&= 
	-\Kin(\omega,\bk) \left[(\nB+1)\: \JMar_\skR - \nB \: \JMar_\skL \right] + \nB\, \Krev(\omega,\bk) \left[\JMar_\skR-\JMar_\skL\right]\ ,\\
\expval{\mathcal{O}_\skL(\omega,\bk)}
&=
	 -\Kin(\omega,\bk) \left[(\nB+1)\: \JMar_\skR - \nB \: \JMar_\skL \right] + (\nB+1) \Krev(\omega,\bk) \left[\JMar_\skR-\JMar_\skL\right]\,.
\end{split}
\end{equation}
Here, we have used the Bose-Einstein identity $\nB\ e^{\beta\omega}=\nB+1$ and  defined  the renormalized reversed Green's function
\begin{equation}\label{eq:KrevM}
\begin{split}
\Krev(\omega,\bk) \equiv \Kin(-\omega,\bk) \,.
\end{split}
\end{equation}

In carrying out the computations it is helpful to note  that the divergence and counterterm structures in this case only involve even powers of $\omega$. Consequently, the counterterms that cancel the divergences of $ \Gin$ also cancels the divergences in $\Grev$ using the observations of \cite{Jana:2020vyx} mentioned above. A useful relation in this regard is 
\begin{equation}\label{eq:Mrevrel}
\mathbb{D}_+\left[ \Gin(-\omega,\bk)e^{-\beta\omega\ctor}\right]
	=\left[\Dz_+ \Gin(\omega,\bk)\right]_{\omega\to-\omega}e^{-\beta\omega\ctor}\,.
\end{equation}   

Let us also record the analog of \eqref{eq:MoneptJ} in the average-difference basis. Taking a linear combination we end up with
\begin{equation}
\begin{split}
\expval{\mathcal{O}_a(\omega,\bk) }
&=
	 -\Kin \, \JMar_a-\left(\nB+\frac{1}{2}\right)\left[\Kin-   \Krev \right]\JMar_d\ ,\\
\expval{\mathcal{O}_d(\omega,\bk) }
&=
	 - \Krev\,\JMar_d\,.
\end{split}
\end{equation}
We see here the characteristic upper triangular structure of two point functions in the average-difference basis with 
the $\expval{\mathcal{O}_a \, \mathcal{O}_d}$ and $\expval{\mathcal{O}_d \mathcal{O}_a}$ corresponding  to retarded and advanced  Green's functions, respectively. The  Keldysh Green's function $\expval{\mathcal{O}_a \mathcal{O}_a}$  is an even function of the frequency whose derivative expansion is given by 
\begin{equation}\label{eq:MKeld}
\begin{split}
\left(\nB+\frac{1}{2}\right)\left[\Kin -   \Krev\right]
& = 
	\frac{d}{2\pi i }   \left(\frac{4\pi}{d\,\beta} \right)^{\ann+1}\,  
	  \left(1+ \left(\frac{4\pi }{d}\right)^2 \frac{\bw^2}{12} \right) \\
&
	\qquad \times
	\left(1 -2 \,\bq^2 \, H_k(\ann,1)- 2\,\bw^2\,  H_\omega(\ann,1) +\cdots\right) .
\end{split}
\end{equation}
We have used the explicit form of the retarded Green's function \eqref{eq:KinMark} to derive the above.
We see that in the Markovian case, the one-point functions are  given by local expressions, i.e., the CFT one-point functions 
at a CFT spacetime point depend on the value of the CFT source and its derivatives at that point.

Alternatively, these one-point functions can also be computed by varying the CFT influence phase with respect to the CFT  sources.
The computation of this influence phase proceeds by generalizing the GKPW method of evaluating the on-shell action to grSK
geometry as described in \cite{Jana:2020vyx}. The generalized Klein-Gordon action with the boundary counterterms  $S[\sen{\ann}]$ defined in \eqref{eq:KGDilCt} evaluates on-shell to a pure boundary term  (cf., the discussion in Section 5.1 of \cite{Jana:2020vyx}). Evaluating this in the boundary Fourier domain we find the on-shell action:\footnote{ We employ the notational shorthand
$\int \frac{d\omega}{2\pi}\, \frac{d^{d-1}\bk}{(2\pi)^{d-1}}  \equiv \int_k$ to keep the expressions compact.}
\begin{equation}\label{eq:Moshellact}
S[\sen{\ann}]\bigg|_\text{on-shell} =
	-\frac{1}{2} \,\lim_{r_c\to \infty} \, \int_k \Bigg[  \sen{\ann}^{\dag}
		\left\{ r^{\ann}\Dz_+ + \ctphi{2} \, r^{\ann-1} \, \sqrt{f}\big(k^2 -\frac{1}{f}\omega^2 \big) +\cdots\right\}  \sen{\ann} 
		\Bigg]^{r=r_c-i0}_{r=r_c+i0} \,.\\
\end{equation}
Using the explicit solution this can be shown to be
\begin{equation}\label{eq:MSeffLR}
\begin{split}
S[\sen{\ann}]\bigg|_\text{on-shell} =
	& -\frac{1}{2}\int_k\, \left(\JMar_\skR-\JMar_\skL\right)^\dag \Kin\bigg((\nB+1)\: \JMar_\skR - \nB \: \JMar_\skL \bigg)\\  
&\qquad 	
	+\frac{1}{2}\int_k\, \bigg(\nB \JMar_\skR -  (\nB+1) \JMar_\skL\bigg)^\dag \Krev \,\bigg(\JMar_\skR-\JMar_\skL\bigg)\\
&=
	 -\int_k\, \left(\JMar_\skR-\JMar_\skL\right)^\dag \Kin\, \bigg((\nB+1)\: \JMar_\skR - \nB \: \JMar_\skL \bigg)\,.
\end{split}
\end{equation}
In the last line, we have redefined $\omega\to-\omega$ in the second integral and have  used the identity $1+\nB(\omega)+\nB(-\omega)=0$.  The last  line represents the quadratic influence phase written in  the advanced-retarded (RA) basis.   In the average-difference or Keldysh basis, we get
\begin{equation}\label{eq:MSeffad}
\begin{split}
S[\sen{\ann}]\bigg|_\text{on-shell} =
& 
	-\int_k\, {\JMar_d}^\dag \Kin\left[\JMar_{a}+\left(\nB+\frac{1}{2}\right)\: \JMar_d \right]\\
&= 
	-\int_k\, \left[{\JMar_d}^\dag \Kin\JMar_{a}+\frac{1}{2}\left(\nB+\frac{1}{2}\right)\:
 	{\JMar_d}^\dag \left(\Kin-\Krev\right) \JMar_d \right]\,.
\end{split}
\end{equation}
This is the standard structure of quadratic influence phase for an open system in contact with thermal bath
\cite{Schwinger:1960qe,Feynman:1963fq}. The explicit expression for the Green's functions has already been recorded in \eqref{eq:KinMark}.

\subsection{Non-Markovian probes}
\label{sec:nMarkprobe}

Let us now turn to the non-Markovian probes. All of the above steps can be repeated in parallel with some minor modifications as elucidated in \cref{sec:trscalar} and \cref{sec:scalarnM}. We analytically continue $\ann$ to $-\ann$ and in this process the CFT sources $\JMar$  of the Markovian probe morph into the long-distance open EFT fields $\snMar$ of the non-Markovian probe. The analog of 
\eqref{eq:SonTeaney} now reads:
\begin{equation}\label{eq:SonTeaneyNM}
\begin{split}
\sen{-\ann}^{\SKs}(\omega,\ctor, \bk) 
&= 
	\GinN \, \left[(\nB+1)\, \snMar_\skR - \nB \, \snMar_\skL \right]
- \GrevN \, \nB \, \left[\snMar_{R}-\snMar_\skL\right]\  e^{\beta \omega(1-\ctor)}\ \\
&= \GinN \,  \snMar_a+\left[\left(\nB+\frac{1}{2}\right)\GinN \,  - \nB \, e^{\beta \omega(1-\ctor)}\,  \GrevN \right]\snMar_d\,.
\end{split}
\end{equation}
where 
\begin{equation}\label{eq:keldnM1}
\snMar_a\equiv  \frac{1}{2} \left( \snMar_{R}+ \snMar_\skL\right)\ ,\qquad
\snMar_d\equiv \snMar_{R}-\snMar_\skL\,.
\end{equation}

One point functions can be computed using the right or left normalizable modes with the appropriate counterterms
determined before in \cref{sec:ccbdynM}. We have
\begin{equation}
\expval{\mathcal{O}_{\skL/\skR} } 
= 
	\lim_{r\to\infty\pm i0} \left\{ \sen{-\ann} - \frac{1}{\sqrt{f}}\left[- \frac{\ctpi{0}}{r} + \frac{\ctpi{2}}{r^3}
		\left(k^2-\frac{1}{f}\omega^2\right)+\cdots\right]\mathbb{D}_+\sen{-\ann}  \right\} .
\end{equation}
The divergence and counterterm structures again involve only even powers of $\omega$,
so the counterterms that cancel the divergences of $\GinN \, $ also cancels the divergences
in $\GrevN $. By explicit computation we verify again the analog of \eqref{eq:pinMcts} in the grSK geometry, 
\begin{equation}\label{eq:nM1ptfn}
\expval{\mathcal{O}_{\skL/\skR} } =\snMar_{\skL/\skR}\ ,
\end{equation}
 viz., the long-distance open EFT fields  $\snMar$ in the CFT can be identified with the dual single-trace primary. 
 A useful identity in deriving the above  relation is
\begin{equation}\label{eq:nMrevrel}
\begin{split}
\lim_{r \to \infty} \, e^{-\beta \omega(1-\ctor)} 
\left\{1 -  \frac{1}{\sqrt{f}}\left[-\frac{\ctpi{0}}{r} +\frac{\ctpi{2}}{r^3}\left(k^2-\frac{1}{f}\omega^2\right)+\cdots\right]\Dz_+  \right\}
	\Big[\GrevN\;e^{\beta \omega(1-\ctor)} \Big]  =1 \,.
\end{split}
\end{equation}

Let us now consider the non-normalizable modes of the non-Markovian designer scalar. These can be computed from the definitions 
\begin{equation}
\begin{split}
\JnMar_\skL &= -  \lim_{r\to\infty + i0}\ \cpen{-\ann} =  \lim_{r\to\infty + i0}\ r^{-\ann}\Dz_+ \sen{-\ann} \,,  \\  
\JnMar_\skR &= -  \lim_{r\to\infty - i0}\ \cpen{-\ann} =  \lim_{r\to\infty - i0}\ r^{-\ann}\Dz_+ \sen{-\ann}  \,, 
\end{split}
\end{equation}
leading to the expressions on the grSK geometry:
\begin{equation}
\begin{split}
\JnMar_\skR 
&= 
	\KinN \left[(\nB+1)\, \snMar_\skR - \nB \: \snMar_\skL \right]  - \nB\KrevN \left[\snMar_{R}-\snMar_\skL\right]\ ,\\
\JnMar_\skL 
&= 
	\KinN \left[(\nB+1)\: \snMar_\skR - \nB \: \snMar_\skL \right] - (\nB+1)\KrevN \left[\snMar_{R}-\snMar_\skL\right]\ ,
\end{split}
\end{equation}
where $\KinN$ is the dispersion function defined in \eqref{eq:Kdispdef}. For completeness we quote here the third order formula derived in \eqref{eq:Kmn3rd}:
\begin{equation}\label{eq:KinnMark}
\begin{split}
\KinN(\omega,\bk)
&=
	b^{\ann-1}\left[-i\, \bw+\frac{ \bq^2}{\ann+1}+  \bw^2\, \Delta(\ann,1) 
	+i \,\bw^3\left( \Delta(\ann,1)^2 - 2\,H_\omega(\ann,1) \right)  \right.\\
&\left.
	\qquad \qquad\quad
		+\,2i \,\frac{\bw\, \bq^2}{\ann+1} \left(\Delta(\ann,1) - (\ann-1)\, H_k(\ann,1) \right) + \cdots \right] 
\end{split}
\end{equation}
It is straightforward to verify, using the values of the gradient expansion functions \eqref{eq:Deltahor} and \eqref{eq:Fhkomhor}, that the above reduces to the  retarded Green's function of the Markovian probe with the replacement $\ann \to -\ann$.

Taking  the average and the difference of the equations above, we obtain the Keldysh basis sources 
\begin{equation}\label{eq:sknMeoms}
\begin{split}
\JnMar_a 
&= 
	\KinN  \snMar_a+\left(\nB+\frac{1}{2}\right)\left[\KinN  -  \KrevN\right]\snMar_d\ ,\\
\JnMar_d 
&=
	\KrevN\snMar_d\,.
\end{split}
\end{equation}
The combination multiplying $\snMar_d$ in the expression for $\JnMar_a$ can be simplified to:
\begin{equation}\label{eq:nMKeld}
\begin{split}
\left(\nB+\frac{1}{2}\right)\left[\KinN  -  \KrevN\right]
&=
 \frac{d}{2\pi i }  \left(\frac{d\,\beta}{4\pi}\right)^{\ann-1} 
 	\left(1+ \left(\frac{4\pi }{d}\right)^2 \frac{\bw^2}{12} \right)\\
&\qquad 
	\times\left(1-2 
	 \left[ \bq^2 \, H_k(-\ann,1)	 + \bw^2 \, H_\omega(-\ann,1)\right]
	 +\cdots\right) ,\\
&=
	 \frac{d}{2\pi i }  \left(\frac{d\,\beta}{4\pi}\right)^{\ann-1}
	 	\left(1+ \left(\frac{4\pi }{d}\right)^2 \frac{\bw^2}{12} \right)\\
&\qquad 
	\times\left(1+2\,  
	\left[
	\bq^2  \frac{ (\ann-1)\, H_k(\ann,1)- \Delta(\ann,1)}{\ann+1} \right. \right. \\
&\left.\left.	 
	\qquad  \qquad \qquad\quad
		+\; \bw^2 \left(H_\omega(\ann,1)-\frac{1}{2}\Delta(\ann,1)^2 \right)\right]
	+\cdots\right) \,.
\end{split}
\end{equation}
The equations that we have derived above should really be thought of as the local equation of motion (or Schwinger-Dyson equations) for the open EFT of the fields $\snMar_{\skR,\skL}$. These equations can be solved for the effective fields $\snMar_{\skR,\skL}$ in terms of the SK sources $\JnMar_{\skR,\skL}$ to yield Schwinger-Keldysh Green's functions.

To wrap up the discussion let us also demonstrate that we might have equivalently evaluated the on-shell bulk action which gives the Wilsonian Influence Functional, the effective  action corresponding to the  above equations of motion. The designer non-Markovian scalar action with the boundary counterterms \eqref{eq:KGDilctnM} evaluates on-shell to a pure boundary term, in analogy with 
\eqref{eq:Moshellact} which in the boundary Fourier domain takes the form 
\begin{equation}\label{eq:nMoshellact}
\begin{split}
&S[\sen{-\ann}] \bigg|_\text{on-shell} 
=
	-\frac{1}{2} \, \lim_{r_c\to \infty} \mathfrak{I} \\
&\mathfrak{I} 
= 
	\int_k\, \Bigg[  \left( \sen{-\ann} + \frac{1}{\sqrt{f}} 
 		\left[ \frac{\ctpi{0}}{ r}-\frac{\ctpi{2}}{r^3}\left(k^2-\frac{1}{f}\omega^2\right)+\cdots\right] \Dz_+\sen{-\ann}  \right)^{\dag}
 		r^{-\ann}\Dz_+ \sen{-\ann}\Bigg]^{r=r_c-i0}_{r=r_c+i0}  \,,\\
\end{split}
\end{equation}
Using the explicit solution for the non-Markovian field on the grSK contour we obtain: 
\begin{equation}\label{eq:nMSeffLR}
\begin{split}
S[\sen{-\ann}] \bigg|_\text{on-shell}
&  =
	-\frac{1}{2}\int_k\, \left[\snMar_{R}-\snMar_\skL\right]^\dag \KinN \left[(\nB+1)\: \snMar_\skR - \nB \: \snMar_\skL \right]\\  
&
	\qquad \qquad
		+ \frac{1}{2}\int_k\, \left[\nB \snMar_\skR -  (\nB+1) \snMar_\skL\right]^\dag \Krev \left[\snMar_{R}-\snMar_\skL\right]\\
&= 
	-\int_k\, \left[\snMar_{R}-\snMar_\skL\right]^\dag \KinN \left[(\nB+1)\: \snMar_\skR - \nB \: \snMar_\skL \right]\,.
\end{split}
\end{equation}
In the last line, we have redefined $\omega\to-\omega$ in the second integral and have  used the identity $1+\nB(\omega)+\nB(-\omega)=0$. This is the expression for the Wilsonian influence functional in the retarded-advanced (RA) basis. In the average-difference or Keldysh  basis, we get 
\begin{equation}\label{eq:nMSeffad}
\begin{split}
S[\sen{-\ann}] \bigg|_\text{on-shell}
&=
	- \int_k\, {\snMar_d}^\dag \,\KinN \left[\snMar_{a}+\left(\nB+\frac{1}{2}\right)\: \snMar_d \right]\\
 &=
 	-\int_k\, \left[{\snMar_d}^\dag \KinN \snMar_{a}+\frac{1}{2}\left(\nB+\frac{1}{2}\right)\:
 {\snMar_d}^\dag \left(\KinN -\KrevN\right) \snMar_d \right]\,.
\end{split}
\end{equation}
By further turning on external sources for the hydrodynamic moduli of the form: 
\begin{equation}\label{eq:}
\int_k \left[\JnMar_\skR  \snMar_\skR^\dag  -  \JnMar_\skL  \snMar_\skL^\dag\right]
= \int_k\left[ \snMar_d^\dag \, \JnMar_a +  \snMar_a \, \JnMar_d^\dag \right] \,,
\end{equation}	
and varying the effective action, we get the hydrodynamic equations of  motion quoted  above in  \eqref{eq:sknMeoms}.
This shows that the bulk  on-shell  action  does indeed  generate the correct hydrodynamic equations   of motion.

\subsection{The Gaussian Wilsonian influence functional}
\label{sec:}

To wrap up the discussion let us consider an example of a probe system which comprises of a non-interacting pair of Markovian and non-Markovian  fields with Markovianity indices $\ann_1$ and $-\ann_2$, respectively. Using the results of the preceding sections we can now write down the final answer for the quadratic approximation to the influence phase. We have 
\begin{equation}\label{eq:WIFscalar}
\begin{split}
\mathcal{S}_\text{WIF}[\JMar_a, \JMar_d, \snMar_a, \snMar_d] 
&= 
	- \int_k  \left\{\JMar_d^\dagger \,K_{_{\ann_1}}^\In \left[\JMar_{a}+\left(\nB+\frac{1}{2}\right)\: \JMar_d 	\right] + {\snMar_d}^\dag \,
		K_{_{-\ann_2}}^\In \left[\snMar_{a}+\left(\nB+\frac{1}{2}\right)\: \snMar_d \right]\right\}
\end{split}
\end{equation}	
where  the retarded Green's function to the third order in the gradient expansion are  given in  \eqref{eq:KinMark}  and \eqref{eq:KinnMark}, respectively. As we deduced earlier these functions for the Markovian and non-Markovian fields have the same functional form.  We record here for completeness the function $\Kin$ with all the factors fixed
\begin{equation}\label{eq:KinMark3ex}
\begin{split}
\Kin(\omega,k)&=
	\left(\frac{4\pi}{d\, \beta}\right)^{\ann+1}
		\left[-i\,\bw-\frac{\bq^2}{\ann-1}
	- \bw^2 \,  \Delta(\ann,1) +   2i  \,  \bw \bq^2 \, H_k(\ann,1) 
	+ 2i\, \bw^3 \, H_\omega(\ann,1)+\cdots
	 \right] \,.
\end{split}
\end{equation}	
with the parameters
\begin{equation}\label{eq: fnsathor}
\begin{split}
\Delta(\ann,1)
&= 
	\frac{1}{d} \left[\psi\left( \frac{\ann+1}{d}\right) - \psi\left(\frac{1-\ann}{d} \right) \right]\\
H_k(\ann,1) 
&= 
	  \frac{1}{d\,(\ann-1)} \left[ \psi\left( \frac{\ann+1}{d}\right) - \psi\left(\frac{2}{d}\right) \right] \\
H_\omega(\ann,1)
&=
	\frac{\Delta(\ann,1)}{2d}\, \harm{\frac{\ann+1}{d}-1} +H_{\omega}^{(2)}(\ann,1)\\
H^{(2)}_\omega(\ann,1) 
&= 
	- \frac{\ann}{d}\, \sum_{n=0}^\infty \,  \frac{\harm{n-1+\frac{2}{d}}}{(n d + \ann-1)  (n d + 1-\ann) }	
\end{split}
\end{equation}	
We have written the result in terms of the digamma function $\psi(x)$ and the related Harmonic number function $\harm{x}$
which has been used before in the fluid/gravity literature.

\section{Time-reversal invariant gauge system}
\label{sec:trgauge}

We now turn to the study of the Maxwell analogue of the designer scalar and explore the  dynamics of an Abelian gauge field coupled to the background geometry and an auxiliary dilaton introduced in \cref{sec:designer}. We will demonstrate below the following  claim: the dynamics of this designer gauge system can be completely encoded, in a gauge invariant manner, in two designer scalars. This statement then  generalizes the observation made earlier in \eqref{eq:maxVpar} for a pure Maxwell theory (i.e., $\ann = d-3$). We will sketch the essence of the argument below, supplementing our discussion with further details in \cref{sec:gaugeapp}. In \cref{sec:gravity} we will see that similar statements apply to linearized gravitational perturbations. 
 
Several authors have attempted in the past to give a Schwinger-Keldysh description for the bulk gauge theory and gravity 
\cite{Nickel:2010pr,Crossley:2015tka,deBoer:2015ija,Glorioso:2018mmw,deBoer:2018qqm}. Our treatment here is substantially different with a focus on the physics of outgoing Hawking modes (the dynamics of infalling quasinormal modes  has been well understood since the early works of \cite{Policastro:2001yc,Policastro:2002se,Policastro:2002tn} and in the fluid/gravity context \cite{Bhattacharyya:2008jc,Hubeny:2011hd}).  As in the scalar problem, we will explicitly keep track of the origin and the effects of the Bose-Einstein distribution in the effective dynamics of the Markovian and non-Markovian modes. 

Let us first highlight the key aspects where our discussion differs   from earlier literature  for systems with bulk gauge symmetry (either abelian gauge symmetry or diffeomorphism). 
\begin{enumerate}[wide,left=0pt]
 \item We eschew the use of  the radial gauge which is commonly employed in AdS/CFT discussions. 
In this gauge outgoing Hawking modes are tricky to explore as the gauge explicitly breaks the $\mathbb{Z}_2$ time reversal isometry. 
The solution to this issue is straightforward: we adopt a gauge invariant scheme for solving the field equations.
 \item Systems with gauge invariance contain both Markovian and non-Markovian modes. Ideally one would like to disentangle them, so that one can integrate out the Markovian modes, whilst keeping the non-Markovian modes off-shell. Fortuitously for us a plane wave decomposition of the perturbation naturally serves to decouple the modes. It furthermore maps them onto the scalar problem we explored earlier.
 \item We will solve the radial Gauss constraints arising from the gauge invariance which are dual to boundary conservation equations. In the non-Markovian sector we turn on appropriate sources  to keep these modes off-shell.
 \end{enumerate}

\subsection{Decomposition of gauge field modes}
\label{sec:gaugedyn}

We start with the designer gauge system introduced in \cref{sec:designer}. The equations of motion arising from the action \eqref{eq:MaxDil}  are simply 
\begin{equation}\label{eq:MaxDileom}
\partial_A \left(\sqrt{-g}\ r^{\ann+3-d}\;\Cm^{AB} \right)=0
\end{equation}	
We can examine these directly for the gauge field strengths, or pass as usual to the conventional parameterization in terms of the gauge potential $\Vm_A$.  We will find it convenient to expand the potential $\Vm$ in terms of plane wave harmonics on $\mathbb{R}^{d-1,1}$. This will have the advantage that we will be easily able to decouple the Markovian and non-Markovian degrees of freedom contained in these equations. We let\footnote{ The notation for the planar harmonic components of the gauge fields being barred is intentional. Later when we discuss gravity dynamics in \cref{sec:gravity} we will employ closely related notation without the  decoration to denote the  gravitational degrees of freedom. \label{fn:gaugenotation}}

\begin{equation}\label{eq:Vharmonic}
\begin{split}
\Vm_r(v, r,\bx) 
&=
	\int_k  \, \sMax_r(r,\omega,\bk)\, \ScS(\omega,\bk|v,\bx)\,, \\
\Vm_v(v, r, \bx) 
&=
	\int_k \,\sMax_v(r,\omega,k)\, \ScS(\omega,\bk|v,\bx) \ ,\quad\\
 \Vm_i (v, r, \bx) 
 &=
 	\int_k \left[\sum_{\ai=1}^{N_V}\, \vMax_\ai(r,\omega,\bk) \, \VV^\ai_i(\omega,\bk|v,\bx)
 	+ i\,\sMax_x(r,\omega,\bk)\,  \ScS_i(\omega,\bk|v,\bx)\right] \\
 &=
 	\int_k \left[\sum_{\ai=1}^{N_V} \, \vMax_\ai(r,\omega,\bk) \,\VV^\ai_i(\omega,\bk|v,\bx)
 		- \sMax_x(r,\omega,\bk)\,  \frac{k_i}{k} \,\ScS(\omega,\bk|v,\bx)\right] .
\end{split}
\end{equation}
Here $\ai = 1, 2, \ldots, N_V=d-2$  labels the different vector polarizations of the gauge field. Our conventions for the harmonics are summarized in \cref{sec:harmonics}. The main point to note that the decomposition is in an orthonormal basis which allows us to decouple the modes of the gauge field.

Plugging in the harmonic decomposition into \eqref{eq:MaxDileom} and exploiting the decoupling  of the transverse vector sector 
from the transverse scalar sector, we find the following:
\begin{itemize}[wide,left=0pt]
\item There is a single equation in the transverse vector sector which can be identified with that of a time-reversal invariant designer scalar with exponent $\ann$. Specifically, all the fields $\vMax_\ai$ satisfy:
\begin{equation}\label{eq:Vveom}
\begin{split}
\frac{1}{r^{\ann}} \Dz_+ \left(r^{\ann}\,\Dz_+\vMax_\ai \right)+\left(\omega^2-k^2f\right)\vMax_\ai &=0\,. 
\end{split}
\end{equation}
Comparing with \eqref{eq:gseom1} we obtain $\vMax_\ai = \sen{\ann}$ for $\ai = 1, 2, \ldots, N_V$.
\item In the scalar sector we find a set of three coupled differential equations for the fields $\sMax_v$, $\sMax_r$, and $\sMax_x$. We introduce the following gauge invariant combinations of these fields: 
\begin{equation}\label{eq:ginvars}
\begin{split}
\psMax_v &\equiv \dv{\sMax_v}{r} + i\omega  \, \sMax_r \,, \qquad
\pJMax_r \equiv \dv{\sMax_x}{r}+ik \,\sMax_r \,, \\ 
\psMax_x &\equiv \Dz_+\sMax_x+ik\,\sMax_v+ik\, r^2 f \,\sMax_r  \equiv r^2 f\, \pJMax_r + \pJMax_v\,,\\
\pJMax_v &\equiv i\, k\, \sMax_v  - i\omega\, \sMax_x \,,
\end{split}
\end{equation}	
in terms of which we find compact expressions for the equations of motion:
\begin{equation}\label{eq:TRIgauge}
\begin{split}
\EMax_v &\equiv 
	\dv{r}(r^{\ann+2} \,\psMax_v ) -ik\, r^{\ann} \,  \pJMax_r  = 0\,,\\
\EMax_x &\equiv
	 \dv{r} (r^{\ann} \,\psMax_x) -i\omega\, r^{\ann} \, \pJMax_r  = 0\,,\\
\EMax_r &\equiv 
	  -i\omega\, r^{\ann+2} \, \psMax_v +ik\, r^{\ann} \,\psMax_x =0\,.
\end{split}
\end{equation}
It is easy to check that the combinations above are closely related to the field strengths in Fourier domain
\begin{equation}\label{eq:FPiMax}
\Cm_{rv} = \psMax_v\,, \qquad \Cm_{ri} = -\frac{k_i}{k} \,\pJMax_r \,, \qquad \Cm_{vi} = -\frac{k_i}{k}\, \pJMax_v \,.
\end{equation}	
\end{itemize}

One can understand our gauge invariant combinations by realizing that the gauge parameter also admits a decomposition in the plane wave harmonics, i.e.,
\begin{equation}
\begin{split}
\Lambda_g(v, r,\bx) =\int_k\, \Lambda(r,\omega,\bk)\, \ScS(\omega,\bk|v,\bx)\,.
\end{split}
\end{equation}
It then follows that gauge transformations leave $\vMax_\ai $ invariant whereas  the scalar sector variables get shifted as 
\begin{equation} \label{eq:gtransfV}
\begin{split}
\sMax_r \mapsto \sMax_r+ \dv{\Lambda}{r}\ ,\quad 
\sMax_v \mapsto \sMax_v -i\omega\,  \Lambda \ ,\quad 
\sMax_x \mapsto \sMax_x -ik\, \Lambda\,,
\end{split}
\end{equation}
which confirms that the combinations defined in \eqref{eq:ginvars} are indeed invariant.

The transverse vector modes, which as we noted above, map simply to a collection of $N_V = d-2$ designer scalars.  Recalling that Markovianity demands $\ann >-1$, and that  in the pure Maxwell case with no dilatonic coupling, $\ann = d-3$, we see that transverse modes are Markovian when $d>2$ and marginally non-Markovian for $d=2$.\footnote{ Note that this is in keeping with the usual expectation. For a Maxwell field in \AdS{d +1} with $d\leq 2$,  the standard boundary conditions are inadmissible \cite{Marolf:2006nd}. The only physically sensible boundary conditions are to freeze the charge on the boundary and leave the currents unconstrained.}  Physically, this sector describes the real electromagnetic waves or photons (which exist when $d>2$) that  fall into the black brane and are, in turn, Hawking radiated out.  The Markovian property tells us that this happens at a short time  scale of order the inverse Hawking temperature $\beta$ of the black brane.  At time scales much larger than $\beta$, one can integrate out the effects of this physics to get a local description for the remaining degrees of freedom.

The equations in the transverse scalar sector themselves comprise a gauge system as they retain all  the characteristics of the underlying gauge field. We will refer to this sector as the \emph{diffusive gauge system} and it has the following additional properties. A characteristic feature of any gauge system is the Bianchi identity. From \eqref{eq:TRIgauge} we  can see the combination
$i\omega\, \EMax_v -ik\, \EMax_x  + \dv{r} \EMax_r = 0$.  In other words, the third equation, $\EMax_r$ can be considered as the  Gauss constraint which is preserved under radial evolution described by the first two equations.  One consequence of the Bianchi identity  is that if the first two equations hold everywhere, then it is sufficient to impose the third equation only at the radial slice at infinity. 

These statements are quite familiar in AdS/CFT. Recall that a bulk gauge symmetry implies a boundary global symmetry by the 
standard rules of AdS/CFT. Up to an overall normalization and counterterms, the Noether charge density and Noether current density of the conserved current are given by 
\begin{equation}\label{eq:bchargecurrent}
\Jcft_v = - \lim_{r\to r_c} r^{\ann+2}\, \psMax_v \,, \qquad
\Jcft_i = \frac{k_i}{k} \lim_{r\to r_c} r^{\ann}\, \psMax_x \,.
\end{equation}	

With this understanding the final equation in \eqref{eq:TRIgauge} is simply that of current conservation $\nabla_\mu (\Jcft)^\mu =0$ on the boundary. This statement in itself implies that we should expect long-lived charge diffusion modes to be present in the system in the scalar sector. This is indeed the case for standard Maxwell dynamics ($\ann = d-3$) in the black hole background \cite{Policastro:2002se,Kovtun:2003wp}, which encodes the physics of ohmic conductivity. 

Before we proceed to establish the connection, we note that the designer gauge system is time reversal invariant. In particular, one can rewrite \eqref{eq:TRIgauge} in a time-reversal invariant form  
\begin{equation}
\begin{split}
\Dz_+\left(r^{\ann+2}\ \psMax_v \right)+ ik\ r^{\ann}\,\pJMax_v &=0\,, \\ 
\Dz_+\left( r^{\ann}\, \psMax_x\right) +i\omega\, r^\ann\,\pJMax_v&=0\,,\\ 
 -i\omega\, r^{\ann+2}\,\psMax_v+ik\, r^{\ann}\,\psMax_x&=0\,.
\end{split}
\end{equation}
In writing the first equation  in the covariant form above, we have used the radial Gauss law constraint. It is helpful to note that $\{r^2f\,\sMax_r +\sMax_v, \sMax_x, \psMax_x\}$ have positive intrinsic time reversal parity while $\{\sMax_v, \pJMax_{v}, \psMax_v\}$ all have negative intrinsic time reversal parity. Details of these facts are explained in \cref{sec:gaugeapp}.

\subsection{The non-Markovian charge diffusion scalar}
\label{sec:scalargauge}

Now that we have explored the basic equations for the diffusive gauge system, we will argue for a solution of \eqref{eq:TRIgauge} in terms of a single non-Markovian scalar degree of freedom. Introduce a field $ \sMaxD$, and fix the variables $\psMax_v,\pJMax_r, \psMax_x$ as follows:
\begin{equation}\label{eq:MasterGaugeInv}
\begin{split}
 r^{\ann} \,\pJMax_r =-\dv{r}(ik\,  \sMaxD)\,, \quad 
 r^{\ann} \,\psMax_x = -i\omega (ik\,  \sMaxD)\,, \quad
  r^{\ann+2}\, \psMax_v  &= -ik (ik\,  \sMaxD )\,.
\end{split}
\end{equation}
This clearly solves \eqref{eq:TRIgauge}. To understand the constraints on $\sMaxD$ we proceed as follows: the first and the second equations  in \eqref{eq:MasterGaugeInv} can be combined to solve for $\psMax_v$, obtaining  
\begin{equation}
\begin{split}
r^{\ann} \,\pJMax_{v} =ik\, \Dz_+ \sMaxD 	\;\; \Longrightarrow \;\;
\sMax_v=r^{-\ann}\Dz_+\sMaxD+\frac{\omega}{k} \sMax_x \,.
\end{split}
\end{equation}
We have used here  the definition of $\pJMax_v$ given in \eqref{eq:ginvars}.   Substituting this back into the third equation 
in \eqref{eq:MasterGaugeInv}, we find:
\begin{equation}
\begin{split}
 r^{\ann+2}\left(\dv{\sMax_v}{r}+i\omega\,\sMax_r\right)
 &= 
 	r^{\ann+2}\left(\dv{r}(r^{-\ann}\Dz_+ \sMaxD ) +\frac{\omega}{k}\left(\dv{\sMax_x}{r}+ik\, \sMax_r\right)\right)\\
 &= r^{\ann+2}\dv{r} (r^{-\ann}\Dz_+ \sMaxD)  -i\omega \,r^2\, \dv{ \sMaxD}{r} \\
  &=  \frac{1}{f}\left(r^{\ann} \Dz_+\left(r^{-\ann}\Dz_+ \sMaxD\right)	
  		+\omega^2 \sMaxD \right)  .
\end{split}
\end{equation}
Comparing back again with the third equation, we conclude that our definitions are mutually consistent only if  $ \sMaxD$ satisfies a homogeneous second order ODE, viz., 
\begin{equation}\label{eq:MaxnMeqn}
\begin{split}
\EMaxD[\sMaxD] \equiv  r^{\ann}\, \Dz_+\left(r^{-\ann}\,\Dz_+ \sMaxD\right)+(\omega^2-k^2 f) \sMaxD= 0 \,.
\end{split}
\end{equation}
We can give a gauge invariant derivation of the same result by first writing the field strengths in terms of $\sMaxD$ using \eqref{eq:FPiMax} and \eqref{eq:MasterGaugeInv} and then
demanding that they satisfy the Bianchi identity. The above result proves our assertion that the scalar sector of the designer gauge field, is indeed encoded in the dynamics of a single non-Markovian scalar degree of freedom, viz., $\sMaxD \equiv \sen{-\ann}$. The statement is not new, the idea of using the underlying gauge invariance to isolate the gauge invariant data has, as we indicated earlier in \cref{sec:desorigin}, a long history. A discussion of gauge fields in gravitational backgrounds can be found for example in \cite{Kodama:2003kk}. 

A physical interpretation of this scalar degree of freedom can be easily understood by noting that the Noether charge current is simply:
\begin{equation}\label{eq:chargecurrent}
\Jcft_\mu \, dx^\mu  = - k^2\, \sMaxD \, dv + \omega \, k_i\, \sMaxD\, dx^i \,.
\end{equation}	
This shows that the constant mode of $\sMaxD$, which is a normalizable mode for $\ann>1$, is  actually the
expectation value of  Noether charge density (up to a factor of $k^2$) in the dual field theory. The time derivative of its gradient is the Noether  current density.

To get the exact form of the original vector potential, we need to choose a gauge and invert  the gauge invariant equations in 
\eqref{eq:MasterGaugeInv}. For example one can solve for $\sMax_r, \sMax_v, \sMax_x$ as:
\begin{equation}
\begin{split}
\sMax_r= - \frac{1}{r^{\ann}}\, \dv{\sMaxD}{r} +\dv{\Lambda}{r}\ , \qquad \sMax_v=  \frac{1}{r^{\ann}}\Dz_+\sMaxD -i\omega\,\Lambda 
\ , \qquad \sMax_x=-ik\,\Lambda \ .
 \end{split}
\end{equation}
One simple gauge choice is to set $\Lambda=0$ in the equation above which sets $\sMax_x=0$. In this case, the solution can be written in the manifestly time reversal invariant form,
\begin{equation}\label{eq:debyevM}
\begin{split}
\sMax_r \,  dr+\sMax_v \, dv-\sMax_x \frac{k_i}{k}\,  dx^i= \frac{1}{r^\ann}\left(  dv\ \Dz_+-dr \, \dv{r}\right)\ \sMaxD \,,
 \end{split}
\end{equation}
which is the form appearing in \eqref{eq:maxVpar}. This is analogous to the Debye gauge used in electromagnetism.

Our  parameterization of the transverse scalar components allows for a generalization. We may modify  the last equation of \eqref{eq:MasterGaugeInv} leaving the  first two equations unchanged: one defines  $r^{\ann+2}\, \psMax_v  = -ik (ik\,  \sMaxD ) + \bmu(x)$,  This deformation corresponds to adding a background charge density to the boundary. We have chosen $\bmu $ to be  a function only of the boundary coordinates so that the radial Gauss constraint remains unmodified (equivalently, $\EMax_v$ implies $\pdv{\bmu}{r}=0$).  Turning on $\bmu$ is akin to turning on a source for \eqref{eq:MaxnMeqn}; with $\bmu \neq 0$, the r.h.s is $f(r)\,\bmu$. The conservation equation $\EMax_r =0$  further demands that $-i\,\omega \bmu =0$. As we are not interested in including additional boundary charges we will work with $\bmu =0$.

\subsection{Maxwell action and Wilsonian  influence phase}
\label{sec:Maxact}

We now have a complete description of the gauge fields in terms of a collection of $N_V$ Markovian scalars  and a single non-Markovian scalar. Let us rewrite this in terms of an action including all the boundary terms and counterterms.  We first isolate the diffusive gauge system from the Markovian 
\begin{equation}\label{eq:diffgauge}
\VmD = \Vm \big|_{\vMax_{\ai} =0} \,, \qquad \CmD = \Cm \big|_{\vMax_{\ai} =0}
\end{equation}	
The action for the designer gauge field takes the form: 
\begin{equation}\label{eq:Maxactgen}
\begin{split}
S_{dv} 
&= 
	 -\frac{1}{4} \, \int \, d^{d+1} x\, \sqrt{-g}\ e^{\dilv} \,\Cm^{AB}\Cm_{AB} +  S_\text{ct} \\
S_\text{ct} 
&= 
	-\frac{\ctV{2} }{4} \, \int d^d x \sqrt{-\gamma} \, e^{\chi_v} \,  \Cm^{\mu\nu} \Cm_{\mu\nu} \,, \qquad \ctV{2} = -\frac{1}{\ann-1}\,.
\end{split}
\end{equation}	
We draw attention to the fact that we are imposing standard Dirichlet boundary conditions for the gauge field and thus require no variational counterterm.\footnote{ This can be checked against the AdS analysis of \cite{Marolf:2006nd,Ishibashi:2004wx} since the asymptotic boundary conditions are dictated purely by the near-boundary behaviour (and thus state independent).} The quadratic counterterm is required to cancel subleading divergences at $\order{\omega^2}$ and $\order{k^2}$ and is naturally built out of the covariant gauge strengths projected onto the boundary.

Using the parameterization in terms of the vector polarization scalars $\vMax_\ai$ and the diffusive gauge field $\VmD$ we decouple the Markovian and non-Markovian parts, viz., 
\begin{equation}\label{eq:Maxred1}
\begin{split}
S_{dv}  
&= 
	S_{dv}^{\text{\tiny{M}}}  + S_{dv}^{\text{\tiny{D}}}  \\
S_{dv}^{\text{\tiny{M}}} 
&=
	-\frac{1}{2} \sum_{\ai=1}^{N_V} \bigg[ \int d^{d+1}x  \, \sqrt{-g}\, e^{\chi_s} \, g^{AB}\,\nabla_A\, \vMax_\ai \nabla_B \vMax_\ai 
	+ \ctV{2}\, \int d^d x \, \sqrt{-\gamma}\, e^{\chi_s}\, \gamma^{\mu\nu} \, \nabla_\mu \vMax_\ai \,\nabla_\nu \vMax_\ai \bigg] \\
S_{dv}^{\text{\tiny{D}}}  
&=
	-\frac{1}{4} \left[ \int d^{d+1}x  \, \sqrt{-g}\, e^{\chi_v} \, (\CmD)^{AB} (\CmD)_{AB} 
	 +\ctV{2}\, \int d^d x \sqrt{-\gamma} \, e^{\chi_v} \,  (\CmD)^{\mu\nu} (\CmD)_{\mu\nu} \right]
\end{split}
\end{equation}	

The Markovian dynamics encoded in $N_V$ fields $\vMax_\ai$ matches our earlier scalar discussion, not just at the level of the equation of motion, but also at the level of the variational principle, and the counterterms. Indeed,  the quadratic counterterm $\ctphi{2}$, given in \eqref{eq:ctcfM}, is inherited in the reduction from the gauge field counterterm $\ctV{2}$. This provides a useful cross-check of our analysis in \cref{sec:ccbdyM}.

The diffusive gauge field $\VmD_A$ and its field strength $\CmD_{AB}$ are parameterized in terms of a single scalar degree of freedom  $\sMaxD$.  Substituting the Debye gauge  vector potential \eqref{eq:debyevM} into $S_{dv}^{\text{\tiny{D}}} $ we can write the action in terms of $\sMaxD$.  The bulk term in $S_{dv}^{\text{\tiny{D}}} $ results in a fourth order action that can be massaged into contributions involving the  equation of motion, $\EMaxD[\sMaxD]$, given in \eqref{eq:MaxnMeqn}:
\begin{equation}\label{eq:}
\begin{split}
S_{dv}^{\text{\tiny{D}}}\bigg|_\text{bulk} 
&= 
 	-\frac{1}{4}  \int d^{d+1}x  \, \sqrt{-g}\, e^{\chi_v} \, (\CmD)^{AB} (\CmD)_{AB}  \\
&=
	\frac{1}{2} \int_k\, dr\, r^{-\ann-2}\, \frac{\EMaxD[\sMaxD]}{f} \left(\frac{\EMaxD[\sMaxD]}{f} +k^2\,  \sMaxD\right) + \frac{1}{2}\, 
	\int_k k^2\, r^{-\ann} \,\sMaxD \Dz_+\sMaxD \,.
\end{split}
\end{equation}
Integrating by parts we can convert the term $ \EMaxD[\sMaxD]  \sMaxD $ into the canonical form involving  $\nabla_A \sMaxD$. We may also  absorb the factors of $k^2$ by writing the expression in terms of  spatial derivatives $\partial_i$ on $\mathbb{R}^{d-1,1}$ and write the action in position  space, instead of the Fourier domain expression above. These steps lead us to\footnote{ We ignore all total derivative terms along the boundary directions, i.e., contributions of the  form $\partial_\mu(\cdot)$ are dropped in our action.}  
\begin{equation}\label{eq:}
\begin{split}	
S_{dv}^{\text{\tiny{D}}}\bigg|_\text{bulk} 
&=	- \frac{1}{2}\, \int d^{d+1}x\, \sqrt{-g}\, r^{-\ann+1-d}\, \partial_i \nabla^A \sMaxD \, \partial_i\nabla_A\sMaxD+ 
	 \int d^d x\, r^{-\ann} \, \partial_i\sMaxD \, \partial_i \Dz_+\sMaxD 	 \\
& 
\qquad \qquad
	+ \frac{1}{2} \int_k\, dr\, r^{-\ann-2}\, \left(\frac{\EMaxD[\sMaxD]}{f} \right)^2	 
\end{split}
\end{equation}
 The contribution proportional to $(\EMaxD[\sMaxD])^2$ can be ignored for the variational principle, since its variation vanishes on-shell.  The first two terms in the last equality above, we recognize, modulo a factor of $k^2$ (from $\partial_i)$, to be precisely the action for the non-Markovian scalar \eqref{eq:nMNeumann}. In particular, we emphasize that the reduction does give directly  the variational boundary term $-\cpen{-\ann} \sen{-\ann}$, required to impose Neumann boundary conditions on the non-Markovian scalar. 

This is quite satisfying. While the boundary conditions on the non-Markovian scalar were inferred from the standard asymptotic analysis earlier, it is useful to recognize that one is not imposing an artificial boundary condition for the gauge field in AdS. While some of these statements are implicit in earlier discussions \cite{Ishibashi:2004wx,Marolf:2006nd}, especially in terms of what modes to freeze and which to allow to fluctuate (in global AdS), it is also useful to have a clear derivation at the level of the variational principle. 

Once we recognize this fact we can also check that the zeroth order counterterm for the non-Markovian scalar $\ctpi{0}$ given in \eqref{eq:ctcfnM} follows from the Maxwell counterterm $\ctV{2}$ at the quadratic order. The counterterm $\ctpi{2}$   corresponds to the contribution from two four-derivative counterterms proportional to  
$e^{\chi_v}\, \gamma^{\alpha \beta} \, \Cm^{\mu\nu} \nabla_\alpha \nabla_\beta\Cm_{\mu\nu}$ 
and $e^{\chi_v}\, \gamma^{\alpha \beta} \, \nabla_\alpha\Cm^{\mu\nu}  \nabla_\beta\Cm_{\mu\nu}$, respectively.

All told we can write down the dynamics of the designer Maxwell field as follows:
\begin{equation}\label{eq:dMaxAct}
S_{dv} = \sum_{\ai=1}^{N_V} \, S_{ds}[\sen{\ann}^{\ai}]  +    S_{ds}[\partial_i \sen{-\ann}] 
\end{equation}	
where the actions for the Markovian and non-Markovian scalars are given in \eqref{eq:KGDilCt} and \eqref{eq:nMNeumann} (see also \eqref{eq:KGDilctnM} for the  relevant counterterms), respectively.

We are then in a position to write down the Wilsonian influence phase for the standard Maxwell field which has Markovianity index, $\ann = d-3$. We let the boundary sources for the Markovian vector modes to be $\JMarQ_a^{\ai}$ and $\JMarQ_d^{\ai}$ in the average-difference basis. These correspond to the magnetic components of the boundary gauge field, written out in our plane wave decomposition. For the non-Markovian sector, we recognize that the hydrodynamic mode is the boundary charge mode $\snMarQ$ and parameterize the Wilsonian influence phase using this field. The sources and the moduli are written in our plane wave basis, so we are using the notation above to just talk about the mode coefficients.  We compute the Wilsonian influence phase via a  Legendre transform dropping the  variational boundary term required to implement the Neumann boundary condition. 

We have from \eqref{eq:WIFscalar} the result expressed in terms of the retarded Green's function:  
\begin{equation}\label{eq:WIFMax}
\begin{split}
\mathcal{S}_\text{WIF}[\JMarQ^\ai_a, \JMarQ^\ai_d, \snMarQ_a, \snMarQ_d] 
&= 
	- \int_k  \left\{ \sum_{\ai=1}^{N_V} \, (\JMarQ_d^\ai)^\dagger \,K_{_{d-3}}^\In \left[\JMarQ_{a}^\ai+\left(\nB+\frac{1}{2}\right)\: \JMarQ_d^\ai 	
	\right] 
	\right.\\
&\left.
\qquad \qquad 
	+\; k^2 \,  {\snMarQ_d}^\dag \,
		K_{_{-d+3}}^\In \left[\snMarQ_{a}+\left(\nB+\frac{1}{2}\right)\: \snMarQ_d \right]\right\}
\end{split}
\end{equation}	
Specializing \eqref{eq:KinMark}  and \eqref{eq:KinnMark} to the case $\ann = d-3$, we have the explicit expression for the probe Maxwell field in a planar \SAdS{d+1} background: 
\begin{equation}\label{eq:KinMark3exMax}
\begin{split}
K_{_{d-3}}(\omega,k)&=
	\left(\frac{4\pi}{d\, \beta}\right)^{d-2}
		\left\{-i\,\bw-\frac{\bq^2}{d-4} - \mathfrak{l}_2(d)\, \bw^2+  \frac{2 \pi i }{d(d-4)}   \cot\left(\frac{2\pi}{d} \right)  \bw \bq^2 
	\right. \\
&	\left.  \qquad \qquad \qquad 
	+ \; i \left[ \frac{1}{d^2} \harm{-\frac{2}{d}}\mathfrak{l}_2(d) +\mathfrak{l}_3(d) \right] \bw^3  +\cdots
	 \right\} \,.
\end{split}
\end{equation}	
and 
\begin{equation}\label{eq:KinnMark3exMax}
\begin{split}
K_{_{3-d}}(\omega,k)&=
	\left(\frac{4\pi}{d\, \beta}\right)^{2-d}
		\left\{-i\,\bw+\frac{\bq^2}{d-2}	+ \mathfrak{l}_2(d)\, \bw^2  
		 +\frac{2i }{d(d-2)}  \left[ \psi\left(\frac{2}{d}\right)  - \psi\left( \frac{4-d}{d}\right) \right]   \,  \bw \bq^2
	\right. \\
&	\left.  \qquad \qquad\qquad 
	+ i\left[- \frac{1}{d^2} \harm{\frac{4-2d}{d}}\mathfrak{l}_2(d) -\mathfrak{l}_3(d) \right] \bw^3  +\cdots
	 \right\} \,.
\end{split}
\end{equation}	
 The two parameters $\mathfrak{l}_2(d)$ and $\mathfrak{l}_3(d)$ introduced above are given by 
\begin{equation}\label{eq:}
\begin{split}
\mathfrak{l}_2(d) 
&= 
	 \Delta(d-3,1) = \frac{1}{d} \left[\harm{-\frac{2}{d}} - \harm{\frac{4-2d}{d}} \right] \\
\mathfrak{l}_3(d) 
&=
	 - \frac{2(d-3)}{d}  \, \sum_{n=0}^\infty\, \frac{\harm{n-1+\frac{2}{d}}}{ (n d +4-d) ( n d+d-2)  }
 \end{split}
\end{equation}

We note that while we have eschewed the study of the marginal case $\ann=-1$ which is relevant for the R-charge diffusion of $\mathcal{N}=4 $ SYM, using the dynamics of Maxwell fields in \SAdS{5}. Nevertheless, we can extract by a pole prescription in a $4-\epsilon$ expansion, the diffusion constant from \eqref{eq:WIFMax}.  One finds that it agrees with the prediction of \cite{Policastro:2002se}, viz., $\mathcal{D}= \frac{1}{2\pi T}$ after an appropriate translation of the variables.

\section{Gravitational perturbations}
\label{sec:gravity}

Let us finally turn to  linearized gravitational perturbations about the  \SAdS{d+1} black hole geometry. We will show that a subset of gravitational modes, viz., the tensor and vector sectors, in the plane wave harmonic decomposition, can be mapped as presaged in \cref{sec:desorigin} onto the designer scalar and gauge field respectively. While we believe that the scalar modes of gravity, which are non-Markovian (as they include the propagating sound mode \cite{Policastro:2002se}) can be similarly dealt with, we will not analyze them in this work and defer them to the future.  Since much of the analysis reduces to that of the previous sections, we will be brief  in our presentation. Details on some of the statements here can be found in \cref{sec:appgravity}.

\subsection{Dynamics of transverse tensor and vector gravitations}
\label{sec:ttvgravitons}

We consider linearized metric perturbations about a planar \SAdS{d+1} black hole, where only transverse tensor and vector type perturbations are turned on. Explicitly, we have:
\begin{equation}\label{eq:GRtv}
\begin{split}
ds^2& = \bigg(g_{AB} + (h_{AB})^{\text{\tiny{Tens}}} + (h_{AB})^{\text{\tiny{Vec}}} +  \cancel{(h_{AB})^{\text{\tiny{Scal}}}} \bigg)dx^A dx^B 
\\
 (h_{AB})^{\text{\tiny{Tens}}} \, dx^A\, dx^B 
&=
	r^2 \, \int_k \; \sum_{\bi=1}^{N_T} \, \tGR_\bi(r,\omega,\bk)\,\TT^\bi_{ij}(\omega,\bk|v,\bx)\ dx^i dx^j \ ,\\
 (h_{AB})^{\text{\tiny{Vec}}} \, dx^A\, dx^B 
 &= 
 	r^2 \, \int_k\,  
  	\sum_{\ai=1}^{N_V} \bigg(2\ (\vGR_r^\ai(r,\omega,\bk) \, dr +\vGR_v^\ai (r,\omega,\bk) dv)\,
  	\VV^\ai_i(\omega,\bk|v,\bx) dx^i \\
  & \qquad \qquad \quad 
  	+ i\, \vGR_x^\ai(r,\omega,\bk)\ \VV_{ij}^\ai\,  dx^i dx^j\bigg),
 \end{split}
\end{equation}
with $N_V = d-2$ transverse vector and $N_T = \frac{d(d-3)}{2} $ transverse tensor polarizations of the gravitons (see \cref{sec:harmonics}). 
 
 What we are after is the dynamics of the modes $\tGR_\bi$ and $\vGR^\ai$. These can be succinctly described with a slight repacking of data. Given a set of  $N_T$ scalar fields $\tGR_\bi$, and $N_V$ vectors $\vGR^\ai$, introduce a collection of auxiliary fields by repackaging the Fourier modes in the harmonic decomposition above as (see \cref{fn:gaugenotation}) 
\begin{equation}\label{eq:defAuxSA}
\begin{split}
\Phi_\bi(v,r,\bx)
&\equiv  
	\int_k \, \tGR_\bi(r,\omega,\bk)\,\ScS(\omega,\bk|v,\bx)\,, \\
\AGR^\ai_B(v,r,\bx) \,dx^B 
&\equiv  \int_k \, 
	 \bigg( (\vGR_r^\ai(r,\omega,\bk) \, dr +\vGR_v^\ai (r,\omega,\bk) dv)\,
  	\ScS(\omega,\bk|v,\bx) \\
  & \qquad \qquad \quad 
  	- i\,\vGR_x^\ai(r,\omega,\bk)\ \ScS_i\,  dx^i \bigg),
 \end{split}
\end{equation}
The set of  1-forms $\AGR^\ai$ are  diffusive Abelian gauge fields with corresponding field strengths $\FGR^\ai_{BC}  = \partial_B \AGR^\ai_C - \partial_C \AGR^\ai_B$. By construction, these auxiliary gauge fields only contains scalar type perturbations  (compare with \eqref{eq:Vharmonic}) and reduce thus to the diffusive gauge field studied in \cref{sec:scalargauge}. As a result its photons are all radially polarized and travel tangentially to the black brane. We will identify these as the non-Markovian momentum diffusion modes which survive to late time and long distances. The radially infalling auxiliary photons (polarized along $\bx$),  which would have had fast Markovian dynamics, are absent. The origin of the gauge symmetry is of course the  underlying diffeomorphisms. Equivalently, one can view the $N_V$ modes as the diffusive momentum modes of the boundary energy-momentum tensor  $(\Tcft)^{\mu\nu}$, as we elaborate in \cref{sec:appgravity}.

With this repackaging, it is actually possible to write down the equations of motion for the linearized gravitational perturbations in a compact form. The linearized Einstein's equations for the parameterization read
\begin{equation}\label{eq:EEqns}
R_{AB} + d\, g_{AB} = 0  \;\ \Longrightarrow\; \; 
\nabla_A\, \nabla^A\, \tGR^\bi = 0 \quad \text{and} \quad \nabla_A\left(r^2\, \FGR^{AB}_\ai \right)  = 0 \,.
\end{equation}	
 We recognize these to be the massless, minimally coupled Klein Gordon equation for the $N_T$ transverse tensor modes, along with a collection of $N_V$ diffusive gauge fields with $\ann = d-1$ as indicated in \cref{sec:desorigin}, see \eqref{eq:grvspar}. 
Both these systems have already been studied in the preceding sections. Hence all results derived heretofore directly apply.

We can further use the  results of \cref{sec:trgauge}  to  rewrite the dynamics of the diffusive auxiliary gauge field $\AGR$ in terms of  a non-Markovian scalar $\sMaxD$. This proves the assertion made in \cref{sec:desorigin} and justifies \eqref{eq:grvspar}.

\subsection{The gravitational action}
\label{sec:sgravity}

One can demonstrate explicitly by a straightforward computation that the Einstein-Hilbert action, together with the Gibbons-Hawking boundary terms, and additional boundary counterterms can be completely mapped to the auxiliary system, up to a time-independent  DC contribution (which originates from the equilibrium free energy of the black hole). The gravitational dynamics is prescribed by\footnote{ To keep the expressions compact we will scale out the usual normalization by $\frac{1}{16\pi G_N}$ in the gravitational action. Boundary CFT results can be obtained by multiplication by $ c_\text{eff} = \frac{\lads^{d-1}}{16\pi G_N}$.\label{fn:ceff}}  
\begin{equation}\label{eq:gravact}
\begin{split}{}
S_\text{grav}  
&= 
	 \int\, d^{d+1}x\, \sqrt{-g}\, \big( R + d(d-1)\big) 
	+ 2\, \int\, d^dx\, \sqrt{-\gamma}\, K + S_\text{ct}\\
S_\text{ct}
 &=
 -\int \, d^dx\, \sqrt{-\gamma} \, \left[ 2(d-1) + \frac{1}{d-2}\, {}^\gamma R\right] 
\end{split}
\end{equation}	
Here $\gamma_{\mu\nu}$ is the timelike induced metric on the boundary.\footnote{We continue to  employ the notation $g_{AB}$ and $\gamma_{\mu\nu}$  for the bulk and the boundary metrics,  respectively. In the gravitational action these include the perturbative corrections, but they are restricted to being just the background values when  we write out the auxiliary system of scalars and vectors. \label{fn:gravSconv}}

We find upon substituting the parameterization \eqref{eq:GRtv} that the dynamics can be repackaged as  
\begin{equation}\label{eq:SgravAux}
\begin{split}
S_\text{grav} & =  
	\sum_{\bi =1}^{N_T}\, S[\tGR_\bi]  + \sum_{\ai=1}^{N_V} \, S[\vGR_\ai]  + 
	 \int\, d^d x\, \sqrt{-\gamma}\, \left[\sqrt{-\gamma_{\mu\nu} \vb{b}^\mu \, \vb{b}^\nu}\ \right]^{-d} \\
 S[\tGR_\bi]
 & =  
 	 -\frac{1}{2}\, \left[ \int\,   d^{d+1} x\, \sqrt{-g}\,    \nabla_A \tGR_\bi \, \nabla^B \tGR_\bi 
 	 + c_{\Phi}\, \int\, d^dx \, \sqrt{-\gamma} \, \nabla_\mu \tGR_\bi \, \nabla^\nu \tGR_\bi  \right]	 \\
 S[\vGR_\ai] 
 &= 
   - \frac{1}{4}\,  \left[ \int\,   d^{d+1} x\, \sqrt{-g}\;   r^2\,   (\FGR^\ai)_{AB} \, (\FGR^\ai)^{AB}+  c_\mathscr{A}\, \int\, d^dx\, \sqrt{-\gamma} \,  r^2 \,   (\FGR_\ai)_{\mu\nu} (\FGR_\ai)^{\mu\nu} \right]
\end{split}
\end{equation}
We explain how this works  in \cref{sec:actionsEH} for completeness.

We can understand \eqref{eq:SgravAux} as follows. We recognize here the bulk action for the auxiliary scalar and gauge system introduced in \eqref{eq:defAuxSA} whose dynamics is given by the equations of motion in \eqref{eq:EEqns}. The counterterm coefficients are fixed by our previous analysis (with $\ann = d-1$) and can also be checked to descend from the gravitational counterterm (the boundary Einstein-Hilbert term) and are given by 
\begin{equation}\label{eq:gcterms}
c_\Phi = c_\mathscr{A} = -\frac{1}{d-2} \,.
\end{equation}	

What remains is the final term. To write this we have introduced the  rescaled thermal vector $\vb{b}^\mu$  which is related to the hydrodynamic thermal vector $\Kbeta^\mu$ \cite{Haehl:2015pja}. For the thermal state on $\mathbb{R}^{d-1,1}$ obtained from the planar \SAdS{d+1} geometry it  is given by $\vb{b}^\mu = b\, (\partial_v)^\mu$ and characterizes the dual boundary fluid configuration in local equilibrium. If we switch off the gravitational perturbations this term is the equilibrium free energy $\sim b^{-d}$  that the standard Gibbons-Hawking computation would give us from the on-shell evaluation of the Einstein-Hilbert action. However, once we turn on the perturbation we obtain additional adiabatic contributions (both hydrostatic and Class L terms in the classification of \cite{Haehl:2015pja}). Since we are working with linearized gravitational perturbations, it is simple to re-express the result in terms of the local  temperature  $T_\text{local}^d$ defined in the local inertial frame set by the timelike vector $\vb{b}^\mu$. We will have more to say about this below.

As noted above the gravitational action can be further repackaged in terms the non-Markovian diffusive scalar, by rewriting the auxiliary gauge system parameterized  by $\AGR$ in terms of  $\sMaxD = \sen{1-d}$. In  \cref{sec:normalizable} we give the explicit expression  for the boundary currents in terms of  these scalars, see  \eqref{eq:TcftScalarpar}.  We will use this expression to compute the Wilsonian influence phase for the energy-momentum tensor components below.

\subsection{The Wilsonian influence phase for momentum diffusion}
\label{sec:wifmom}

We now have all the pieces in place to write down the Wilsonian influence phase for the transverse tensor and vector graviton modes. We will write the expression in terms of the sources $\JMarP^\bi_a$ and $\JMarP^\bi_d$ that couple to the transverse tensor polarizations,  and the diffusive hydrodynamic moduli, $\snMarP^\ai_a$ and $\snMarP^\ai_d$.  The former are the transverse tensor components (i.e., the magnetic components) of the boundary metric, though we will assume a harmonic decomposition and not write out the index structure to avoid notational clutter. On the other hand  $\snMarP^\ai$ are the momentum flux vectors and capture the shear modes corresponding to momentum diffusion.

We use the result for the designer scalar system  \eqref{eq:WIFscalar}  and express the influence phase in terms of the retarded Green's function of the corresponding modes. The final expression we seek, reads at the quadratic order in amplitudes as, in field theory conventions (see \cref{fn:ceff})
\begin{equation}\label{eq:WIFEin}
\begin{split}
\mathcal{S}_\text{WIF}[\JMarP^\ai_a, \JMarP^\ai_d, \snMarP_a, \snMarP_d] 
&= \mathcal{S}_\text{ideal}
  -  c_\text{eff} \int_k  \left\{ \sum_{\bi=1}^{N_T} \, (\JMarP_d^\bi)^\dagger \,K_{_{d-1}}^\In \left[\JMarP_{a}^\bi+\left(\nB+\frac{1}{2}\right)\: \JMarP_d^\bi   
  \right] 
  \right.\\
&\left.
\hspace{3.2cm}
  +\; k^2 \,  \sum_{\ai=1}^{N_V}  (\snMarP^\ai_d)^\dag \,
    K_{_{-d+1}}^\In \left[\snMarP^\ai_{a}+\left(\nB+\frac{1}{2}\right)\: \snMarP^\ai_d \right]\right\}
\end{split}
\end{equation}  
where $\mathcal{S}_\text{ideal}$ is the background thermal contribution to the Wilsonian influence phase arising from the local free energy derived above. We write this more naturally in the LR basis as
\begin{equation}\label{eq:Sideal}
\mathcal{S}_\text{ideal} =    c_\text{eff} \, \int\, d^d x\, \sqrt{-\gamma}\, \left[\sqrt{-(\gamma_\skR)_{\mu\nu} \vb{b}_\skR^\mu \, \vb{b}_\skR^\nu}\ \right]^{-d}
  - c_\text{eff}\,  \int\, d^d x\, \sqrt{-\gamma}\, \left[\sqrt{-(\gamma_\skL)_{\mu\nu} \vb{b}_\skL^\mu \, \vb{b}_\skL^\nu}\ \right]^{-d}
\end{equation}  
The retarded Green's function data entering the expression above can be obtained from the designer scalar analysis. Specializing \eqref{eq:KinMark}  and \eqref{eq:KinnMark} to the case $\ann = \pm(d-1)$, one finds surprisingly compact formulae for the parameters as several of the coefficients simplify significantly. To wit,
\begin{equation}\label{eq:KinMark3exEin}
\begin{split}
K_{_{d-1}}(\omega,k)
&=
  \left(\frac{4\pi}{d\, \beta}\right)^{d}
    \left\{-i\,\bw-\frac{\bq^2}{d-2} + \frac{1}{d} \harm{\frac{2-2d}{d}} \, \bw^2 + i\, \mathfrak{h}_3(d)\, \bw^3   \right. \\
& \left.  \qquad \qquad \qquad 
    -  \frac{2 i }{d(d-2)}   \harm{\frac{2-d}{d}} \bw \bq^2   +\cdots
   \right\} \,.
\end{split}
\end{equation}  
and 
\begin{equation}\label{eq:KinnMark3exEin}
\begin{split}
K_{_{1-d}}(\omega,k)&=
  \left(\frac{4\pi}{d\, \beta}\right)^{2-d}
    \left\{-i\,\bw+\frac{\bq^2}{d}  - \frac{1}{d} \harm{\frac{2-2d}{d}} \,  \bw^2  
     -\frac{2i }{d(d-2)}  \,  \bw \bq^2
  \right. \\
& \left.  \qquad \qquad\qquad 
  + \; i \left[ \frac{1}{d^2}\left[ \harm{\frac{2-2d}{d}}\right]^2 -\mathfrak{h}_3 (d)\right] \bw^3  +\cdots
   \right\} \,.
\end{split}
\end{equation}  
 The parameters $\mathfrak{h}_3$ introduced above is given by the infinite sum:
\begin{equation}\label{eq:dis3par}
\begin{split}
\mathfrak{h}_3(d) 
&=
   - \frac{2(d-1)}{d^3}  \, \sum_{n=0}^\infty\, \frac{\harm{n-1+\frac{2}{d}}}{(n -1+\frac{2}{d})} \, \frac{1}{n+1} \,.
 \end{split}
\end{equation}
It should be possible to resum this expression in terms of polylogs (see example, \cite{Diles:2019uft} for results in $ d=3$), but we will settle for quoting the numeric values in special cases.

\subsection{Comparison with fluid/gravity}
\label{sec:flugra}

The Wilsonian influence phase for the stress tensor components can be compared directly with the predictions of hydrodynamics. We will focus first on the dispersion relations which have been discussed extensively in the literature, and then turn to the Green's functions of the energy-momentum tensor components.
 
\paragraph{Shear dispersion:}  From \eqref{eq:KinnMark3exEin} we see that the dispersion relation we derive by setting $K_{_{1-d}}(\omega,k) =0$ gives
\begin{equation}\label{eq:dshear}
\begin{split}
0
&= 
	-i\omega +\, \frac{1}{4\pi T}\, k^2 - \frac{1}{4\pi T} \, \harm{\frac{2-2d}{d}} \, \omega^2 
	-\frac{2i}{(d-2)} \,\frac{d}{(4\pi T)^2}\,\omega\,k^2 
\\
& \qquad 
	+ i\, \frac{d^2}{(4\pi T)^2} 
	\left[ \frac{1}{d^2}\left[ \harm{\frac{2-2d}{d}}\right]^2 -\mathfrak{h}_3 (d)\right] \omega^3
	+ \frac{d^3}{(4\pi T)^3}\, \mathfrak{h}_{0,4}(d) \, k^4+ \cdots
\end{split}
\end{equation}  
Apart from the terms computed before, we have included in the above a quartic contribution proportional to $k^4$ with a coefficient 
$\mathfrak{h}_{0,4}(d)$ which is necessary for obtaining $\omega(k)$ accurate to quartic order. This expression recovers the familiar expression for the shear diffusion constant:
\begin{equation}\label{eq:etas}
\mathcal{D}= \frac{1}{4\pi T} \;\; \Longrightarrow \;\; \frac{\eta}{s} = \frac{1}{4\pi}
\end{equation}  
Per se, this is not a surprise, since the computation one is doing to derive $\Kin$ is the standard quasinormal mode analysis that was first carried out for gravitons in \cite{Policastro:2001yc}. Specializing to $\mathcal{N}=4 $ SYM we can write the dispersion relation as
\begin{equation}\label{eq:N4shear}
0 = -i\,\omega + \frac{1}{4\pi T}\, k^2 - \frac{1-\ln 2}{2\pi T} \, \omega^2 - i\,\frac{1}{4 (\pi T)^2}\, \omega\, 
  \left[k^2  - \left((1-\ln 2)^2 -4\,\mathfrak{h}_3(4)\right) \omega^2 \right]+  \frac{d^3}{(4\pi T)^3}
  \, \mathfrak{h}_{0,4}(4) \, k^4
\end{equation}  
with $\mathfrak{h}_3(4) = -0.432$ when evaluated numerically.  

Solving the shear dispersion relation to quartic  order one finds for $\omega(k)$  the expression
\begin{equation}\label{eq:}
\omega(k) = -i\frac{1}{4\pi T}\, k^2  -i \, \frac{1}{(4\pi T)^3} 
	\left[\harm{\frac{2}{d}-2} -\frac{2d}{d-2} + d^3\, \mathfrak{h}_{0,4}(d) \right] k^4 +\cdots
\end{equation}  
Dispersion relations in this form were the first signal of hydrodynamic behaviour from AdS/CFT \cite{Policastro:2001yc} who used the quadratic piece in the dispersion to obtain the shear viscosity. 
The quartic term was also computed in \cite{Policastro:2001yc} for $\mathcal{N}=4$ SYM, while 
\cite{Diles:2019uft}  obtained the analogous expression for $d=3$ (ABJM plasma). The dispersions accurate to quartic order obtained in these references are 
\begin{equation}\label{eq:}
\begin{split}
d=4: & \quad 
	\omega(k) = -i\frac{1}{4\pi T}\, k^2 - i\frac{1-\ln 2}{32 \pi^3\, T^3} \, k^4\,,   \\
d=3: & \quad 
	\omega(k) =-i\frac{1}{4\pi T}\, k^2 -i\frac{9+\sqrt{3}\pi-9\,\ln 3}{384\pi^3\, T^3} \, k^4 \,.
\end{split}
\end{equation}	

It is easy to check that we need to retain the $\mathfrak{h}_{0,4}\,  \bq^4$ term to reproduce these results. In fact, they predict  $\mathfrak{h}_{0,4}(4) = \frac{1}{16}$ and $\mathfrak{h}_{0,4}(3) = \frac{1}{6}$, respectively.  This point was already emphasized in \cite{Baier:2007ix}: to get the quartic piece of the dispersion one needs not just the $\order{\omega\, k^2}$ term that we have derived here, but also the $\order{k^4}$ term. Specifically, the higher order corrections in shear dispersion not only include the second order hydrodynamic transport coefficient $\tau_{\pi}$ but also include contributions from third (and potentially fourth) order transport data. We will not attempt here to get the $\order{\bq^4}$ contribution to the dispersion, though it appears straightforward to do so by using the  solutions to the function $J_k$ \eqref{eq:Jeqns}.\footnote{ We thank Temple He and Julio Virrueta for useful discussions on this issue.}  For now we simply note the fact that there are corrections with simple coefficients predicted above.  We note in passing that the dispersion relation has been used to obtain certain third order transport coefficients in  \cite{Diles:2019uft,Grozdanov:2015kqa}, but we find this puzzling in light of the aforementioned mixing with fourth order transport data.\footnote{ Explicit expressions for  tensor structures that could appear at third order in the stress tensor are given in the recent work of \cite{Diles:2019uft} which updates the original computation of \cite{Grozdanov:2015kqa}. However, these structures need to be further constrained by the second law, eliminating terms that comprise the forbidden class $H_F$ in the terminology of \cite{Haehl:2015pja}. It would be interesting to reorganize the data obtained in \cite{Diles:2019uft,Grozdanov:2015kqa} within the eightfold classification scheme.}

\paragraph{Stress tensor correlators:} Let us use the Wilsonian influence phase to write out the stress tensor correlation functions for the polarizations we have studied in the paper. We will focus on the correlation functions for $\mathcal{N}= 4$ SYM, and to keep expressions simple  pick the momentum to point along the $z$-direction, $\bk = k\, \hat{e}_z$. The overall normalization for $SU(N)$ gauge group is simply $c_\text{eff}= \frac{N^2}{8\pi^2} $.

Consider first the tensor polarization of gravitons, which map to minimally coupled massless scalars in the \SAdS{5} geometry
and is the Markovian sector of the stress tensor.  The retarded Green's function for the corresponding component of the boundary stress tensor $T_{xy}$ can be read off directly from  \eqref{eq:KinMark3exEin}. Specializing to $\mathcal{N}=4$ SYM one finds the retarded correlator 
\begin{equation}\label{eq:N4TenRet}
\begin{split}
&\expval{\Tcft_{xy} (-\omega,-\bk)\, \Tcft_{xy}(\omega,\bk)}^\text{Ret} \\
&
\qquad
=  i \,c_\text{eff}\,  \left(\pi T\right)^4 \left[1 -i\, \bw - \frac{\bq^2}{2} +\frac{1-\ln 2}{2}\, \bw^2 +i \, \frac{\ln 2}{2}\bw\,\bq^2 +i \mathfrak{h}_3(4)\,\bwt^3+\cdots\right]
\end{split}
\end{equation}  
This expression includes the pressure term, which is the spatio-temporally constant, background term, which is required by Kubo formulae  analysis \cite{Baier:2007ix}. The expression  above should be compared with Eq.~(4.8) of \cite{Baier:2007ix}   -- we see perfect agreement  at quadratic  order.\footnote{ The dimensionless frequencies and momenta used  in \cite{Baier:2007ix} differs from the one we use by a factor of  two.} We do also include here  the cubic order corrections which is a new result. 

The Keldysh correlator corresponding to the fluctuations using \eqref{eq:MKeld}.  Here we find  
\begin{equation}\label{eq:N4TenKel}
\begin{split}
\expval{\Tcft_{xy} (-\omega,-\bk)\, \Tcft_{xy}(\omega,\bk)}^\text{Kel}
= \frac{c_\text{eff}}{\pi}\,    \left(\pi T\right)^4 \, 
    \left(1+ \frac{\pi^2}{12} \,\bw^2 \right) \left(1 -\frac{\ln 2}{2}\,\bq^2  -  \mathfrak{h}_3(4)\,\bw^2 +\cdots\right) .
\end{split}
\end{equation}  
This  shows that the correlators obey the  KMS relation  to the quadratic order  in gradients, 
\begin{equation}\label{eq:}
\expval{\Tcft_{xy} (-\omega,-\bk)\, \Tcft_{xy}(\omega,\bk)}^\text{Kel}  = \frac{1}{2}\, \coth\left(\frac{\beta\omega}{2}\right)
\, \Re\left[\expval{\Tcft_{xy} (-\omega,-\bk)\, \Tcft_{xy}(\omega,\bk)}^\text{Ret} \right] .
\end{equation}  
The first principles derivation of this expression has not appeared in the literature hitherto, though given the retarded Green's function and the KMS condition one could have easily written it down. Note that we computed the third order gradient terms in the infalling Green's function mainly to get the first non-trivial terms which are not fixed by the Bose-Einstein statistics.
The above expressions may be written more directly in the average-difference basis from the effective action.

For the transverse vector polarization of gravitons with momentum along $\bk =k\,\hat{e}_z$ we will focus on the momentum density $\Tcft_{vx}$ and momentum current $\Tcft_{zx}$. These currents are in turn expressed in terms of the non-Markovian scalar $\sen{1-d}$ in $d$-dimensions. To obtain the shear sector momentum flux correlators one needs to invert the non-Markovian inverse Green's functions $K_{_{1-d}}(\omega,k)$ which is straightforward. Accounting for the translation we have the following result for the non-contact or shear part of the stress tensor correlators:
\begin{equation}\label{eq:TTshear}
\begin{split}
\expval{\Tcft_{vx} (-\omega,-\bk)\, \Tcft_{vx}(\omega,\bk)}_\text{shear} &=
    c_\text{eff}^2\, k^4\,  \expval{\sen{-3}(-\omega,-\bk)   \ \sen{-3}(\omega,\bk) } \,, \\
\expval{\Tcft_{vx} (-\omega,-\bk)\, \Tcft_{zx}(\omega,\bk)}_\text{shear}  &= -
    c_\text{eff}^2\,k^3\,\omega \,  \expval{  \sen{-3}(-\omega,-\bk)   \ \sen{-3}(\omega,\bk) } \,, \\
\expval{\Tcft_{zx} (-\omega,-\bk)\, \Tcft_{zx}(\omega,\bk)}_\text{shear}  &=
  c_\text{eff}^2\, k^2\,\omega^2\,  \expval{  \sen{-3}(-\omega,-\bk)  \ \sen{-3}(\omega,\bk)} \,.
\end{split}
\end{equation}  
In addition we also have a background contact, or ideal contribution which can be obtained directly from $\mathcal{S}_\text{ideal}$, though this only contributes to the retarded Green's function. One has
\begin{equation}\label{eq:TTideal}
\begin{split}
\expval{\Tcft_{vx} (-\omega,-\bk)\, \Tcft_{vx}(\omega,\bk)}_\text{ideal}^\text{Ret} &=
3\, c_\text{eff}\, (\pi T)^4 \\
\expval{\Tcft_{vx} (-\omega,-\bk)\, \Tcft_{zx}(\omega,\bk)}_\text{ideal}^\text{Ret} &=0\\
\expval{\Tcft_{zx} (-\omega,-\bk)\, \Tcft_{zx}(\omega,\bk)}_\text{ideal}^\text{Ret}  &=
  c_\text{eff}\, (\pi T)^4
\end{split}
\end{equation}  
The shear part of retarded Green's function can be obtained by inverting the dispersion function, modulo a factor of $k^2$. We thus find:
\begin{equation}\label{eq:N4VecRet}
\begin{split}
 &\expval{  \sen{-3}(-\omega,-\bk)  \sen{-3}(\omega,\bk)}_\text{Ret}  
 = 
 - \frac{i}{ c_\text{eff}} \, \frac{1}{\bq^2}\, \\
& \qquad \qquad
\times 
  \left(-i\bw + \frac{1}{4}\, \bq^2-\frac{1-\ln2}{2}\, \bw^2 -i\frac{\bwt}{4}
    \left[\bq^2-\left((1-\ln 2)^2 - 4\, \mathfrak{h}_3(4)\right) \bw^2\right] \right)^{-1}
\end{split}
\end{equation}  

These expression should be compared with the shear sector correlators derived in \cite{Arnold:2011ja} extending the  early work of \cite{Policastro:2002se}. In that work they choose to expand the correlator about the shear pole, which leads to a non-local expression. 
However, as we have argued the presence of hydrodynamic moduli can be handled more effectively by invoking a suitable Legendre transform to write the inverse Green's function given in \eqref{eq:KinnMark3exEin}. 

The Keldysh correlator for the shear modes can similarly be obtained from \eqref{eq:nMKeld}. Specializing again to $\ann=d-1$ and $d=4$ we find the following
\begin{equation}\label{eq:N4VecKel}
\begin{split}
&\expval{  \sen{-3}(-\omega,-\bk)  \sen{-3}(\omega,\bk)}_\text{Kel}\\
\qquad\quad  &=
   \frac{1}{\pi  c_\text{eff}} \, \frac{1}{\bq^2}  
    \left(1+ \frac{\pi^2 }{12} \bw^2  \right)  \left(1+\frac{1}{4} \, \bq^2  +   \left( \mathfrak{h}_3(4) -\frac{(1-\ln 2 )^2}{4}  \right) \, \bw^2 
  +\cdots\right) \\
&\qquad\quad
\times   \abs{-i\bw + \frac{1}{4}\,\bq^2-\frac{1-\ln2}{2}\, \bw^2 -i\frac{\bwt}{4}
    \left[\bq^2-\left((1-\ln 2)^2 - 4\, \mathfrak{h}_3(4)\right) \bw^2\right] }^{-2}  
\end{split}
\end{equation}
Once again we see the KMS relations satisfied to the order we have computed,  viz.,
\begin{equation}\label{eq:}
\expval{  \sen{-3}(-\omega,-\bk)  \sen{-3}(\omega,\bk)}_\text{Kel} =
\frac{1}{2}\coth\left(\frac{\beta \omega}{2}\right) \Re\left[\expval{  \sen{-3}(-\omega,-\bk)  \sen{-3}(\omega,\bk)}_\text{Ret} \right]
\end{equation}  

\paragraph{Hydrodynamic effective actions:}
As a final application to the fluid/gravity correspondence, we consider the non-dissipative contributions to the stress-tensor correlations. In \cite{Haehl:2015pja} some of us had conjectured a hydrodynamic effective action for this sector, the so called Class L Lagrangian. This conjecture relied on the structure of non-dissipative contributions to hydrodynamic stress tensor and the transport data for holographic fluids. In particular, it had been argued that the non-dissipative contribution should be captured by the following boundary Lagrangian density (cf., Eq.~(14.37) of \cite{Haehl:2015pja}):
\begin{equation}\label{eq:classL}
\mathcal{L}^\mathcal{W} = 
c_\text{eff} \left(\frac{4\pi T}{d}\right)^d - c_\text{eff} \left(\frac{4\pi T}{d}\right)^{d-2} \left[ \frac{{}^\mathcal{W}R}{d-2}
+ \frac{1}{2} \omega_\text{vor}^2 + \frac{1}{d}\, \mathrm{Harmonic}\left(\frac{2}{d}-1\right) \, \sigma_\text{sh}^2\right]
\end{equation}  
This action is written in  Weyl covariant hydrodynamic variables ${}^\mathcal{W}R$ is the Weyl covariant curvature scalar on the boundary, $\omega_\text{vor}$ is the fluid vorticity, and $\sigma_\text{sh}$ is the shear tensor of the fluid. This action was written down by reverse engineering known fluid transport and was conjectured to hold non-linearly in amplitudes, but in a boundary gradient expansion.

It is a simple to check that our results  reproduce this structure at quadratic order in amplitudes and gradients. All we need is the  defining property of harmonic number function, viz.,  $\harm{x+1} =\harm{x} + \frac{1}{x+1}$, to rewrite
\begin{equation}\label{eq:}\begin{split}
 \Delta(d-1,1) = -\frac{1}{d-2} - \frac{1}{d}\, \mathrm{Harmonic}\left(\frac{2}{d}-1\right) \,.
\end{split}
\end{equation}  
This  basically says that the contribution from the non-dissipative terms at $\order{\omega^2}$ arises from a combination of the Weyl curvature and shear squared terms in the Class L action. At a numerological level this is not a strong check and one might argue that it was guaranteed by the fact that we get the correct Green's functions. 

However, the reason for our optimism here is that unlike in fluid/gravity, we have put the gravitons in the bulk on-shell, evaluated the on-shell action and recovered directly the Class L action. In fact, we get a clear prediction thanks to the grSK geometry. The non-dissipative part of the hydrodynamic effective action obtained from Einstein-Hilbert dynamics directly gives us two copies of Class L, for the R/L
degrees of freedom, consistent with the general discussion in \cite{Haehl:2015pja}. This is a promising start and hints at where in the gravitational dynamics one can find the appropriate data to prove this conjecture. The generalization to the dissipative sector also works as expected and is consistent with field theory analyses. However,  as noted in  \cref{fn:hydroactions}  the structure at quadratic order in amplitudes is not a strong test, but provides a useful sanity check that one is on the right track.

\section{Discussion}
\label{sec:discuss}

We have initiated here a study of open quantum systems with memory.  One imagines coupling some probe degrees of freedom to conserved currents of a thermal field theory. While most modes of the conserved currents relax back to equilibrium quickly (typically in a time-scale of order the inverse temperature), conservation ordains the presence of long-lived modes. Any coupling to such modes will retain long-term memory leading to challenges of constructing a open effective description. While one can eschew the standard low energy gradient expansion and come up with alternate ways to tackle the problem, we propose to take inspiration from the Wilsonian effective field theory approach, and construct instead a local effective action for the long-lived moduli fields.

The key idea we employed is to exploit holography and use the bulk gravity as a guidepost for constructing such an effective description. Taking inspiration from studies of systems with gauge invariance, we furnished a canonical template in the form of a designer scalar system that captures the essential physical content. This description involved a simple characterization in terms of a single Markovianity index parameter, and gives a universal picture of the open effective field theory. Our primary thesis is that for systems with both fast and slow modes, one should decouple their dynamics at leading quadratic order and understand each in their own terms. In particular, for the slow modes which pertain to  memory one should isolate the subspace of low energy Goldstone-like degrees of freedom, the hydrodynamic moduli space, and use them instead of the background sources to parameterize the effective action.

Motivated by the above reasoning we argued that the contribution from a particular point in hydrodynamic moduli space is characterized by the Wilsonian influence functional  $\mathcal{S}_\text{WIF}$ which  we have chosen to parameterize in terms of the data $\{\JMar_\skR, \JMar_\skL, \snMar_\skR, \snMar_\skL\}$, the sources for the Markovian (fast) modes, and the fields (or vevs) for the non-Markovian (slow) modes. We established that this object can be effectively computed using the holographic dual description and used it to describe a universal framework to understand charge and momentum diffusion. 

Having the Wilsonian influence functional at hand, we can go back and understand how to construct the generating function of conserved current correlators. This can be achieved following the usual construction of the quantum effective action. One can obtain the (non-local) Schwinger-Keldysh  influence functional in terms of the sources of both Markovian and non-Markovian fields, by first solving for the non-Markovian sources in terms of the hydrodynamic moduli and then performing a constrained functional integral over  these light fields.  

From a certain perspective, it is quite amazing that the analysis of the bulk theory instructs us to directly focus on computing the Wilsonian influence functional for the non-Markovian fields. Recall that, one usually wants to freeze the classical sources (non-normalizable modes) while functionally integrating over the normalizable modes. For Markovian fields, quasinormal modes are non-normalizable while Hawking modes are normalizable. Non-Markovian fields, on the other hand, are characterized by normalizable long-lived quasinormal modes and non-normalizable Hawking radiation that is turned on only in situations with open boundary conditions. As we saw in our discussion freezing the non-normalizable modes results in localizing  the normalizable modes to the hydrodynamic moduli space. However, if we focus on solving the non-Markovian dynamics with alternate (Neumann) boundary conditions without additional variational counterterms, one directly lands on the Wilsonian influence functional. 

It is rather remarkable that, despite the many years of familiarity with the AdS/CFT correspondence,  that the gravitational description naturally implements these considerations by allowing for a simple change of boundary conditions to implement these Legendre transforms. At a technical level there is an added advantage that most of the non-Markovian data can be trivially recovered from the Markovian ones by analytic continuation of the Markovianity index from positive to negative values. 

At a conceptual level it is satisfying to see elements of the fluid/gravity correspondence emerge directly for the diffusive sector from an effective action. We emphasize that while earlier discussions either focused solely on the retarded response to get hydrodynamic correlators from the study of quasinormal modes, or left the conservation equations off-shell as in the fluid/gravity correspondence, we are explicitly solving all of the dynamical equations and evaluating the boundary observables. In the sub-sectors we studied we see clear glimmers of the structure of hydrodynamic effective field theories. Whether the localization to the hydrodynamic moduli space can be understood broadly in terms of the vision outlined in \cite{Haehl:2015foa} remains however to be checked.

At a more pragmatic level, our focus has been on diffusive dynamics. We found that energy dynamics and the physics of sound involves additional technical complications. Essentially the scalar type linearized perturbations of gravitons which involve the plane waves $\{\ScS(\omega,\bk|v,\bx), \ScS_i(\omega,\bk|v,\bx), \ScS_{ij}(\omega,\bk|v,\bx)\}$ have been ignored in our discussion.  To count the number of components that have been ignored, we begin by noting that the linearized metric perturbations are characterized by a symmetric tensor in AdS$_{d+1}$ which has $\frac{1}{2}(d+1)(d+2)=N_T+ 3N_V+7$ components. Thus, apart from $\{\tGR_\bi, \vGR_{\ai,r},\vGR_{\ai,v},\vGR_{\ai,x}\}$, we have $7$ more functions worth of metric perturbations which are not accounted for in the discussion above. These modes are dual in the long distance (i.e., small $\{\omega, k_i\}$) limit to sound modes of the CFT fluid.  The leading physics here is not diffusive but wave-like and the leading dispersion is linear: $\omega\sim c_s k - i\, \Gamma_s \, k^2$. In this sense, they are analogous to Goldstone fields in the symmetry broken phases  and are different structurally from the modes that we have described so far. For this reason, we did not discuss these modes here. 

 We however believe that the elements of our framework should suffice to broadly understand energy transport as well.  We hope to report on the timbre of Hawking gravitons in AdS in the not too distant future.

\acknowledgments
It is a pleasure to thank  Arpan Bhattacharyya, Bidisha Chakrabarty, Temple He, Veronika Hubeny, Hewei Frederic Jia,  Chandan Jana, Shiraz Minwalla, Shivam K.~Sharma,  Krishnendu Ray, Julio Virrueta, and Spenta Wadia for helpful discussions.  We would also like to thank Alex Miranda for helpful correspondence regarding some details of gravitational perturbations of  the \SAdS{} geometry. RL, SGP, and AS would also like to acknowledge their debt to the people of India for their steady and generous support to research in the basic sciences. JKG acknowledges postdoctoral program at ICTS for funding support through the Department of Atomic Energy, Government of India, under project no. RTI4001 and partially supported through the Simons Foundation Targeted Grant to ICTS-TIFR ``Science without Boundaries".
MR was supported  by U.S.\ Department of Energy grant DE-SC0009999 and by funds from the University of California. 

\appendix

\section{Designer scalar analysis}
\label{sec:descanal}

The designer scalar wave equation \eqref{eq:gseom1} can be written in several equivalent forms that are useful not only for  comparison with similar expressions that exist in literature, but also to intuit some of the general features by quick inspection. We record a few of these for the benefit of the reader: 
\begin{itemize}[wide,left=0pt]
\item 
A trivial rewriting that follows from \eqref{eq:gseom1} is
\begin{equation}\label{eq:gseom2}
\begin{split}
 (\Dz_+)^2 \sen{\ann}+\ann\, r\, f\, \Dz_+\sen{\ann}+\left(\omega^2-k^2\, f\right) \sen{\ann}=0\,. 
\end{split}
\end{equation}
which when converted to the explicit radial derivatives using \eqref{eq:Dz} leads to:
\begin{equation}\label{eq:gseom3}
\begin{split}
\dv{r}(r^{\ann+2}\,f\, \dv{\sen{\ann}}{r} )- i\,\omega\left[\dv{r} (r^{\ann}\,\sen{\ann} )+r^{\ann}\, \dv{\sen{\ann}}{r} \right]
-k^2 \, r^{\ann-2} \, \sen{\ann}=0\ .
\end{split}
\end{equation}
\item Usual study of linearized fields in {}a black hole background is carried out in the Schwarzschild time coordinate rather than the ingoing time $v$. We can recover this form by a simple field redefinition,  $\sen{\ann}^{\text{\tiny {Schw}}}  \equiv  e^{\frac{\beta\omega}{2}\ctor}\, \sen{\ann}$, which leads to 
\begin{equation}\label{eq:gseomSchw}
\begin{split}
\left[\left(r^2 \dv{r}\right)^2+\left(\ann+\frac{d(1-f)}{f}\right)r^3 \dv{r}+\frac{\omega^2-k^2f}{f^2}\right] \sen{\ann}^{\text{\tiny {Schw}}}= 0\ ,
\end{split}
\end{equation}
Time reversal for these modes are simply achieved by $\omega\mapsto -\omega$: this is evident from the fact that their equations have no linear terms in $\omega$.
\item A Schr\"odinger-like form in the regular  tortoise coordinate can be obtained by defining $\sen{\ann}^{\text{\tiny {Schr}}}\equiv  
r^{\frac{\ann}{2}} \, \sen{\ann}^{\text{\tiny {Schw}}} = r^{\frac{\ann}{2}} \, e^{\frac{\beta\omega}{2}\ctor} \, \sen{\ann}$. One obtains a standard eigenvalue equation: 
\begin{equation}\label{eq:gseomSchr}
\begin{split}
\left[\left(r^2 f\dv{r}\right)^2 -V_{_\ann} + \omega^2 \right] \sen{\ann}^{\text{\tiny {Schr}}} = 0\ ,
\end{split}
\end{equation}
with
\begin{equation}\label{eq:gsschrpot}
\begin{split}
V_{_\ann}\equiv f\left[k^2+\frac{d}{2}\ \ann\ r^2(1-f)+\frac{1}{4}\ann(\ann+2)\  r^2 f\right] \,.
\end{split}
\end{equation}
This form is often useful to ascertain stability of the solution to linearized perturbations and argue for the self-adjointness of the linearized fluctuation operator. In the  In the tensor and vector sector of Einstein equations, these coincide with the Regge-Wheeler-Zerelli master fields (a characterization we avoid in the text) studied in the literature as can be inferred from Eq.~(3.10) and Eq.~(3.13) of \cite{Morgan:2009pn}. The precise identifications are:
$\sen{T}^{\text{\tiny {RWZ}}} =\sen{d-1}^{\text{\tiny {Schr}}}\ ,\quad \sen{V}^{\text{\tiny {RWZ}}}=\sen{1-d}^{\text{\tiny {Schr}}} 
$. We furthermore note that the differential operator 
\begin{equation}\label{eq:orbitlap}
\nabla^2_\text{o} = \left(r^2 f\pdv{r}\right)^2   - \pdv[2]{t}
\end{equation}	
is the so-called orbit-space Laplacian which is used in the parameterization of the gravitational perturbations \cite{Kodama:2003jz}
\end{itemize}

\subsection{Gradient expansion of the Green's function}
\label{sec:appphigradexp}

In the main text we have considered the gradient expansion of the designer scalar to quadratic order in the frequency and momenta. We now give a general analysis in particular highlighting some of the central features and explicit results accurate to third order, viz., $\order{\omega^3, k^2\,\omega}$. Firstly, we note that parameterizing the ingoing solution as
\begin{equation}\label{eq:Gingen}
\Gin(\omega,r, \bk)\equiv  e^{-i \,\bw \, F(\ann,\xi)} \, \Xi(\ann, \br) \,, 
\end{equation}	
we can then rewrite \eqref{eq:gsrad} in terms of the following recursion relation for $\Xi(\ann,\xi)$:
\begin{equation}
\begin{split}
\br^{\ann+2-d}(\br^d-1) \, \dv{\br}\Xi(\ann,\br)
&=2 i \, \bw\, \left[\Xi(\ann,\br)-\Xi(\ann,1)\right]\\
&\qquad+\int_{1}^{\br}\,dy \; \Xi(\ann,y)  \left[\bq^2 \,y^{\ann-2}+ \bw^2 \,\dv{\Delta(\ann,y)}{y}\right] \ .
\end{split}
\end{equation}
As before, we have performed one integral and fixed the constant of integration to remove the pole at the horizon. This recursion relation
can then be used to readily give the appropriate first order ODEs at any order in derivative expansion.  Till the third order in the derivative expansion, we can parametrize the ingoing solution as in \eqref{eq:Ginpar3}. 

\paragraph{Solution up to quadratic order:} Upon expanding out we find the differential equations for the functions $\{F, H_k\}$ to be given quite simply since the leading order contribution comes from $\Gin(\omega=0,r,\bk=0) =1$.  This then implies that these functions satisfy simple differential equations that can be integrated up once immediately. One finds them to satisfy. 
\begin{equation}\label{eq:FHkeq}
\begin{split}
\dv{F(\ann,\br)}{\br} +\br^{d-2}\, \frac{\br^{\ann}-1}{\br^\ann(\br^d-1)}
&=0\,, \\
\dv{H_k(\ann,\br)}{\br}+\frac{\br^{d-2}}{\ann-1}\, \frac{\br^{\ann-1}-1}{\br^\ann(\br^d-1)} 
&=0 \,
\end{split}
\end{equation}
As advertised,  we have performed the subtraction needed to make the derivatives of the functions analytic near $\br=1$. 
Comparing these equations with the defining equation for the incomplete beta function \eqref{eq:ibfode}  we see that we can immediately write down the solution as advertised in \eqref{eq:FHkSol}.

The function $H_\omega(\ann,\br)$ obeys a second order ODE sourced by $F(\ann,\br)$ which can also be written down:
\begin{equation}\label{eq:Hweq2}
\begin{split}
\dv{\br}(\br^{\ann+2-d}(\br^d-1)\, \dv{H_\omega(\ann,\br)}{\br} )+(\br^{\ann}+1)\dv{F(\ann,\br)}{\br}
=0\,.
\end{split}
\end{equation}
Before solving this equation, it is useful to attend to the function $\Delta(\ann,\br)$ introduced in \eqref{eq:Deltadef}. 
This function satisfies  a first order ODE
\begin{equation}\label{eq:DeltaEq}
\begin{split}
\dv{\Delta(\ann,\br)}{\br} - (\br^{\ann}+1)\, \dv{F(\ann,\br)}{\br}=\dv{\Delta(\ann,\br)}{\br}+\br^{d-2-\ann}\, \frac{\br^{2\ann}-1}{\br^d-1}=0\,.
\end{split}
\end{equation}
A simple calculation then reveals that we can massage \eqref{eq:Hweq2} into a more tractable form of a first order ODE using our definition of $\Delta(\ann,\br)$ using \eqref{eq:DeltaEq}. We have
 \begin{equation}\label{eq:Hweq}
\begin{split}
\frac{dH_\omega(\ann,\br)}{d\br}+\br^{d-2}\frac{\widehat{\Delta}(\ann,\br)}{\br^\ann(\br^d-1)}=0\ .
\end{split}
\end{equation} 
where as mentioned in the main text around \eqref{eq:hatfns} we see an explicit occurrence of a hatted function in the source term for $H_\omega$. Using the explicit parameterization of $\hat{\Delta}(\ann,\br)$ in terms of the incomplete beta functions written in their defining series form \eqref{eq:ibfser} we can again trivially integrate \eqref{eq:Hweq} to arrive at \eqref{eq:HwSol}.

In addition to the differential equations and the explicit solutions in terms of incomplete beta functions, it is also helpful to have at hand the asymptotic form of these functions which will play an important role in our analysis of normalizability and boundary conditions. We can use the defining series of the incomplete beta functions \eqref{eq:ibfser} and write down immediately,
\begin{equation}\label{eq:AsympExpFHkDelta}
\begin{split}
F(\ann,\br)
&= 
	- \sum_{n=0}^\infty\frac{1}{(nd+1+\ann)\br^{nd+1+\ann}}+ \sum_{n=0}^\infty  \frac{1}{(nd+1)\br^{nd+1}}\ ,\\
(\ann-1)H_k(\ann,\br)
&= 	
	 - \sum_{n=0}^\infty\frac{1}{(nd+1+\ann)\br^{nd+1+\ann}}+\sum_{n=0}^\infty  \frac{1}{(nd+2)\br^{nd+2}}\ ,\\
\Delta(\ann,\br)
&= 
	 - \sum_{n=0}^\infty\frac{1}{(nd+1+\ann)\br^{nd+1+\ann}}+\sum_{n=0}^\infty\frac{1}{(nd+1-\ann)\br^{nd+1-\ann}}\  .
 \end{split}
\end{equation}
Similarly, a double series expansion can be written down for $H_\omega$.

\paragraph{The functions at third order in gradients:} At the next order we obtain for our functions $I_k$ and $I_\omega$, the differential equations 
 \begin{equation}\label{eq:Ieqns}
\begin{split}
&
	\dv{I_k(\ann,\br)}{\br}+2\, \br^{d-2}\, \frac{\widehat{H}_k(\ann,\br)}{\br^\ann(\br^d-1)} =0\ ,\\
&
	\dv{I_\omega(\ann,\br)}{\br}+2\, \br^{d-2}\frac{\widehat{H}_\omega(\ann,\br)}{\br^\ann(\br^d-1)} = 0\,.\\
\end{split}
\end{equation}
We have employed the hat decoration to simplify the presentation of the source terms.  The solution to the functions $I_a(\ann,\br)$  can be obtained using the lower order functions we have derived. At each order the trick is to use series representation of the incomplete beta function \eqref{eq:ibfser}  to convert the final integral into the ODE for the incomplete beta function \eqref{eq:ibfode}.   Since $H_k(\ann,\br)$ was a simple combination of incomplete beta functions, it follows that $I_k$ will be given a single series representation (like $H_\omega$ in \eqref{eq:HwSol}). On the other hand $I_\omega$ will be written in terms of a double series representation. We will outline the structure of these solutions elsewhere, but for now will focus on obtaining the results for the non-Markovian sector $\ann < -1$ assuming that we have solved the equations for the Markovian sector $\ann >-1$.  

To obtain the solutions in the non-Markovian case, especially with the view towards determining their asymptotic behaviour, we will exploit a strategy similar to the one used to determine the functions $F(-\ann,\br)$, $H_\omega(-\ann,\br)$ and $H_k(-\ann,\br)$.
The idea is to define judicious combinations that simplify the analytic continuation from $\ann \to -\ann$.  Consider the following functions:
\begin{equation}\label{eq:Delkomdef}
\begin{split}
\Delta_k(\ann,\br) 
&= 
	(\ann+1) \left[ I_k(-\ann, \br) + \widehat{\Delta}(-\ann, \br) \, \widehat{H}_k(-\ann,\br) - \Delta(-\ann,1) \, H_k(-\ann,1)\right] \\ 
& \qquad 
	+ (\ann-1) \left[ I_k(\ann, \br) + \widehat{\Delta}(\ann, \br) \, \widehat{H}_k(\ann,\br) - \Delta(\ann,1) \, H_k(\ann,1)\right]	 \\
\Delta_\omega(\ann,\br) 	
&=
	I_\omega(-\ann,\br) + \widehat{\Delta}(-\ann,\br) \, \widehat{H}_{\omega}(-\ann,\br) - \Delta(-\ann,1)\, H_\omega(-\ann,1) \\
& \qquad 	
	+ I_\omega(\ann,\br) + \widehat{\Delta}(\ann,\br) \, \widehat{H}_{\omega}(\ann,\br) - \Delta(\ann,1)\, H_\omega(\ann,1) \\
& \qquad \quad
	+ \frac{1}{2(\ann-1)} \, \Delta_k(\ann,\br) - \frac{1}{6}  \left[ \widehat{\Delta}(\ann,\br)^3 + \Delta(\ann,1)^3 \right]	
\end{split}
\end{equation}
By explicit computation we can check that these functions satisfy the following ODEs:
 \begin{equation}\label{eq:Delkweqns}
\begin{split}
	\dv{\Delta_k(\ann,\br)}{\br}
		+ 2\,  \frac{\br^{d-3}}{(\br^d-1)} \, \widehat{\Delta}(\ann,\br)  &=0	\,, \\
	\dv{\Delta_\omega(\ann,\br)}{\br} + 	\frac{\br^{d-2}}{\br^\ann\, (\br^d-1)} \, \widehat{\Delta}(\ann,\br) \left[\widehat{\Delta}(\ann,\xi) + \frac{\xi^{\ann-1}}{\ann-1}\right] &=0 
\end{split}
\end{equation}
We can  solve these equations  at large $\br$ in a Taylor expansion and use the freedom of picking the integration constant to demand the asymptotic expansion
\begin{equation}\label{eq:Delasym3rd}
\begin{split}
\lim_{\br\to\infty}\left\{\Delta_k(\ann,\br)- 2\frac{\br^{\ann-3}}{(\ann-1)(\ann-3)}\right\} & =0\,, \\
\lim_{\br \to\infty}  \Delta_\omega(\ann,\br)  &=0\,.
\end{split}
\end{equation}	
The asymptotic growth of the functions $I_k(-\ann,\br)$  and $I_\omega(-\ann,\br)$ can then be found by using the asymptotic solution for the Markovian sector, and inverting \eqref{eq:Delkomdef}.

\paragraph{At fourth order:} 
Continuing thus, if we parameterize the fourth order contribution to  the function $\Xi(\ann,\br)$ as 
$\Xi(\ann,\br) =  \bq^4\,  J_k(\ann,\br)+ \bw^4\, J_\omega(\ann,\br)+ \bw^2 \bq^2 \, J_{\omega k}(\ann,\br)$ 
we obtain the equations for the functions $J_a$ to be
 \begin{equation}\label{eq:Jeqns}
\begin{split}
&
	\dv{J_k(\ann,\br)}{\br}+ \frac{\br^{d-2}}{\br^\ann(\br^d-1)} \, \int_{1}^{\br}\, dy\, H_k(\ann,y) \, y^{\ann-2}  
	= 0\ ,\\
&
	\dv{J_\omega(\ann,\br)}{\br}+2\,  \frac{\br^{d-2}}{\br^\ann(\br^d-1)}  
	\left[\widehat{I}_\omega(\ann,\br) +\int_{1}^{\br} \, dy\, H_\omega(\ann,y) \dv{\Delta(\ann,y)}{y}\right]  =0\ ,\\
&
	\dv{J_{\omega k}(\ann,\br)}{\br}+2\, \frac{\br^{d-2}}{\br^\ann\, (\br^d-1)}  
	\left[ \widehat{I}_k(\ann,\br)+\int_{1}^{\br}\, dy\, \left(H_\omega(\ann,y) \,y^{\ann-2} + H_k(\ann,y) \, \dv{\Delta(\ann,y)}{y} \right) \right]=0\ .
\end{split}
\end{equation}
It is clear that these can again be tackled as detailed above. We hope to report on useful explicit parameterizations of the functions at a later date.

\subsection{Non-Markovian ingoing Green's function at third order}
\label{sec:nMGin3}

We can now put together all the functions up to third derivative order and write the ingoing solution for the non-Markovian
case in a manner similar to \eqref{eq:GinNmain} and obtain a conveniently factorized form:
\begin{equation}
\begin{split}
\GinN &= 
	e^{-i \,\bw F(\ann,\br) } \left[1-\frac{\KinN }{b^{\ann-1}} \, \XiNN(\ann,\br) \right]   	\bigg[1+\frac{\ann-1}{\ann+1}\, \bq^2\, H_k(\ann,\br)+\bw^2\,  H_\omega(\ann,\br)  \\
 &\hspace{4cm}
 	 - i \bw\,  \bq^2\,  \frac{\ann-1}{\ann+1} \,  I_k(\ann,\br) 
	-i \, \bw^3 \, I_\omega(\ann,\br)  +\cdots\bigg]  .
\end{split}
\end{equation}
We have written this expression in terms of the Markovian data using \eqref{eq:Deltadef} and \eqref{eq:Delkweqns} and parameterized it further in terms of a particular mode function $\XiNN(\ann,\br)$ which is a non-normalizable mode function and a dispersion function $\KinN$.  These two pieces of data are given by:
\begin{equation}\label{eq:Kmn3rd}
\begin{split}
\KinN(\omega,\bk)
&\equiv 
	b^{\ann-1}	 \left[-i\, \bw+\frac{\bq^2}{\ann+1}  - \bw^2\, \Delta(-\ann,1) \right.\\
&\left.
\qquad \qquad 
	+\, 2i\,  \bw  \left[ \bq^2\,  H_k(-\ann,1)+ \bw^2 \, H_\omega(-\ann,1)\right]+\cdots \right] \\
&=
	b^{\ann-1}\left[-i\, \bw+\frac{ \bq^2}{\ann+1}+  \bw^2\, \Delta(\ann,1) 
	+i \bw^3\left( \Delta(\ann,1)^2 - 2\,H_\omega(\ann,1) \right)  \right.\\
&\left.
	\qquad \qquad
		+\,2i \,\frac{\bw\, \bq^2}{\ann+1} \left(\Delta(\ann,1) - (\ann-1)\, H_k(\ann,1) \right) + \cdots \right]  ,
\end{split}
\end{equation} 
and  
\begin{equation}\label{eq:XiNN3rd}
\begin{split}
\XiNN(\ann,\br)
&\equiv  
	\Delta(\ann,\br)-2\left[\frac{\ann-1}{\ann+1} \, \bq^2\, H_k(\ann,\br)+ \bw^2\,  H_\omega(\ann,\br)\right] \widehat{\Delta}(\ann,\br) \\
&
\qquad \qquad 
	+	 \left(\frac{\bq^2}{\ann+1}-\frac{\bw^2}{2(\ann-1)}\right)\Delta_k(\ann,\br)
	 + \bw^2\, \Delta_\omega(\ann,\br)+\cdots \,, 
\end{split}
\end{equation}
respectively. 

We note that, unlike the Markovian case (where there was no ingoing normalizable mode), we now have ingoing normalizable modes
for $(\omega,\bk)$ satisfying the dispersion relation $\KinN(\omega,\bk)=0$. In other words, there exists a codimension-1 locus,  a hypersurface, in $(\omega,\bk)$  space where ingoing normalizable modes exist and are given by writing 
\begin{equation}\label{eq:GinNnorm3rd}
\begin{split}
\GinNt(\omega,r,\bk)  
&= \GinN(\omega,r,\bk)  \bigg|_{\KinN(\omega,\bk) =0}\\
&=
	e^{-i b \omega F(\ann,\br) }  \left[1+\frac{\ann-1}{\ann+1}\,  \bq^2\, H_k(\ann,\br)+ \bw^2\,  H_\omega(\ann,\br)\right.\\
&\qquad\qquad \qquad\quad 
	\left.-i \,\frac{\ann-1}{\ann+1} \,\bw \, \bq^2 \, I_k(\ann,\br)-i\, \bw^3 \, I_\omega(\ann,\br)  +\cdots\right]  .
\end{split}
\end{equation} 
This ingoing normalizable mode (which only exists  on the locus $ \KinN(\omega,\bk)= 0$)   bears a close resemblance to the ingoing non-normalizable mode in the Markovian case, cf., \eqref{eq:Ginpar3}. As mentioned in the text there are some differences in the normalization of the various functions appearing at each order in the gradient expansion, but the overall structure is closely related. One can indeed given the Markovian Green's function guess at the non-Markovian normalizable mode.

When $ \KinN \neq 0$, we get a non-normalizable mode $\XiNN(\ann,\br)$ whose dominant growth at large $\br$ is given by $\Delta(\ann,\br)$. Using \eqref{eq:AsympExpFHkDelta} we see that this function grows as $\br^{\ann-1}$.  While  the functions $H_\omega(-\ann,\br),
I_k(-\ann,\br)$ and $I_\omega(-\ann,\br)$ have higher powers of $\Delta(\ann,\br)$ in their parameterization,  the powers higher than unity all cancel against the contributions of higher powers of  $F(-\ann,\br)$. This is a good sanity check and consistent with the requirement familiar in other holographic examples that  the power of $r$ giving the dominant growth  of  non-normalizable mode is independent of  $(\omega,\bk)$ and hence cannot change as we go to higher orders in derivative expansion.  

For the computation  of the conjugate momentum and counterterms  we record that asymptotically
\begin{equation}\label{eq:Xiasym}
\lim_{r\to \infty} \XiNN(\omega,r,\bk)  \sim  -\frac{r^{\ann-1}}{\ann-1} + \order{r^{\ann-2}} 
\end{equation}	
which follows from the asymptotics of the function $\Delta(\ann,\br)$, cf., \eqref{eq:AsympExpFHkDelta}.  Likewise 
\begin{equation}\label{eq:GinNhasym}
\lim_{r\to \infty} \GinNt(\omega,r,\bk) \sim 1+ \order{r^{-\ann-1}}
\end{equation}	
%

\subsection{Horizon values and transport data}
\label{sec:horvalues}
 
 In addition to the specific form of the functions appearing in the gradient expansion it will also be useful to record the values of these functions at the horizon $\br =1$. These determine the coefficients in our Green's functions and in fact directly parameterize the transport data for the hydrodynamic moduli. 

 To obtain them we can use the fact that the difference of two incomplete beta functions we encountered, has a finite limit as we take $\xi \to 1$, cf., \eqref{eq:ibflim1}. This allows to extract the following values for the functions appearing in the gradient expansion
\begin{equation}\label{eq:Deltahor}
 \Delta(\ann,1) = \od \bigg[ \psi\left( \od(\ann+1)\right) - \psi\left(\od (1-\ann) \right) \bigg]
\end{equation}	
while
\begin{equation}\label{eq:Fhkomhor}
\begin{split}
F(\ann,1)  
&= 
	 \od \left[ \psi\left( \od(\ann+1)\right) - \psi(\od) \right] \\
H_k(\ann,1)  
&= 
	 \frac{\od}{\ann-1} \left[ \psi\left( \od(\ann+1)\right) - \psi(2\od) \right] \\	 
H_\omega(\ann,1)  
&= 
	 \frac{\od^2}{2}\, \psi\left(\od(\ann+1)\right) \left[ \psi\left(\od(\ann+1)\right)  - \psi\left( \od(1-\ann)\right) \right] \\
&  \qquad \qquad
	+ \left[ \frac{1}{n+ \od(1+\ann)} - \frac{1}{n + \od(1-\ann)}\right] \psi\left(n+2\od\right)
\end{split}
\end{equation}
\begin{table}[ht!]
\begin{center}
\begin{tabular}{|c | c|c|}
\hline
\shadeR{Function} 	&	 \shadeR{$\ann  = d-3$} & \shadeR{$\ann = d-1$} \\
\hline\hline
$\Delta(\ann,1)$ &  $\frac{1}{d} \left[\harm{-\frac{2}{d}} - \harm{\frac{4-2d}{d}} \right]$  & $ -\frac{1}{d}  \harm{\frac{2-2d}{d}}  $\\
$H_k(\ann,1)$ & $\frac{\pi}{d(d-4)}\, \cot(\frac{2\pi}{d})$ & $-\frac{1}{d(d-2)} \harm{\frac{2-d}{d}}$\\
$H_\omega^{(1)}(\ann,1)$ & $\frac{1}{2d^2} \, \harm{-\frac{2}{d}} \left[\harm{-\frac{2}{d}} - \harm{\frac{4}{d}-2} \right]$ & $0$ \\
\hline
\end{tabular}
\\
\vspace{5mm}
\begin{tabular}{|c | c|c|}
\hline
\shadeR{Function} 	&	 \shadeR{$\ann  = 3-d$} & \shadeR{$\ann = 1-d$} \\
\hline\hline
$\Delta(\ann,1)$ &  $-\frac{1}{d} \left[\harm{-\frac{2}{d}} - \harm{\frac{4-2d}{d}} \right]$  & $\frac{1}{d}  \harm{\frac{2-2d}{d}}  $\\
$H_k(\ann,1)$ & $\frac{1}{d(d-2)} \left[\harm{\frac{2}{d} -1} - \harm{\frac{4-2d}{d}}\right]$ & $-\frac{1}{d(d-2)} $\\
$H_\omega^{(1)}(\ann,1)$ & $-\frac{1}{2d^2} \, \harm{\frac{4-2d}{d}} \left[\harm{-\frac{2}{d}} - \harm{\frac{4}{d}-2} \right]$ & $
\frac{1}{2d^2} \left[\harm{\frac{2-2d}{d}}\right]^2$ \\
\hline
\end{tabular}
\end{center}
\caption{Horizon values of the special functions appearing in the Wilsonian influence phase and determining transport data for Maxwell fields and gravitons.}
\label{tab:horval}
\end{table}

We can simplify $H_\omega(\ann,1)$ using the harmonic sum representation of the digamma function
\begin{equation}\label{eq:}
\psi(z) = \sum_{m=0}^\infty \left(\frac{1}{m+1} - \frac{1}{m+z} \right) - \gE\,.
\end{equation}	
Using this rewriting we may reassemble the contributions to $H_\omega(\ann,1)$ as 
\begin{equation}\label{eq:}
\begin{split}
H_\omega(\ann,1) &= H_\omega^{( 1)}(\ann,1) + H_\omega^{( 2)}(\ann,1) \\
H_\omega^{( 1)}(\ann,1) 
&=
	 \frac{s}{2} \, \harm{-1+\od(1+\ann) }\, \Delta(\ann,1)	 \\
H_\omega^{( 2)}(\ann,1) 
&=
	- \sum_{n=0}^\infty \, \frac{\ann\, \od^3 \, \harm{n+2\od-1}}{(n + \od(\ann-1))  (n  + \od(1-\ann)) }	
 \end{split}
\end{equation}	
Note that $H_\omega^{(2)(\ann,1)} = - H_\omega^{(2)}(-\ann,1)$. One should be able to find an expression for this quantity in terms of polygamma values, but we will not attempt to do so here. 

Of interest to us are the special cases $\ann=\pm(d-3)$ relevant for probe Maxwell fields, and $\ann = \pm(d-1)$ relevant for probe gravitons. In those cases we record the horizon values of the functions that enter the Wilsonian influence phase in \cref{tab:horval}. In writing the expressions above we have employed the Harmonic number function that has appeared before in the transport data of holographic plasmas \cite{Bhattacharyya:2008mz} which is defined by 
\begin{equation}\label{eq:}
\harm{x-1} = \gE + \psi(x) \,, \qquad \psi(x) = \frac{\Gamma'(x)}{\Gamma(x)} \,.
\end{equation}

\section{Further details for the gauge system}
\label{sec:gaugeapp}

The equations of motion for the designer Maxwell probe \eqref{eq:MaxDil} given in \eqref{eq:MaxDileom} can be explicitly written out in terms of the covariant field strengths as:
\begin{equation}\label{eq:MaxV1}
\begin{split}
\partial_r (r^{\ann+2} \Cm_{rv} ) +\partial_i (r^{\ann}   \Cm_{ri})  &= 0\,,\\
\partial_r\left(r^{\ann}(r^2f \,\Cm_{ri}+ \Cm_{vi})\right)+ r^{\ann}\,  \partial_v  \Cm_{ri} - r^{\ann-2} \,\partial_j \Cm_{ij} &= 0\,,\\
  r^{\ann+2}\, \partial_v  \Cm_{rv}- r^{\ann}\, \partial_i\left(r^2f \Cm_{ri}+ \Cm_{vi}\right)&=0\,.
\end{split}
\end{equation}
We have used here $\sqrt{-g}=r^{d-1}$ as well as the relations
\begin{equation}
\begin{split}
\Cm^{vr} =\Cm_{rv}\,,\quad 
\Cm^{ir} =-r^{-2}(\, \Cm_{vi}+r^2f\, \Cm_{ri})\,,\quad
\, \Cm^{iv} =-r^{-2}\, \Cm_{ri}\,,\quad \, \Cm^{ij}=r^{-4}\, \Cm_{ij}\ .
\end{split}
\end{equation}
In terms of the potentials, the equations of motion take the form 
\begin{equation}\label{eq:MaxV2}
\begin{split}
\pdv{r} (r^{\ann+2}\, \left(\pdv{\Vm_v}{r}-\pdv{\Vm_r}{v}\right) )+r^{\ann}  \pdv{x^i} (\pdv{\Vm_i}{r}-\pdv{\Vm_r}{x^i}) 
&= 0\,,\\
\pdv{r} (r^{\ann}\left(\Dz_+ \Vm_i-\partial_i(r^2 f \,\Vm_r+\Vm_v)\right)) + r^{\ann}\, \pdv{v}(\pdv{\Vm_i}{r}-\pdv{\Vm_r}{x^i} ) 
&=0 \,, \\ 
\qquad-r^{\ann-2}(\partial_i \partial_j-\partial_k\partial_k \delta_{ij})\Vm_j&= 0\,,\\
 r^{\ann+2} \pdv{v} (\pdv{\Vm_v}{r}-\pdv{\Vm_r}{v})+ r^{\ann}\,\pdv{x^i}(\Dz_+ \Vm_i-\pdv{x^i} (r^2 f \, \Vm_r+\Vm_v) )&=0\,.
\end{split}
\end{equation}
These equations are invariant under the gauge redundancy $\Vm_A\mapsto \Vm_A +\partial_A \overline{\Lambda}$ as can be verified directly. Using the harmonic plane wave decomposition \eqref{eq:Vharmonic} one can the infer the equations \eqref{eq:Vveom} and \eqref{eq:TRIgauge} quoted in the main text.

\subsection{Action of \texorpdfstring{$\mathbb{Z}_2$}{Z2} time reversal}
\label{sec:z2gauge}

The time reversal $\mathbb{Z}_2$ isometry $v\mapsto i\beta\ctor-v,\omega\mapsto -\omega$  leaves the background metric invariant. 
The transverse vector equation of motion \eqref{eq:Vveom} has been explicitly demonstrated to be invariant under this involution. 
We now claim that the designer gauge system in \eqref{eq:TRIgauge} is also invariant under this  $\mathbb{Z}_2$. The action of time reversal on the fields may be determined by the transformation of the 1-form
\begin{equation}\label{eq:}
\Vm_{s} = \sMax_r  \,dr+\sMax_v \, dv-\sMax_x \frac{k_i}{k} dx^i \,.
\end{equation}	
One finds:
\begin{equation}
\begin{split}
\sMax_r(\omega, r, \bk) &\mapsto e^{-\beta\omega\ctor}
	\left(\sMax_r(-\omega, r, \bk) +\frac{2 }{r^2 f}\,\sMax_v(-\omega, r, \bk) \right),\quad \\
\sMax_v(\omega, r, \bk)  &\mapsto -e^{-\beta\omega\ctor} \; \sMax_v(-\omega, r, \bk)  ,\quad \\ 
\sMax_x(\omega, r, \bk)  &\mapsto e^{-\beta\omega\ctor}\; \sMax_x(-\omega, r, \bk) \,.
\end{split}
\end{equation}

Furthermore, using the relation $ -\beta\omega \frac{d\ctor}{dr} =2\frac{i\omega}{r^2 f}$ one can check that the following transformations hold:
\begin{equation}
\begin{split}
\pJMax_r(\omega,r,\bk) & \mapsto
e^{-\beta\omega\ctor} \left( \pJMax_r(-\omega,r,\bk) +\frac{2}{r^2f} \,\pJMax_{v}(-\omega, r,\bk) \right) ,\\
\pJMax_{v}(\omega,r,\bk)  &\mapsto - e^{-\beta\omega\ctor} \, \pJMax_{v}(-\omega,r,\bk)\,,\\
\psMax_v(\omega,r,\bk) & \mapsto - e^{-\beta\omega\ctor}  \, \psMax_v(-\omega,r,\bk) \,.
\end{split}
\end{equation}
The last two combinations, $\pJMax_{v}$ and $\psMax_{v}$, transform covariantly under $\mathbb{Z}_2$ with an odd time reversal parity. A third $\mathbb{Z}_2$ covariant, even time reversal parity combination can be formed from the first two and implies that 
\begin{equation}
\begin{split}
\psMax_x(\omega, r, \bk) &\mapsto e^{-\beta\omega\ctor}\, \psMax_x(-\omega, r,\bk) \,.
\end{split}
\end{equation}
%

\subsection{Radial gauge analysis of the gauge system}
\label{sec:radgauge}

In the standard discussion of gauge systems in AdS spacetime, one often tends to a-priori pick a gauge. In this context the gauge choice that is most natural for the ingoing mode analysis is the radial gauge. We have discussed the potential advantages of our gauge invariant formalism in the main text, but for completeness let us examine now the solution of the ingoing modes in the radial gauge and recover the standard story of diffusion therefrom. 

We set $\sMax_r=0$ in the gauge system \eqref{eq:TRIgauge} to get the dynamical equations for the remaining components 
$\sMax_v$ and $\sMax_x$ 
\begin{equation}\label{eq:Maxradialeom}
\begin{split}
\dv{r}[\frac{1}{r^{\ann}} \, \dv{r}\left(r^{\ann+1}\, \sMax_v\right) ]-\frac{ik}{r}\, \dv{\sMax_x}{r} 
	&=0\ ,\\
\frac{1}{r^\ann} \Dz_+ \left[r^{\ann} \Dz_+\sMax_x\right]+\omega^2 \sMax_x
	+ik \, r^2 f \, \times \frac{1}{r^\ann} \dv{r}\left(r^{\ann}\,\sMax_v\right)&=0\ ,\\
 -i\omega\ r^{\ann+2} \, \dv{\sMax_v}{r}+ik\, r^{\ann}\left(\Dz_+\sMax_x+ik\, \sMax_v\right)
 	&=0\,.
\end{split}
\end{equation}
where we have used  the identity    $\dv{r}[\frac{1}{r^{\ann}} \, \dv{r}(r^{\ann+1}\,\sMax_v)]=
\frac{1}{r^{\ann+1}}\, \dv{r}[r^{\ann+2}\dv{\sMax_v}{r}]$  to simplify the equations. For now, we will ignore the third equation, which is the  radial Gauss constraint, and solve the first two equations in derivative expansion.

To second order in derivative expansion, the most general ingoing or analytic solution in this radial gauge is given by:
\begin{equation}
\begin{split}
\sMax_v
&= C_v- \muN \left(\frac{1}{\br^{\ann+1}}  -(\ann+1)\, \bq^2 \int_{\br}^\infty \frac{dy}{y^{\ann+2}} \, H_k(-\ann,y) \right)\\
&
\qquad \qquad 
	+  \bq \left[\bq\, (\muN-C_v) +  \bw \,  C_x\right]
	\int_{\br}^\infty \frac{dy}{y^{\ann+2}}\, F(-\ann,y) +\cdots\ ,\\
\sMax_x
&= 
	C_x+ i \,\bq\, \muN \left[ F(\ann+1,\br)-i \bw \, \widetilde{H}_k(\ann,\br) \right] 
		\\
&
	\qquad\quad 
	-i\left[ \bq\, (\muN-C_v) +  \bw\,  C_x\right]
		\left[ F(\ann,\br)- i \bw \left( H_\omega(\ann,\br) + \frac{1}{2} \, F(\ann,\br)^2 \right)\right]
			+\cdots\,.
\end{split}
\end{equation}
where we introduced the combination
\begin{equation}
\begin{split}
\widetilde{H}_k(\ann,\br)
\equiv
	\int_{\br}^\infty \, 
	\frac{y^{d-2}\,dy}{y^\ann(y^d-1)} \left[
	y^\ann F(\ann+1,y)- F(\ann+1,1) + (\ann+1) \widehat{H}_k(-\ann,y) \right] .
\end{split}
\end{equation}
The remaining functions $\{F,H_k,H_\omega\}$ are the familiar ones which we encountered in the scalar field gradient expansion analysis in \cref{sec:scalarM}.

The coefficients $C_v$ and $C_x$ are the two  non-normalizable modes  (both of which are analytic) whereas $\muN$ is the unique analytic normalizable mode. The coefficients $C_v,C_x$  are fixed by the Dirichlet conditions
\begin{equation}
\begin{split}
\lim_{r\to\infty} \sMax_v &= C_v \ ,\qquad \lim_{r\to\infty} \sMax_x = C_x\ ,
\end{split}
\end{equation}
and they correspond to the boundary source perturbations. One can verify all these statements using the asymptotic expansions in \cref{sec:appphigradexp}. 

The normalizable mode can be related to the Noether charge density via
\begin{equation} 
(\Jcft)^v \equiv \lim_{r\to \infty} \left[r^{\ann+2}\left(\frac{d\sMax_v}{dr}+i\omega\sMax_r\right)-k\frac{r^{\ann-1}}{\ann-1}(k\, C_v-\omega\, C_x)\right] =(\ann+1)\frac{ \muN}{b^{\ann+1}} 
\end{equation}
 We  subtracted a temperature-independent, gauge invariant counterterm to get a finite result to this order in derivative expansion above. The corresponding  Noether current density, after an analogous counterterm subtraction, is given by
\begin{equation}\label{eq:Jidefrad}
\begin{split}
(\Jcft)^i 
&\equiv
	 \frac{k_i}{k}\, \lim_{r\to \infty}\left[ r^{\ann}\left(\Dz_+ \sMax_x+ik\,  \sMax_v+ik \, r^2 f \, \sMax_r\right)-\omega\, 
	 \frac{r^{\ann-1}}{\ann-1}(k\, C_v-\omega \, C_x)\right] \\
&= 
	-i (bk_i)\,  \bigg[ 1+i \, \bw \left(F(\ann+1,1) - (\ann-1)  \, H_k(\ann,1)\right) +\cdots\bigg] \frac{\muN}{b^{\ann+1}} \\
&\qquad \qquad  
	+\frac{1}{b^\ann}\bigg[1-i \, \bw \left( F(\ann,1)-F(-\ann,1)\right)+\cdots\bigg] \, \frac{ik_i}{k} \, (k\,C_v-\omega \,C_x)+\cdots
\,.
\end{split}
\end{equation}
Here we have used \eqref{eq:NMtoMfns} to write the functions corresponding to the exponent $-\ann$ in terms of the functions corresponding to the exponent $\ann$ .  In the final expression for $J^i$, the penultimate line in \eqref{eq:Jidefrad} gives the diffusion current while  the final line denotes the drift current due to the external applied field. The coefficients here are  the (frequency dependent) diffusion constant and conductivity respectively.

The qualitative structure of the ingoing solution in the generalized gauge system is now clear. Unlike the  Markovian
sectors, we have here a long-lived generalized charge mode in the CFT indicated by the presence of an analytic normalizable mode. The physics here is the generalized diffusion of this Noether charge density  with generalized chemical potential $\muN$ (Fick's law)  along with a drift in the charge due to external forcing by potentials $C_v$ and $C_x$ (Ohm's law). 

The normalizable mode $\mu_{_{\ann}}$ however is not arbitrary. Substituting our solution into the radial Gauss constraint, which we recall we have left off-shell, we get a radius independent relation between the normalizable and  the non-normalizable modes (as we should). We obtain the following constraint:
 \begin{equation}
\begin{split}
&
	\left\{-i\omega\, (\ann+1)+bk^2\left[1+i \, \bw\ \left(F(\ann+1,1)- (\ann-1)\,  H_k(\ann,1)\right)+\cdots\right]\right\}\mu_{_{\ann}} \\
&\qquad \quad 
	-  \bq\left\{1-i\, \bw
	\left[ F(\ann,1)-F(-\ann,1)\right]+\cdots
		\right\} (k \, C_v-\omega \, C_x)\ =0\ .
\end{split}
\end{equation}
The above equation is the generalized diffusion equation  in the Fourier domain describing the diffusion of CFT charge density corresponding to $\mu_{_{\ann}}$ along with an ohmic drift  due to external field produced by  $C_v$ and $C_x$. If  we set the external forcing to zero (i.e., fix $C_v =C_x - 0$) we get the inverse of the diffusion Green function
 \begin{equation}
G_{\ann,\text{diff}}^{-1} \equiv
	-i\, \omega\, (\ann+1) + bk^2\ \bigg( 1+i \, \bw \left[F(\ann+1,1)- (\ann-1)\, H_k(\ann,1)\right]+\cdots\bigg)\ .
\end{equation}
This is a diffusion equation with a frequency dependent  diffusion constant 
\begin{equation}
\mathcal{D} 
	= \frac{1}{\ann+1}\, \frac{d \,\beta}{4\pi} \, \left(1+ \frac{\beta\omega}{2\pi}\, d \left[F(\ann+1,1)- (\ann-1)\, H_k(\ann,1)\right]+\cdots
	\right)
\end{equation}
Thus, in the absence of forcing we get a charge density that slowly diffuses  and equilibrates over the planar \SAdS{d+1} black hole horizon.

When we do having external sources  forcing the system, we get a superposition of the free  normalizable diffusion mode with the normalizable drift mode due to external driving (parametrized as heretofore mentioned by $C_v$ and $C_x$):
 \begin{equation}
\muN
	=  \muN^\text{diff} +\bq \,G_{\ann,\text{diff}} 
		\left[1-i \, \bw \left( F(\ann,1)-F(-\ann,1)\right )+\cdots\right]   (k\, C_v-\omega\,  C_x) \,.
\end{equation}

The analysis in the radial gauge has some useful pointers for our general solution. As we see above, the moment we impose the Gauss constraint,  the third equation of \eqref{eq:Maxradialeom}, we see that the gradient expansion breaks down. However, the physics of this breakdown is simple it is associated with the presence of long-lived, long wavelength modes that we are integrating out. 

Thus, once the radial Gauss constraint is fully solved for, we get a breakdown of derivative expansion and  the Markovian approximation,  but only via the diffusion pole in $G_{\ann,\text{Diff}}$.

\section{On the gravitational perturbations}
\label{sec:appgravity}

In this appendix we give some of the details regarding the reduction of gravitational dynamics encoded in the Einstein-Hilbert action onto the designer scalar and gauge fields that we have discussed. The main aim is to show that the transverse tensor polarizations of gravitons are a Markovian scalar (a minimally coupled massless scalar with $\ann = d-1$) while the transverse vector  polarizations leads to a diffusive gauge field with $\ann = 1-d$. We do not discuss the scalar polarization, which are expected to be non-Markovian, since they comprise the low energy sound mode.

\subsection{Linearized  diffeomorphisms and abelian  gauge symmetries}
\label{sec:lindiffeosb}

 Let us begin by quickly motivating the abelian gauge symmetry encountered in the auxiliary gauge system in \eqref{eq:defAuxSA}. This symmetry is inherited from vector diffeomorphisms of the form
\begin{equation}\label{eq:}
 x^i\mapsto x^i + \int_k\,\sum_{\ai=1}^{N_V}\Lambda_\ai(r,\omega,\bk) \, \VV_i^\ai (\omega,\bk|v,\bx)\ .
\end{equation}	 
This is the only allowed diffeomorphism once we ignore  terms with scalar plane waves in the harmonic decomposition. To see how this works, we feed the shift 
\begin{equation} 
\begin{split}
dx^i &\mapsto 
		dx^i + \int_k\,\sum_{\ai=1}^{N_V}\,\left[
	 	\dv{\Lambda_\ai}{r} \,    dr 
		+ \Lambda_\ai(r,\omega,\bk)\   ( dv\, \partial_v+ dx^j\ \partial_j)  \right]\VV^i_\ai (\omega,\bk|v,\bx) \,,
 \end{split}
 \end{equation}
into  the background black brane metric \eqref{eq:efads}. Retaining terms linear in $\Lambda_\ai$ and ignoring  further the modifications coming from the perturbations to the geometry parameterized in \eqref{eq:GRtv} (which are non-linear effects), we can read off the effect of the diffeomorphism on the fields parameterizing $h_{AB}$. One finds, 
\begin{equation}
\begin{split}
\vGR_r^\ai 
	\mapsto \vGR_r + \dv{\Lambda_\ai}{r}\ ,\quad 
\vGR_v^\ai \mapsto \vGR_v -i\omega\  \Lambda_\ai \ ,\quad 
\vGR_x^\ai \mapsto \vGR_x  -ik\,\Lambda_\ai\,,
\end{split}
\end{equation}
which is of course equivalent to (cf., \eqref{eq:gtransfV})
\begin{equation}\
\AGR_B^\ai\, dx^B \mapsto \AGR_B^\ai\, dx^B + \partial_B\, \int_k\, \Lambda_\ai \, \ScS(\omega,\bk|v,\bx)\, dx^B\,,
\end{equation}	
Thus to this leading order we are justified in thinking of the transverse vector gravitons as an auxiliary gauge field.
 
An equivalent way to think about the emergence of this auxiliary gauge field is from the dual CFT: we know that the shear sector of the CFT sector carries a divergence free momentum density. We can then expand such a momentum density (the $vi$ components  of the energy-momentum tensor $(\Tcft)^{\mu\nu}$) in the basis of vector plane waves 
\begin{equation}
\begin{split}
(\Tcft)^{vi}(v,\bx) &\equiv \int_k\,   \sum_{\ai=1}^{N_V}  \,  \TcftP^\ai(\omega,\bk)\,\VV^i_\ai (\omega,\bk|v,\bx)\ .
 \end{split}
\end{equation}
The diffusion of this momentum density is then equivalent to the diffusion of $N_V=d-2$ charges defined via
\begin{equation}
\begin{split}
\int_k\,  \TcftP^\ai(\omega,k) \, \ScS(\omega,\bk|v,\bx)\,.
 \end{split}
\end{equation}
In holography, as we saw in \cref{sec:trgauge}  the diffusion of a charge density is dual to the physics of radially polarized photons traveling tangentially to the boundary. This then naturally leads to a  construction of $N_V=d-2$ radially polarized gauge fields in the bulk, as we have described above.

\subsection{Graviton dynamics repackaged}
\label{sec:actionsEH}

The Einstein's equations \eqref{eq:EEqns} are of course derived from the Einstein-Hilbert action, along with the Gibbons-Hawking  boundary term. In addition, one has to specific boundary covariant counterterms that regulate the UV divergences and give finite boundary observables, viz., the CFT stress tensor $(\Tcft)^{\mu\nu}$ and its correlation functions. This is given in \eqref{eq:gravact}. 

The auxiliary system of scalars and gauge fields introduced in \eqref{eq:defAuxSA}  and their resulting dynamics in \eqref{eq:EEqns} can be derived from the following action:
\begin{equation}\label{eq:auxacts}
\begin{split}
S_{\text{Aux}}
	&= 
		- \int \, d^{d+1}x\, \sqrt{-g} \bigg[\frac{1}{2}  \,\sum_{\bi=1}^{N_T} \, \nabla_A \tGR_\bi \, \nabla^A \tGR_\bi
		+ \frac{r^2}{4}\  
		\sum_{\ai=1}^{N_V}  \, g^{AC}\, g^{BD}\, \FGR^\ai_{AB} \, \FGR^\ai_{CD}
		\bigg] + S_{\text{Aux, ct}} \\
S_{\text{Aux, ct}}
&= 
	  \frac{1}{d-2} \, \int\, d^d x\ \sqrt{-\gamma}\,
	  \bigg[ \frac{1}{2}\sum_{\bi=1}^{N_T}\, \gamma^{\mu\nu} \, \partial_\mu \tGR_\bi \, \partial_\nu \tGR_\bi
		+  \frac{r^2}{4}\,
		 \sum_{\ai=1}^{N_V}\,\gamma^{\mu\sigma}\ \gamma^{\nu \lambda} \FGR^\ai_{\mu\nu} \FGR^\ai_{\sigma\lambda} \bigg]
\end{split}
\end{equation}

The boundary  counterterms  in $S_{\text{Aux, ct}}$  are fixed by the following requirements:  they should  be diagonal in the auxiliary fields since the equations of motion do not mix the fields. They also ought to obey the original symmetries of $S_{\text{Aux}}$: 
the shift invariance in $\tGR_\bi$ and gauge invariance of $\vGR^\ai_{B}$. The terms written above respect these conditions and furthermore only include terms that are at most quadratic in the auxiliary fields. The coefficients are fixed by demanding that the counterterms cancel the divergences of  $S_{\text{Aux}}$ when we evaluate the boundary observables.

The induced metric can be obtained by taking the boundary limit of \eqref{eq:GRtv}, leading to\footnote{We will write the induced metric with explicit coordinate $r$ for simplicity. It is to be understood that we are considering the metric induced at a fixed radial cut-off $r=r_c$ and interested in the limit $r_c\to \infty$. This also applies below when we discuss the boundary counterterms.}
\begin{equation}\label{eq:bdyGRtvs}
\begin{split}
ds^2_\text{bdy} 
&= 
	\left(\gamma_{\mu\nu} + \left({}^h\gamma_{\mu\nu}\right)_\text{Tens} +  \left({}^h\gamma_{\mu\nu}\right)_\text{Vec} \right)\, dx^\mu\, dx^\nu  \\
\gamma_{\mu\nu}\, dx^\mu\, dx^\nu 
&= 
	-r^2\, f(r)\, dv^2 + r^2 \, d\bx^2\,,\\
 \left({}^h\gamma_{\mu\nu}\right)_\text{Tens} \,dx^\mu\, dx^\nu
 &=
 	  r^2 \, \int_k\, \sum_{\bi=1}^{N_T} \, \tGR_\bi\, \TT_{ij}^\bi(\omega,\bk|v,\bx)\, dx^i dx^j \,,\\
 \left({}^h\gamma_{\mu\nu}\right)_\text{Vec} \, dx^\mu\, dx^\nu
&=
	  r^2 \int_k\,   \sum_{\ai=1}^{N_V} \bigg[2\, \vGR^\ai_v (r,\omega,\bk)  \, \VV^\ai_i(\omega,\bk|v,\bx)\, dv dx^i 
		+ i\,\vGR^\ai_x(r,\omega,\bk)\, \VV_{ij}(\omega,\bk|v,\bx) \, dx^i\, dx^j \bigg] \,.
\end{split}
\end{equation}	
We further note that the auxiliary action \eqref{eq:auxacts} is defined with only the background metric, while the Einstein-Hilbert action \eqref{eq:gravact} includes the tensor and vector perturbations as indicated in  \eqref{eq:GRtv}.

For the purposes of computing the two-point function of the energy-momentum tensor we need an expression for the on-shell action of  the gravitational system. One quick way to obtain this is to use the auxiliary  fields and evaluate their on-shell action. We find:
 \begin{equation}
\begin{split}
S_{\text{Aux}}|_{_{\text{On-shell}}}
= -\frac{1}{2}\, \int d^dx\, \bigg[ r^{d-1} \sum_{\bi=1}^{N_T}\,  \tGR_\bi\mathbb{D}_+\Phi_\bi+
r^{d+1} \sum_{\ai=1}^{N_V}\, g^{BC}\, \AGR^\ai_{B}( \FGR^\ai_{Cv}+r^2 f \, \FGR^\ai_{Cr})\bigg] \,.
\end{split}
\end{equation}

We now explain how one obtains the dynamics captured by  the auxiliary system  directly  from the Einstein-Hilbert action along with  the Gibbons-Hawking and boundary counterterms in \eqref{eq:gravact}. Plugging in the perturbation ansatz \eqref{eq:GRtv} one finds by  direct evaluation: 
 \begin{equation}
\begin{split}
S_\text{EH} &\equiv \int\, d^{d+1}x\, \sqrt{-g}\, \big( R + d(d-1)\big)  
	=S_{\text{Aux}} -\int d^dx\, L_{\text{EH},bdy} +\order{h^3}\,.
\end{split}
\end{equation}
The fact that the equations of motion  agree in fact guarantee  that the  above equality should hold, up  to  the boundary terms  in $L_{\text{EH},bdy}$. In obtaining this result we have dropped all the total derivatives along the boundary -- terms  of the form $\partial_\mu F$ are not included in the above. The boundary term $L_{\text{EH},bdy}$ comes from total derivatives along the radial direction and can  be computed to  be
 \begin{equation}
\begin{split}
L_{\text{EH},bdy}
&=  
	2\, r^d\left\{f -\frac{f}{2}\pdv{r}( r\left[\sum_{\bi=1}^{N_T} \, \tGR_\bi^2 +r^2 \, \sum_{\ai=1}^{N_V}\,
		 \gamma^{\mu\nu}\,\AGR^\ai_\mu \,\AGR^\ai_\nu\right] )\right.\\
&
	\left. \qquad \qquad 
	+ \sum_{\ai=1}^{N_V}\; 	\left[\frac{d}{4f}(1+f)\,  (\AGR^\ai_v+ r^2 f\, \AGR^\ai_r)^2 
			+\frac{d}{4f}\, (1-f)\, (\AGR^\ai_v)^2\right] \right\} .
\end{split}
\end{equation}
A good consistency check is that this total radial derivative term should be canceled by the standard  Gibbons-Hawking term. We indeed find plugging in the ansatz \eqref{eq:GRtv} that 
 \begin{equation}
\begin{split}
S_\text{GH} &  \equiv 2\, \int d^dx\, \sqrt{-\gamma}\, K =  \int\, d^dx\ \left(L_{\text{EH},bdy}+L_{\text{ideal}}\right)\,, \\
L_{\text{ideal}} &= 
	  r^d\left\{(d+(d-2)f) \left(1-\frac{1}{2}\sum_{\bi=1}^{N_T} \tGR_\bi^2-\frac{1}{2}\sum_{\ai=1}^{N_V} \mathcal{A}^\ai_i \mathcal{A}^\ai_i\right)+(d-1)\sum_{\ai=1}^{N_V} (\mathcal{A}^\ai_v)^2\right\}.
\end{split}
\end{equation}

$L_\text{ideal}$ is by itself still divergent, but these UV divergences are canceled by the gravitational counterterms encoded in  
$S_\text{ct}$ given in \eqref{eq:gravact}. Once again computing this quantity by plugging in \eqref{eq:bdyGRtvs} we find we can express the answer in terms of the counterterms evaluated for the auxiliary system of scalars and gauge fields along with an additional piece.
The final answer is given as
\begin{equation}\label{eq:gravcts}
\begin{split}
S_\text{ct} &=S_{\text{Aux, ct}} + \int d^dx\, L_\text{ideal, ct} \\ 
L_\text{ideal,ct} &=
	 2(d-1)\, r^d \, \sqrt{f} \left[-1+\frac{1}{2}\sum_{\bi=1}^{N_T} \tGR_\bi^2+\frac{1}{2}\sum_{\ai=1}^{N_V} \mathcal{A}^\ai_i \mathcal{A}^\ai_i-\frac{1}{2f} \sum_{\ai=1}^{N_V} (\mathcal{A}^\ai_v)^2\right] \,.
\end{split}
\end{equation}	
With this parameterization we find the finite combination for the piece we have characterized by the adjective `ideal': 
\begin{equation}\label{eq:idealct}
\begin{split}
\lim_{r \to \infty} \, \left( L_\text{ideal} + L_\text{ideal, ct}\right) 
&= 
	- \lim_{r \to \infty} \, \frac{1}{b^d} \left[- 1+\frac{1}{2}\,\sum_{\bi=1}^{N_T} \, \tGR_\bi^2
	+\frac{1}{2}\, \sum_{\ai=1}^{N_V}  \left(\mathcal{A}^\ai_i \,\mathcal{A}^\ai_i+  
		(d-1)\; (\mathcal{A}^\ai_v)^2\right)\right] \\
&= 
	\int d^dx\, \sqrt{-\gamma}\, \left[\sqrt{-\gamma_{\mu\nu} \,\vb{b}^\mu \,\vb{b}^\nu} \; \right]^{-d}
\end{split}
\end{equation}	
where we introduce the thermal vector $\vb{b}^\mu$, which in the limit where we probe the equilibrium state of the static \SAdS{d+1} black hole  is given by $\vb{b}^\mu \partial_\mu = b\, \partial_v $. 
The adjective `ideal' is now easily explained: the finite contribution appearing in \eqref{eq:idealct} is the ideal fluid free energy in the thermal state on the boundary. This action is explicitly invariant under the Weyl transformations of the CFT metric.

\section{Conserved currents: gauge theory and gravity}
\label{sec:normalizable}

In this appendix, we would like to describe the normalizable modes in the gauge theory and gravity and relate it to normalizable modes of the auxiliary scalar fields. By standard AdS/CFT dictionary, these correspond  to the CFT expectation  values of the  global current and the energy-momentum tensor respectively. We will show that the normalizable modes in the original gauge or gravity description matches with the normalizable modes in the designer scalar description. This is perhaps expected given the reduction to the designer scalars
described in the main text.

We remind the reader that the definition of normalizable modes needs counterterms built out of non-normalizable modes. The corresponding statement about counterterms would be that  the known counterterms for gauge/gravity variables reproduce the required counterterms for the scalar fields. 

We begin with the time-reversal invariant gauge system. Including the boundary Maxwell counterterm corrections (see \eqref{eq:Maxactgen}) and the Markovian contributions into \eqref{eq:bchargecurrent}, we obtain
\begin{equation} \label{eq:BdryNonMarkOps}
\begin{split}
\Jcft_v &=
	-\lim_{r\to \infty}r^{\ann}\left\{
	r^2\Cm_{rv}+\ctV{2}\partial_i \frac{\Cm_{vi}}{r\sqrt{f}} \right\} \\
&= 
	-\int_k  k^2 \lim_{r \to \infty}\left\{ \sMaxD+  \frac{\ctpi{0}}{r \sqrt{f}} \Dz_+ \sMaxD \right\} \ , \\
\Jcft_i 
&=
	 - \lim_{r\to \infty}r^{\ann}\left\{r^{2}f\Cm_{ri}+ \Cm_{vi}+\ctV{2}  \frac{\partial_{v}\Cm_{vi} - f \,\partial_j \Cm_{ji}}{r\sqrt{f}} \right\}\\
&=
	- \lim_{r\to \infty}\sum_{\ai=1}^{N_V}\int_k \VV^\ai_i \left\{ r^{\ann}\Dz_+ \vMax_\ai  +\ctphi{2}  \, \sqrt{f}\ r^{\ann-1} 
	\left(k^2-\frac{1}{f}\omega^2\right)\vMax_\ai\right\}\\
&
	\qquad +\int_k \omega k_i \lim_{r \to \infty}\left\{ \sMaxD+  \frac{\ctpi{0}}{r \sqrt{f}} \Dz_+ \sMaxD \right\}\ ,
\end{split}
\end{equation}
where we have used the parametrization given in \eqref{eq:MasterGaugeInv}.  We have also used the fact that the counterterm
coefficient  $\ctV{2}$ is equal to the leading counterterms $\ctphi{2}$ and $\ctpi{0}$ for the corresponding scalars. The derivation above shows that 
the normalizable modes of the gauge theory map directly to the normalizable mode of the corresponding designer scalars.
Further, note that the CFT current appears in a form where the current conservation is automatic,  even after the couterterm is added.
We can go further and add in the four-derivative counterterms proportional to  $e^{\chi_v}\, \gamma^{\alpha \beta} \, \Cm^{\mu\nu} \nabla_\alpha \nabla_\beta\Cm_{\mu\nu}$ and $e^{\chi_v}\, \gamma^{\alpha \beta} \, \nabla_\alpha\Cm^{\mu\nu}  \nabla_\beta\Cm_{\mu\nu}$
 in order to reproduce  $\ctpi{2}$ counterterm for the non-Markovian contribution.

Once the map between the normalizable modes is established, we can directly import the results from the subsections \S\ref{sec:Markprobe} and \S\ref{sec:nMarkprobe}. The expectation value of the CFT current is given by adopting \eqref{eq:MoneptJ} and \eqref{eq:nM1ptfn} to the gauge problem. This then gives the CFT current in terms of Markovian sources and the non-Markovian effective fields parameterizing the hydrodynamic moduli space to the SK one-point functions :
\begin{equation}
\begin{split}
\expval{\Jcft_{v,\skR} } &=
	-\int_k k^2\snMarQ_{\skR}\,, \qquad
		\expval{\Jcft_{v,\skL} } =-\int_k k^2\snMarQ_{\skL}\,, \\
\expval{\Jcft_{i,\skR} } &=
	-\sum_{\ai=1}^{N_V}\int_k \, \VV^\ai_i 
	\left(\Kin \left[(\nB+1)\: \JMarQ_\skR^\ai - \nB \: \JMarQ_\skL^\ai \right] + \nB\, \Krev\left[\JMarQ_\skR^\ai-\JMarQ_\skL^\ai\right]\right)\\
&\qquad \qquad 
	+ \int_k \, \omega\,  k_i \, \snMarQ_{\skR}\,, \\
\expval{\Jcft_{i,\skL} } &=
	-\sum_{\ai=1}^{N_V}\int_k \, \VV^\ai_i 
		\left(\Kin\left[(\nB+1)\: \JMarQ_\skR^\ai - \nB \: \JMarQ_\skL^\ai \right] + (\nB+1) \Krev  \left[\JMarQ_\skR^\ai-\JMarQ_\skL^\ai \right]\right)\\
&\qquad\qquad
	+ \int_k\, \omega\,  k_i\, \snMarQ_{\skL}\,. 
\end{split}
\end{equation}
Here $\JMarQ$ denotes the source for the Markovian sector (magnetic part of the CFT current source) whereas
$\snMarQ$ denotes the effective fields in  the non-Markovian sector (the charge diffusion mode).  It can be readily checked that these expressions coincide with expectation values obtained by varying the influence phase computed in the main text.

The above discussion can be extended to gravity. The analysis goes through with minor modifications. The energy momentum tensor is given by the boundary limit of the  Brown-York tensor with appropriate counterterms:
\begin{equation}\label{eq:TCFT}
\begin{split}
\sqrt{-g^{\text{\tiny{CFT}}}}\, \Tcft_{\mu\nu}
&= c_{\text{eff}} \lim_{r\to\infty} r^{-2}\sqrt{-\gamma}\ \left(2K\gamma_{\mu\nu}-2K_{\mu\nu}-2(d-1)\gamma_{\mu\nu} + \frac{2}{d-2}\,
{}^\gamma G_{\mu\nu} \right),
 \end{split}
\end{equation}
where ${}^\gamma G_{\mu\nu}$ is the Einstein tensor of the induced boundary metric and $ c_\text{eff}= \frac{\lads^{d-1}}{16\pi G_N}$. This expression follows from the on-shell variation of the gravity action with its counterterms.

As a warm-up, let us compute the CFT stress tensor for the background black-brane. This gives
\begin{equation}
\begin{split}
\Tcft_{\mu\nu}\ dx^\mu dx^\nu &=\frac{ c_{\text{eff}}}{b^d}\left[(d-1) dv^2 +  dx_i dx_i\right]= T_{\mu\nu}^{\text{Ideal}}\ dx^\mu dx^\nu\,.
 \end{split}
\end{equation}
We recognize here the energy-momentum tensor of an ideal  conformal fluid at rest with a pressure $p=c_\text{eff} \, b^{-d}$ and an energy density $\varepsilon=(d-1)p=(d-1) \,c_{\text{eff}}\, b^{-d}$ . We remind the reader that the ideal fluid stress tensor for a fluid with  an energy density $\varepsilon$, a pressure $p$ and a spacetime velocity field $u^\mu$ is given by 
 \begin{equation}
\begin{split}
T^{\mu\nu}_{\text{Ideal}}=\varepsilon\ u^\mu u^\nu + p \,(\eta^{\mu\nu}+u^\mu u^\nu)=p\, (\eta^{\mu\nu}+ d\, u^\mu u^\nu)\ .
\end{split}
\end{equation}
In the last step we have used the relation $\varepsilon=(d-1)p$ arising from the fact that $T^{\mu\nu}$ should be trace free in a CFT.

Next, we turn on the tensor/vector perturbations to ask how the CFT stress tensor changes under such a deformation. Since we are not turning on any scalar perturbations, the contribution to the energy density $ \Tcft_{vv}$ due to these perturbations 
is zero. The components  $ T^{\text{CFT}}_{vi}$ and  $ T^{\text{CFT}}_{ij}$ take the form
\begin{equation}
\begin{split}
\Tcft_{vi, \text{Non-Ideal}}
&= 
	- c_{\text{eff}}\sum_{\ai=1}^{N_V}\int_k k^2 \VV_{i}^\ai  \lim_{r \to \infty}\left\{ \sMaxD+  \frac{\ctpi{0}}{r \sqrt{f}} \Dz_+ \sMaxD \right\} \ , \\
\Tcft_{ij, \text{Non-Ideal}}&=
	  - c_{\text{eff}}\sum_{\bi=1}^{N_T}\int_k \TT^\bi_{ij}  \lim_{r\to \infty}\left\{ r^{d-1}\Dz_+\tGR_\bi +\ctphi{2}  \, \sqrt{f}\ r^{d-2} 
	\left(k^2-\frac{1}{f}\omega^2\right)\tGR_\bi\right\}\\
&\qquad 
	+ c_{\text{eff}}\sum_{\ai=1}^{N_V}\int_k\omega \left(k_i \VV_{j}^\ai+k_j \VV_{i}^\ai\right)  \lim_{r \to \infty}\left\{ {\sEinD}^{\ai}+  \frac{\ctpi{0}}{r \sqrt{f}} \Dz_+ {\sEinD}^{\ai} \right\}\ .
\end{split}
\end{equation}
These expressions are the gravitational analogues of the gauge theory expressions that were derived above.
The counterterm coefficients in the above equation are evaluated by setting $\ann=d-1$. The normalizable modes in the non-ideal part again map  to the normalizable mode of the corresponding designer scalars and the energy-momentum conservation is automatic. To get  the four-derivative counterterms in the non-Markovian sector $\ctpi{2}$ with $\ann=d-1$, we need to add in the corresponding Riemann square counterterms in gravity. 

Evaluating the above expression on our solution, we can derive the SK one point functions of the CFT energy momentum tensor as
\begin{equation}\label{eq:TcftScalarpar}
\begin{split}
\expval{\Tcft_{vi,\skR} } &=
	-\sum_{\ai=1}^{N_V}\int_k \, k^2 \, \snMarP_{\skR}^\ai\ , \qquad
		\expval{\Tcft_{vi,\skL} } =
				-\sum_{\ai=1}^{N_V}\int_k \, k^2\, \snMarP_{\skL}^\ai\ , \\
\expval{\Tcft_{ij,\skR} } 
	&=-\sum_{\bi=1}^{N_T}\int_k  \, \TT^\bi_{ij} \left( K_{_{d-1}}^\In \left[(\nB+1)\: \JMarP_\skR^\bi - \nB \: \JMarP_\skL^\bi \right] + \nB\, K_{_{d-1}}^\Rev \left[\JMarP_\skR^\bi-\JMarP_\skL^\bi\right]\right) \\
&\qquad
	+\sum_{\ai=1}^{N_V} \int_k\, \omega\,  \left(k_i \VV_{j}^\ai+k_j \VV_{i}^\ai\right) \snMarP_{\skR}^\ai\  , \\
\expval{\Tcft_{ij,\skL} } &=
	-\sum_{\bi=1}^{N_T}\int_k  \, \TT^\bi_{ij}\left( K_{_{d-1}}^\In \left[(\nB+1)\: \JMarP_\skR^\ai - \nB \: \JMarP_\skL^\ai \right] + (\nB+1) K_{_{d-1}}^\Rev  \left[\JMarP_\skR^\ai-\JMarP_\skL^\ai \right]\right)\\
&\qquad
	+\sum_{\ai=1}^{N_V} \int_k\, \omega \left(k_i \VV_{j}^\ai+k_j \VV_{i}^\ai\right)\snMarP_{\skL}^\ai\  . \\
\end{split}
\end{equation}
Here $\JMarP$ denotes the source for the Markovian sector (the magnetic part of the CFT metric) whereas
$\snMarP$ denotes the effective fields in  the non-Markovian sector (the momentum diffusion or shear mode).
 It can be readily checked that these expressions coincide with expectation values obtained by varying the influence phase computed in the main text. We will leave for future  a detailed comparison of the  above expressions to the results from fluid/gravity correspondence.

\section{On some incomplete Beta functions}
\label{sec:ibf}

We briefly review the properties of the subclass of  incomplete Beta functions that have been used to parameterize the solutions in the gradient expansion. We start by noting that their definition in terms of the hypergeometric series expansion and outline a few identities that are helpful in verifying some of our statements. We recall, \cite[Eq.~8.17.E7]{NIST:DLMF}
\begin{equation}\label{eq:ibfser}
\begin{split}
\ibf{\alpha,0;z} \equiv \frac{z^\alpha}{\alpha}\  \, \tFo{1, \alpha,1+\alpha;z}= \sum_{m=0}^\infty  \frac{z^{m+\alpha}}{m+\alpha}\ .
 \end{split}
\end{equation}
This series can be written down for any $\alpha$ except when  $\alpha$ is a negative integer. It is absolutely convergent for $|z|<1$, as can be easily shown by ratio test.  We conclude that this series gives a well-defined function for any $\alpha\notin \mathbb{Z}_-$ and for all $|z|<1$. Further absolute convergence legitimizes   term by term differentiation, integration etc., of this series representation.

The central fact relevant for our purpose  is that via term by term differentiation, it can be shown that the incomplete Beta function solves the following inhomogeneous first order ODE:
\begin{equation}\label{eq:ibfode}
\begin{split}
\xi^2 \left(1-\xi^{-d}\right) \frac{d}{d\xi}\left[\frac{1}{d}  \, \ibf{\frac{k+1}{d},0;\frac{1}{\xi^d}} \right]+\frac{1}{\xi^{k}}=0\ .
 \end{split}
\end{equation}
This can also be equivalently shown via the following integral representation valid for $\alpha>0$ and $|z|<1$:
\begin{equation}
\begin{split}
\ibf{\alpha,0;z} =\int_0^z \frac{t^{\alpha-1}dt}{1-t}\,.
 \end{split}
\end{equation}
One can check that a binomial expansion of the denominator in the integrand gives us back the original series. The differential equation 
above uniquely determines  the  incomplete Beta function up to an additive constant. For $\alpha>0$, the additive constant can be fixed by demanding that 
the function vanishes as $z\to 0$. The $\alpha<0$ case can then be thought of as an analytic continuation from $\alpha>0$ case.

The incomplete Beta function has a logarithmic branch cut at $z=1$: the series expansion as we take  $z\to 1^-$ is given by 
\begin{equation}\label{eq:ibfint}
\begin{split}
-\ibf{\alpha,0;z} =
	\gE+\psi(\alpha)+\ln(1-z)+\sum_{n=1}^\infty \frac{(-)^n}{n}\frac{\Gamma(\alpha)}{n!\ \Gamma(\alpha-n)}(1-z)^n\ .
 \end{split}
\end{equation}
Here $\gE$ is the Euler constant and $\psi(\alpha)$ is the digamma function. The presence of a branch cut is easily seen when $\alpha$ is a positive integer. In this case, the defining series can be recognized as the Madhava-Taylor series of the logarithm with the first few terms dropped:
\begin{equation}\label{eq:ibfcut}
\begin{split}
\ibf{\alpha,0;z} &=\ln(1-z)+\sum_{n=1}^{\alpha-1} \frac{z^k}{k}\quad\text{for}\quad \alpha\in\mathbb{Z}_+\ .
 \end{split}
\end{equation}
The logarithmic branch cut is evident in this case. Based  on the above note  in particular that the difference of two incomplete Beta functions is analytic near $z=1$, and is given in terms of the digamma function $\psi(x) = \dv{x} \log \Gamma(x)$
\begin{equation}\label{eq:ibflim1}
\begin{split}
\lim_{z\to 1^-}\left[\ibf{\alpha_1,0;z}-\ibf{\alpha_2,0;z}\right] &=\psi(\alpha_2)-\psi(\alpha_1)= \sum_{n=0}^\infty\left[\frac{1}{n+\alpha_2}-\frac{1}{n+\alpha_1} \right] .
 \end{split}
\end{equation}
%

\section{Plane wave harmonics}
\label{sec:harmonics}

We give a quick summary of our conventions for harmonic decomposition on $\mathbb{R}^{d-1,1}$.  We will classify the harmonics to be scalar, vector and tensor, based on their transformation of the spatial $SO(d-2)$ rotation group transverse to a fixed momentum vector $\bk$. This is standard in most of the literature, see for example \cite{Kodama:2003jz} for a general discussion. We note that a more compact presentation would have used $SO(d-1)$ representation theory as is employed in the fluid/gravity literature \cite{Bhattacharyya:2008jc,Bhattacharyya:2008mz}. Our discussion differs in a minor way from the bulk of the literature in that we directly work in  $\mathbb{R}^{d-1,1}$ including the time-frequency dependence in our definition.

\begin{itemize}[wide,left=0pt]
\item \textbf{Scalar plane waves}: These are simply scalar plane waves 
\begin{equation}\label{eq:sharm}
\ScS(\omega,\bk|v, \bx)\equiv e^{i \, \bk\cdot \bx-i\, \omega v}
\end{equation}	
We will also need its spatial derivatives in what follows, which we denote by $\ScS_i$ and $\ScS_{ij}$.
\item \textbf{Vector plane waves}: These are transverse vector plane waves, denoted  $\VV_i^\ai$.  They are transverse to the momentum vector ${\bf k}$, satisfying $k_i\, \VV_i^\ai=0$ and transform in a spin-one (vector) representation of $SO(d-2)$, which is  the rotation group in spatial  directions normal to $\bk$. They will be taken to furnish an orthonormal basis for the transverse vectors and are $N_V = d-2$ in number.
\item \textbf{Tensor plane waves}: These are transverse, symmetric, trace-free, tensor plane waves, obeying
$k_i\, \TT^\bi_{ij} =0$  and $\TT_{ii}^\bi =0$. There are clearly $N_T = \frac{d(d-3)}{2}$ such modes which furnish a spin-2 representation of $SO(d-2)$.
 \end{itemize}

All these  functions $\{\ScS,\VV^\ai_i,   \TT^\bi_{ij} \}$ appearing above are, as noted,  plane waves, i.e., eigenfunctions of $\{i\,\pdv{v} ,-i\pdv{x^i}\}$ with eigenvalue  $\{\omega, k_i\}$. On the plane waves we define orthonormality  with respect to the flat measure on $\mathbb{R}^{d-1,1}$. Let
\begin{equation}\label{eq:L2norm}
\expval{\mathcal{P}(\omega_1, \bk_1|v,\bx) ,\mathcal{Q}(\omega_2,\bk_2|v,\bx) } = 
\int\, d^d x\, \mathcal{P}(\omega_1, \bk_1|v,\bx) \,\mathcal{Q}(\omega_2,\bk_2|v,\bx) \,,
\end{equation}	
where we adopt the notational shorthand
\begin{equation}\label{eq:dxshort}
d^d x = dv \, d^{d-1}\bx \,, \qquad \delta^d(k_1+k_2) =  \delta(\omega_1+\omega_2)\times  \delta^{d-1}(\bk_1+\bk_2)\,.
\end{equation}	
The statement of orthonormality is then simply the standard plane wave normalization in flat spacetime dressed with Kronecker deltas for the discrete labels, viz.,
\begin{equation}
\begin{split}
\expval{ \ScS(\omega_1,\bk_1|v,\bx) ,\ScS(\omega_2,\bk_2|v,\bx) }
 	 &=  (2\pi)^d\,  \delta^d(k_1+k_2) \\
 \expval{ \VV^{\ai_1}_i(\omega_1,\bk_1|v,\bx) , \VV^{\ai_2}_i(\omega_2,\bk_2|v,\bx) }
 &  
 	= \delta_{\ai_1\ai_2}\times(2\pi)^d\,  \delta^d(k_1+k_2)\,, \\
 \frac{1}{2}\, \expval{  \TT^{\bi_1}_{ij}(\omega_1,\bk_1|v,\bx)  , \TT^{\bi_2}_{ij}(\omega_2,\bk_2|v,\bx) }
 &
 	= \delta_{\bi_1 \bi_2}\times(2\pi)^d\,  \delta^d(k_1+k_2)\ .
 \end{split}
\end{equation}

The basis of vector and tensor plane waves are also orthonormal with respect to derived objects obtained by differentiating scalar or vector plane waves.  Define:
 \begin{equation}
\begin{split}
\ScS_i(\omega,\bk|v,\bx) 
&\equiv 
	\frac{1}{k}\partial_i  \ScS(\omega,\bk|v,\bx)\ ,\\
\ScS_{ij}(\omega,\bk|v,\bx) 
&\equiv 
	\frac{1}{k^2}\partial_i \partial_j \ScS(\omega,\bk|v,\bx)\ ,\\
\VV^{\ai}_{ij}(\omega,\bk|v,\bx) 
&\equiv 
	\frac{1}{k} \left( \partial_i \VV^\ai_j(\omega,\bk|v,\bx)+\partial_j \VV^\ai_i(\omega,\bk|v,\bx) \right) \,,
 \end{split}
\end{equation}
These derived wave harmonics are also normalized as 
\begin{equation}
\begin{split}
\expval{ \ScS_i(\omega_1,\bk_1|v,\bx) ,\ScS_i(\omega_2,\bk_2|v,\bx) }
  	&=  (2\pi)^d\,  \delta^d(k_1+k_2) \\ 
 \frac{1}{2}\, \expval{ \VV^{\ai_1}_{ij}(\omega_1,\bk_1|v,\bx) ,\VV^{\ai_2}_{ij}(\omega_2,\bk_2|v,\bx) }
 &= 
 	 \delta_{\ai_1 \ai_2}\times(2\pi)^d\,  \delta^d(k_1+k_2)\ ,\\
 \frac{1}{2}\, \expval{ \ScS_{ij}(\omega_1,\bk_1|v,\bx),\ScS_{ij}(\omega_2,\bk_2|v,\bx) }
&= 
  	\frac{1}{2} \times(2\pi)^d\,  \delta^d(k_1+k_2)\ .	
 \end{split}
\end{equation}
The relative orthogonality between the derived plane waves and the vector/tensor plane waves is simply the statement that

\begin{equation}
\begin{split}
 \expval{ \VV^{\ai}_i(\omega_1,\bk_1|v,\bx) ,\ScS_i(\omega_2,\bk_2|v,\bx) }
 &=0\,,\\
\expval{  \TT^{\bi}_{ij}(\omega_1,\bk_1|v,\bx) ,\VV^{\ai}_{ij}(\omega_2,\bk_2|v,\bx) }
 &= 0\,,\\
\expval{  \TT^{\bi}_{ij}(\omega_1,\bk_1|v,\bx) ,\ScS_{ij}(\omega_2,\bk_2|v,\bx) }
 &= 0\,,\\
  \expval{ \VV^{\ai}_{ij}(\omega_1,\bk_1|v,\bx) , \ScS_{ij}(\omega_2,\bk_2|v,\bx) }
&= 0\,.
  \end{split}
\end{equation}

These vector plane waves may be familiar to the reader as the plane wave solutions for massless vector fields in $\mathbb{R}^{d-1,1}$ and they  fill out a  $(d-2)$ dimensional representation of the little group  $SO(d-2)$. From the point of AdS/CFT, long distance (i.e., small $\{\omega, k_i\}$) physics of these modes on the gravity side is dual to the physics of the shear modes in the dual CFT fluid.  Shear modes are a diffusive branch of solutions of Navier-Stokes equations where the fluid velocity or momentum density is divergence-free. They describe the shear viscosity driven diffusion of momentum density transverse to the wave vector direction. Because of the divergence free property, these modes occur both in compressible as well as incompressible fluids.

These tensor plane waves may be familiar to the reader as the plane wave solutions for massless spin 2 (graviton) fields in $\mathbb{R}^{d-1,1}$. There are $\frac{d(d-3)}{2}$ such graviton polarizations in $d$ spacetime dimensions. In AdS/CFT, these modes do not survive the 
long distance (i.e., small $\{\omega, k_i\}$) limit. They are Markovian and hence do not have a dual CFT fluid analogue.


\begin{thebibliography}{10}

\bibitem{Vishveshwara:1970cc}
C.~Vishveshwara, \emph{{Stability of the schwarzschild metric}},
  \href{https://doi.org/10.1103/PhysRevD.1.2870}{\emph{Phys. Rev. D} {\bfseries
  1} (1970) 2870}.

\bibitem{Press:1971wr}
W.H.~Press, \emph{{Long Wave Trains of Gravitational Waves from a Vibrating
  Black Hole}}, \href{https://doi.org/10.1086/180849}{\emph{Astrophys. J.
  Lett.} {\bfseries 170} (1971) L105}.

\bibitem{Hawking:1974sw}
S.~Hawking, \emph{{Particle Creation by Black Holes}},
  \href{https://doi.org/10.1007/BF02345020}{\emph{Commun. Math. Phys.}
  {\bfseries 43} (1975) 199}.

\bibitem{Horowitz:1999jd}
G.T.~Horowitz and V.E.~Hubeny, \emph{{Quasinormal modes of AdS black holes and
  the approach to thermal equilibrium}},
  \href{https://doi.org/10.1103/PhysRevD.62.024027}{\emph{Phys. Rev.}
  {\bfseries D62} (2000) 024027}
  [\href{https://arxiv.org/abs/hep-th/9909056}{{\ttfamily hep-th/9909056}}].

\bibitem{Policastro:2001yc}
G.~Policastro, D.T.~Son and A.O.~Starinets, \emph{{The Shear viscosity of
  strongly coupled N=4 supersymmetric Yang-Mills plasma}},
  \href{https://doi.org/10.1103/PhysRevLett.87.081601}{\emph{Phys. Rev. Lett.}
  {\bfseries 87} (2001) 081601}
  [\href{https://arxiv.org/abs/hep-th/0104066}{{\ttfamily hep-th/0104066}}].

\bibitem{Policastro:2002se}
G.~Policastro, D.T.~Son and A.O.~Starinets, \emph{{From AdS / CFT
  correspondence to hydrodynamics}},
  \href{https://doi.org/10.1088/1126-6708/2002/09/043}{\emph{JHEP} {\bfseries
  09} (2002) 043} [\href{https://arxiv.org/abs/hep-th/0205052}{{\ttfamily
  hep-th/0205052}}].

\bibitem{Policastro:2002tn}
G.~Policastro, D.T.~Son and A.O.~Starinets, \emph{{From AdS / CFT
  correspondence to hydrodynamics. 2. Sound waves}},
  \href{https://doi.org/10.1088/1126-6708/2002/12/054}{\emph{JHEP} {\bfseries
  12} (2002) 054} [\href{https://arxiv.org/abs/hep-th/0210220}{{\ttfamily
  hep-th/0210220}}].

\bibitem{Bhattacharyya:2007vs}
S.~Bhattacharyya, S.~Lahiri, R.~Loganayagam and S.~Minwalla, \emph{{Large
  rotating AdS black holes from fluid mechanics}},
  \href{https://doi.org/10.1088/1126-6708/2008/09/054}{\emph{JHEP} {\bfseries
  09} (2008) 054} [\href{https://arxiv.org/abs/0708.1770}{{\ttfamily
  0708.1770}}].

\bibitem{Bhattacharyya:2008jc}
S.~Bhattacharyya, V.E.~Hubeny, S.~Minwalla and M.~Rangamani, \emph{{Nonlinear
  Fluid Dynamics from Gravity}},
  \href{https://doi.org/10.1088/1126-6708/2008/02/045}{\emph{JHEP} {\bfseries
  02} (2008) 045} [\href{https://arxiv.org/abs/0712.2456}{{\ttfamily
  0712.2456}}].

\bibitem{Hubeny:2011hd}
V.E.~Hubeny, S.~Minwalla and M.~Rangamani, \emph{{The fluid/gravity
  correspondence}},  in \emph{{Black holes in higher dimensions}},
  pp.~348--383, 2012 [\href{https://arxiv.org/abs/1107.5780}{{\ttfamily
  1107.5780}}].

\bibitem{Berti:2009kk}
E.~Berti, V.~Cardoso and A.O.~Starinets, \emph{{Quasinormal modes of black
  holes and black branes}},
  \href{https://doi.org/10.1088/0264-9381/26/16/163001}{\emph{Class. Quant.
  Grav.} {\bfseries 26} (2009) 163001}
  [\href{https://arxiv.org/abs/0905.2975}{{\ttfamily 0905.2975}}].

\bibitem{Morgan:2009pn}
J.~Morgan, V.~Cardoso, A.S.~Miranda, C.~Molina and V.T.~Zanchin,
  \emph{{Gravitational quasinormal modes of AdS black branes in d spacetime
  dimensions}},
  \href{https://doi.org/10.1088/1126-6708/2009/09/117}{\emph{JHEP} {\bfseries
  09} (2009) 117} [\href{https://arxiv.org/abs/0907.5011}{{\ttfamily
  0907.5011}}].

\bibitem{Jana:2020vyx}
C.~Jana, R.~Loganayagam and M.~Rangamani, \emph{{Open quantum systems and
  Schwinger-Keldysh holograms}},
  \href{https://doi.org/10.1007/JHEP07(2020)242}{\emph{JHEP} {\bfseries 07}
  (2020) 242} [\href{https://arxiv.org/abs/2004.02888}{{\ttfamily
  2004.02888}}].

\bibitem{deBoer:2008gu}
J.~de~Boer, V.E.~Hubeny, M.~Rangamani and M.~Shigemori, \emph{{Brownian motion
  in AdS/CFT}},
  \href{https://doi.org/10.1088/1126-6708/2009/07/094}{\emph{JHEP} {\bfseries
  07} (2009) 094} [\href{https://arxiv.org/abs/0812.5112}{{\ttfamily
  0812.5112}}].

\bibitem{Son:2009vu}
D.T.~Son and D.~Teaney, \emph{{Thermal Noise and Stochastic Strings in
  AdS/CFT}}, \href{https://doi.org/10.1088/1126-6708/2009/07/021}{\emph{JHEP}
  {\bfseries 07} (2009) 021} [\href{https://arxiv.org/abs/0901.2338}{{\ttfamily
  0901.2338}}].

\bibitem{Glorioso:2018mmw}
P.~Glorioso, M.~Crossley and H.~Liu, \emph{{A prescription for holographic
  Schwinger-Keldysh contour in non-equilibrium systems}},
  \href{https://arxiv.org/abs/1812.08785}{{\ttfamily 1812.08785}}.

\bibitem{Son:2002sd}
D.T.~Son and A.O.~Starinets, \emph{{Minkowski space correlators in AdS / CFT
  correspondence: Recipe and applications}},
  \href{https://doi.org/10.1088/1126-6708/2002/09/042}{\emph{JHEP} {\bfseries
  09} (2002) 042} [\href{https://arxiv.org/abs/hep-th/0205051}{{\ttfamily
  hep-th/0205051}}].

\bibitem{Herzog:2002pc}
C.P.~Herzog and D.T.~Son, \emph{{Schwinger-Keldysh propagators from AdS/CFT
  correspondence}},
  \href{https://doi.org/10.1088/1126-6708/2003/03/046}{\emph{JHEP} {\bfseries
  03} (2003) 046} [\href{https://arxiv.org/abs/hep-th/0212072}{{\ttfamily
  hep-th/0212072}}].

\bibitem{Skenderis:2008dh}
K.~Skenderis and B.C.~van Rees, \emph{{Real-time gauge/gravity duality}},
  \href{https://doi.org/10.1103/PhysRevLett.101.081601}{\emph{Phys. Rev. Lett.}
  {\bfseries 101} (2008) 081601}
  [\href{https://arxiv.org/abs/0805.0150}{{\ttfamily 0805.0150}}].

\bibitem{Skenderis:2008dg}
K.~Skenderis and B.C.~van Rees, \emph{{Real-time gauge/gravity duality:
  Prescription, Renormalization and Examples}},
  \href{https://doi.org/10.1088/1126-6708/2009/05/085}{\emph{JHEP} {\bfseries
  05} (2009) 085} [\href{https://arxiv.org/abs/0812.2909}{{\ttfamily
  0812.2909}}].

\bibitem{vanRees:2009rw}
B.C.~van Rees, \emph{{Real-time gauge/gravity duality and ingoing boundary
  conditions}},
  \href{https://doi.org/10.1016/j.nuclphysbps.2009.07.078}{\emph{Nucl. Phys.
  Proc. Suppl.} {\bfseries 192-193} (2009) 193}
  [\href{https://arxiv.org/abs/0902.4010}{{\ttfamily 0902.4010}}].

\bibitem{Chakrabarty:2019aeu}
B.~Chakrabarty, J.~Chakravarty, S.~Chaudhuri, C.~Jana, R.~Loganayagam and
  A.~Sivakumar, \emph{{Nonlinear Langevin dynamics via holography}},
  \href{https://arxiv.org/abs/1906.07762}{{\ttfamily 1906.07762}}.

\bibitem{Loganayagam:2020eue}
R.~Loganayagam, K.~Ray and A.~Sivakumar, \emph{{Fermionic Open EFT from
  Holography}},  \href{https://arxiv.org/abs/2011.07039}{{\ttfamily
  2011.07039}}.

\bibitem{Loganayagam:2020iol}
R.~Loganayagam, K.~Ray, S.K.~Sharma and A.~Sivakumar, \emph{{Holographic KMS
  relations at finite density}},
  \href{https://arxiv.org/abs/2011.08173}{{\ttfamily 2011.08173}}.

\bibitem{Chakrabarty:2020ohe}
B.~Chakrabarty and P.M.~Aswin, \emph{{Open effective theory of scalar field in
  rotating plasma}},  \href{https://arxiv.org/abs/2011.13223}{{\ttfamily
  2011.13223}}.

\bibitem{Chamblin:1999ya}
H.~Chamblin and H.~Reall, \emph{{Dynamic dilatonic domain walls}},
  \href{https://doi.org/10.1016/S0550-3213(99)00520-9}{\emph{Nucl. Phys. B}
  {\bfseries 562} (1999) 133}
  [\href{https://arxiv.org/abs/hep-th/9903225}{{\ttfamily hep-th/9903225}}].

\bibitem{Charmousis:2010zz}
C.~Charmousis, B.~Gouteraux, B.S.~Kim, E.~Kiritsis and R.~Meyer,
  \emph{{Effective Holographic Theories for low-temperature condensed matter
  systems}}, \href{https://doi.org/10.1007/JHEP11(2010)151}{\emph{JHEP}
  {\bfseries 11} (2010) 151} [\href{https://arxiv.org/abs/1005.4690}{{\ttfamily
  1005.4690}}].

\bibitem{Iizuka:2011hg}
N.~Iizuka, N.~Kundu, P.~Narayan and S.P.~Trivedi, \emph{{Holographic Fermi and
  Non-Fermi Liquids with Transitions in Dilaton Gravity}},
  \href{https://doi.org/10.1007/JHEP01(2012)094}{\emph{JHEP} {\bfseries 01}
  (2012) 094} [\href{https://arxiv.org/abs/1105.1162}{{\ttfamily 1105.1162}}].

\bibitem{Kodama:2003jz}
H.~Kodama and A.~Ishibashi, \emph{{A Master equation for gravitational
  perturbations of maximally symmetric black holes in higher dimensions}},
  \href{https://doi.org/10.1143/PTP.110.701}{\emph{Prog. Theor. Phys.}
  {\bfseries 110} (2003) 701}
  [\href{https://arxiv.org/abs/hep-th/0305147}{{\ttfamily hep-th/0305147}}].

\bibitem{Kodama:2003kk}
H.~Kodama and A.~Ishibashi, \emph{{Master equations for perturbations of
  generalized static black holes with charge in higher dimensions}},
  \href{https://doi.org/10.1143/PTP.111.29}{\emph{Prog. Theor. Phys.}
  {\bfseries 111} (2004) 29}
  [\href{https://arxiv.org/abs/hep-th/0308128}{{\ttfamily hep-th/0308128}}].

\bibitem{Haehl:2015pja}
F.M.~Haehl, R.~Loganayagam and M.~Rangamani, \emph{{Adiabatic hydrodynamics:
  The eightfold way to dissipation}},
  \href{https://doi.org/10.1007/JHEP05(2015)060}{\emph{JHEP} {\bfseries 05}
  (2015) 060} [\href{https://arxiv.org/abs/1502.00636}{{\ttfamily
  1502.00636}}].

\bibitem{Baier:2007ix}
R.~Baier, P.~Romatschke, D.T.~Son, A.O.~Starinets and M.A.~Stephanov,
  \emph{{Relativistic viscous hydrodynamics, conformal invariance, and
  holography}},
  \href{https://doi.org/10.1088/1126-6708/2008/04/100}{\emph{JHEP} {\bfseries
  04} (2008) 100} [\href{https://arxiv.org/abs/0712.2451}{{\ttfamily
  0712.2451}}].

\bibitem{Arnold:2011ja}
P.~Arnold, D.~Vaman, C.~Wu and W.~Xiao, \emph{{Second order hydrodynamic
  coefficients from 3-point stress tensor correlators via AdS/CFT}},
  \href{https://doi.org/10.1007/JHEP10(2011)033}{\emph{JHEP} {\bfseries 10}
  (2011) 033} [\href{https://arxiv.org/abs/1105.4645}{{\ttfamily 1105.4645}}].

\bibitem{Faulkner:2010tq}
T.~Faulkner and J.~Polchinski, \emph{{Semi-Holographic Fermi Liquids}},
  \href{https://doi.org/10.1007/JHEP06(2011)012}{\emph{JHEP} {\bfseries 06}
  (2011) 012} [\href{https://arxiv.org/abs/1001.5049}{{\ttfamily 1001.5049}}].

\bibitem{Nickel:2010pr}
D.~Nickel and D.T.~Son, \emph{{Deconstructing holographic liquids}},
  \href{https://doi.org/10.1088/1367-2630/13/7/075010}{\emph{New J. Phys.}
  {\bfseries 13} (2011) 075010}
  [\href{https://arxiv.org/abs/1009.3094}{{\ttfamily 1009.3094}}].

\bibitem{Crossley:2015tka}
M.~Crossley, P.~Glorioso, H.~Liu and Y.~Wang, \emph{{Off-shell hydrodynamics
  from holography}}, \href{https://doi.org/10.1007/JHEP02(2016)124}{\emph{JHEP}
  {\bfseries 02} (2016) 124}
  [\href{https://arxiv.org/abs/1504.07611}{{\ttfamily 1504.07611}}].

\bibitem{deBoer:2015ija}
J.~de~Boer, M.P.~Heller and N.~Pinzani-Fokeeva, \emph{{Effective actions for
  relativistic fluids from holography}},
  \href{https://doi.org/10.1007/JHEP08(2015)086}{\emph{JHEP} {\bfseries 08}
  (2015) 086} [\href{https://arxiv.org/abs/1504.07616}{{\ttfamily
  1504.07616}}].

\bibitem{deBoer:2018qqm}
J.~de~Boer, M.P.~Heller and N.~Pinzani-Fokeeva, \emph{{Holographic
  Schwinger-Keldysh effective field theories}},
  \href{https://doi.org/10.1007/JHEP05(2019)188}{\emph{JHEP} {\bfseries 05}
  (2019) 188} [\href{https://arxiv.org/abs/1812.06093}{{\ttfamily
  1812.06093}}].

\bibitem{Kovtun:2012rj}
P.~Kovtun, \emph{{Lectures on hydrodynamic fluctuations in relativistic
  theories}}, \href{https://doi.org/10.1088/1751-8113/45/47/473001}{\emph{J.
  Phys. A} {\bfseries 45} (2012) 473001}
  [\href{https://arxiv.org/abs/1205.5040}{{\ttfamily 1205.5040}}].

\bibitem{Grozdanov:2013dba}
S.~Grozdanov and J.~Polonyi, \emph{{Viscosity and dissipative hydrodynamics
  from effective field theory}},
  \href{https://doi.org/10.1103/PhysRevD.91.105031}{\emph{Phys. Rev. D}
  {\bfseries 91} (2015) 105031}
  [\href{https://arxiv.org/abs/1305.3670}{{\ttfamily 1305.3670}}].

\bibitem{Kovtun:2014hpa}
P.~Kovtun, G.D.~Moore and P.~Romatschke, \emph{{Towards an effective action for
  relativistic dissipative hydrodynamics}},
  \href{https://doi.org/10.1007/JHEP07(2014)123}{\emph{JHEP} {\bfseries 07}
  (2014) 123} [\href{https://arxiv.org/abs/1405.3967}{{\ttfamily 1405.3967}}].

\bibitem{Haehl:2014zda}
F.M.~Haehl, R.~Loganayagam and M.~Rangamani, \emph{{The eightfold way to
  dissipation}},
  \href{https://doi.org/10.1103/PhysRevLett.114.201601}{\emph{Phys. Rev. Lett.}
  {\bfseries 114} (2015) 201601}
  [\href{https://arxiv.org/abs/1412.1090}{{\ttfamily 1412.1090}}].

\bibitem{Crossley:2015evo}
M.~Crossley, P.~Glorioso and H.~Liu, \emph{{Effective field theory of
  dissipative fluids}},
  \href{https://doi.org/10.1007/JHEP09(2017)095}{\emph{JHEP} {\bfseries 09}
  (2017) 095} [\href{https://arxiv.org/abs/1511.03646}{{\ttfamily
  1511.03646}}].

\bibitem{Haehl:2015uoc}
F.M.~Haehl, R.~Loganayagam and M.~Rangamani, \emph{{Topological sigma models
  \textbackslash{}\& dissipative hydrodynamics}},
  \href{https://doi.org/10.1007/JHEP04(2016)039}{\emph{JHEP} {\bfseries 04}
  (2016) 039} [\href{https://arxiv.org/abs/1511.07809}{{\ttfamily
  1511.07809}}].

\bibitem{Jensen:2017kzi}
K.~Jensen, N.~Pinzani-Fokeeva and A.~Yarom, \emph{{Dissipative hydrodynamics in
  superspace}}, \href{https://doi.org/10.1007/JHEP09(2018)127}{\emph{JHEP}
  {\bfseries 09} (2018) 127}
  [\href{https://arxiv.org/abs/1701.07436}{{\ttfamily 1701.07436}}].

\bibitem{Haehl:2018lcu}
F.M.~Haehl, R.~Loganayagam and M.~Rangamani, \emph{{Effective Action for
  Relativistic Hydrodynamics: Fluctuations, Dissipation, and Entropy Inflow}},
  \href{https://doi.org/10.1007/JHEP10(2018)194}{\emph{JHEP} {\bfseries 10}
  (2018) 194} [\href{https://arxiv.org/abs/1803.11155}{{\ttfamily
  1803.11155}}].

\bibitem{Jensen:2018hse}
K.~Jensen, R.~Marjieh, N.~Pinzani-Fokeeva and A.~Yarom, \emph{{A panoply of
  Schwinger-Keldysh transport}},
  \href{https://doi.org/10.21468/SciPostPhys.5.5.053}{\emph{SciPost Phys.}
  {\bfseries 5} (2018) 053} [\href{https://arxiv.org/abs/1804.04654}{{\ttfamily
  1804.04654}}].

\bibitem{Chen-Lin:2018kfl}
X.~Chen-Lin, L.V.~Delacr\'etaz and S.A.~Hartnoll, \emph{{Theory of diffusive
  fluctuations}},
  \href{https://doi.org/10.1103/PhysRevLett.122.091602}{\emph{Phys. Rev. Lett.}
  {\bfseries 122} (2019) 091602}
  [\href{https://arxiv.org/abs/1811.12540}{{\ttfamily 1811.12540}}].

\bibitem{Colin-Ellerin:2020mva}
S.~Colin-Ellerin, X.~Dong, D.~Marolf, M.~Rangamani and Z.~Wang,
  \emph{{Real-time gravitational replicas: Formalism and a variational
  principle}},  \href{https://arxiv.org/abs/2012.00828}{{\ttfamily
  2012.00828}}.

\bibitem{NIST:DLMF}
``{\it NIST Digital Library of Mathematical Functions}.''
  http://dlmf.nist.gov/, Release 1.0.26 of 2020-03-15.

\bibitem{Susskind:1998dq}
L.~Susskind and E.~Witten, \emph{{The Holographic bound in anti-de Sitter
  space}},  \href{https://arxiv.org/abs/hep-th/9805114}{{\ttfamily
  hep-th/9805114}}.

\bibitem{Hartnoll:2016apf}
S.A.~Hartnoll, A.~Lucas and S.~Sachdev, \emph{{Holographic quantum matter}},
  \href{https://arxiv.org/abs/1612.07324}{{\ttfamily 1612.07324}}.

\bibitem{Kanitscheider:2008kd}
I.~Kanitscheider, K.~Skenderis and M.~Taylor, \emph{{Precision holography for
  non-conformal branes}},
  \href{https://doi.org/10.1088/1126-6708/2008/09/094}{\emph{JHEP} {\bfseries
  09} (2008) 094} [\href{https://arxiv.org/abs/0807.3324}{{\ttfamily
  0807.3324}}].

\bibitem{Caldarelli:2013aaa}
M.M.~Caldarelli, J.~Camps, B.~Gout\'eraux and K.~Skenderis,
  \emph{{AdS/Ricci-flat correspondence}},
  \href{https://doi.org/10.1007/JHEP04(2014)071}{\emph{JHEP} {\bfseries 04}
  (2014) 071} [\href{https://arxiv.org/abs/1312.7874}{{\ttfamily 1312.7874}}].

\bibitem{Gouteraux:2011qh}
B.~Gouteraux, J.~Smolic, M.~Smolic, K.~Skenderis and M.~Taylor,
  \emph{{Holography for Einstein-Maxwell-dilaton theories from generalized
  dimensional reduction}},
  \href{https://doi.org/10.1007/JHEP01(2012)089}{\emph{JHEP} {\bfseries 01}
  (2012) 089} [\href{https://arxiv.org/abs/1110.2320}{{\ttfamily 1110.2320}}].

\bibitem{Regge:1957td}
T.~Regge and J.A.~Wheeler, \emph{{Stability of a Schwarzschild singularity}},
  \href{https://doi.org/10.1103/PhysRev.108.1063}{\emph{Phys. Rev.} {\bfseries
  108} (1957) 1063}.

\bibitem{Zerilli:1970se}
F.J.~Zerilli, \emph{{Effective potential for even parity Regge-Wheeler
  gravitational perturbation equations}},
  \href{https://doi.org/10.1103/PhysRevLett.24.737}{\emph{Phys. Rev. Lett.}
  {\bfseries 24} (1970) 737}.

\bibitem{Zerilli:1971wd}
F.~Zerilli, \emph{{Gravitational field of a particle falling in a schwarzschild
  geometry analyzed in tensor harmonics}},
  \href{https://doi.org/10.1103/PhysRevD.2.2141}{\emph{Phys. Rev. D} {\bfseries
  2} (1970) 2141}.

\bibitem{Kovtun:2005ev}
P.K.~Kovtun and A.O.~Starinets, \emph{{Quasinormal modes and holography}},
  \href{https://doi.org/10.1103/PhysRevD.72.086009}{\emph{Phys. Rev. D}
  {\bfseries 72} (2005) 086009}
  [\href{https://arxiv.org/abs/hep-th/0506184}{{\ttfamily hep-th/0506184}}].

\bibitem{Michalogiorgakis:2006jc}
G.~Michalogiorgakis and S.S.~Pufu, \emph{{Low-lying gravitational modes in the
  scalar sector of the global AdS(4) black hole}},
  \href{https://doi.org/10.1088/1126-6708/2007/02/023}{\emph{JHEP} {\bfseries
  02} (2007) 023} [\href{https://arxiv.org/abs/hep-th/0612065}{{\ttfamily
  hep-th/0612065}}].

\bibitem{Gubser:2007nd}
S.S.~Gubser and S.S.~Pufu, \emph{{Master field treatment of metric
  perturbations sourced by the trailing string}},
  \href{https://doi.org/10.1016/j.nuclphysb.2007.08.015}{\emph{Nucl. Phys. B}
  {\bfseries 790} (2008) 42}
  [\href{https://arxiv.org/abs/hep-th/0703090}{{\ttfamily hep-th/0703090}}].

\bibitem{Diles:2019uft}
S.M.~Diles, L.A.~Mamani, A.S.~Miranda and V.T.~Zanchin, \emph{{Third-order
  relativistic hydrodynamics: dispersion relations and transport coefficients
  of a dual plasma}},
  \href{https://doi.org/10.1007/JHEP05(2020)019}{\emph{JHEP} {\bfseries 05}
  (2020) 019} [\href{https://arxiv.org/abs/1909.05199}{{\ttfamily
  1909.05199}}].

\bibitem{Ishibashi:2004wx}
A.~Ishibashi and R.M.~Wald, \emph{{Dynamics in nonglobally hyperbolic static
  space-times. 3. Anti-de Sitter space-time}},
  \href{https://doi.org/10.1088/0264-9381/21/12/012}{\emph{Class. Quant. Grav.}
  {\bfseries 21} (2004) 2981}
  [\href{https://arxiv.org/abs/hep-th/0402184}{{\ttfamily hep-th/0402184}}].

\bibitem{Gubser:1998bc}
S.~Gubser, I.R.~Klebanov and A.M.~Polyakov, \emph{{Gauge theory correlators
  from noncritical string theory}},
  \href{https://doi.org/10.1016/S0370-2693(98)00377-3}{\emph{Phys. Lett. B}
  {\bfseries 428} (1998) 105}
  [\href{https://arxiv.org/abs/hep-th/9802109}{{\ttfamily hep-th/9802109}}].

\bibitem{Witten:1998qj}
E.~Witten, \emph{{Anti-de Sitter space and holography}},
  \href{https://doi.org/10.4310/ATMP.1998.v2.n2.a2}{\emph{Adv.\ Theor.\ Math.\
  Phys.} {\bfseries 2} (1998) 253}
  [\href{https://arxiv.org/abs/hep-th/9802150}{{\ttfamily hep-th/9802150}}].

\bibitem{Klebanov:1999tb}
I.R.~Klebanov and E.~Witten, \emph{{AdS / CFT correspondence and symmetry
  breaking}}, \href{https://doi.org/10.1016/S0550-3213(99)00387-9}{\emph{Nucl.
  Phys. B} {\bfseries 556} (1999) 89}
  [\href{https://arxiv.org/abs/hep-th/9905104}{{\ttfamily hep-th/9905104}}].

\bibitem{Witten:2001ua}
E.~Witten, \emph{{Multitrace operators, boundary conditions, and AdS / CFT
  correspondence}},  \href{https://arxiv.org/abs/hep-th/0112258}{{\ttfamily
  hep-th/0112258}}.

\bibitem{Heemskerk:2010hk}
I.~Heemskerk and J.~Polchinski, \emph{{Holographic and Wilsonian
  Renormalization Groups}},
  \href{https://doi.org/10.1007/JHEP06(2011)031}{\emph{JHEP} {\bfseries 06}
  (2011) 031} [\href{https://arxiv.org/abs/1010.1264}{{\ttfamily 1010.1264}}].

\bibitem{Faulkner:2010jy}
T.~Faulkner, H.~Liu and M.~Rangamani, \emph{{Integrating out geometry:
  Holographic Wilsonian RG and the membrane paradigm}},
  \href{https://doi.org/10.1007/JHEP08(2011)051}{\emph{JHEP} {\bfseries 08}
  (2011) 051} [\href{https://arxiv.org/abs/1010.4036}{{\ttfamily 1010.4036}}].

\bibitem{Bhattacharyya:2008mz}
S.~Bhattacharyya, R.~Loganayagam, I.~Mandal, S.~Minwalla and A.~Sharma,
  \emph{{Conformal Nonlinear Fluid Dynamics from Gravity in Arbitrary
  Dimensions}},
  \href{https://doi.org/10.1088/1126-6708/2008/12/116}{\emph{JHEP} {\bfseries
  12} (2008) 116} [\href{https://arxiv.org/abs/0809.4272}{{\ttfamily
  0809.4272}}].

\bibitem{Uhlenbeck:1963lec}
G.E.~Uhlenbeck, G.W.~Ford, G.W.~Ford and E.W.~Montroll, \emph{Lectures in
  statistical mechanics}, vol.~1, Proquest/Csa Journal Division (1963).

\bibitem{Schwinger:1960qe}
J.S.~Schwinger, \emph{{Brownian motion of a quantum oscillator}},
  \href{https://doi.org/10.1063/1.1703727}{\emph{J. Math. Phys.} {\bfseries 2}
  (1961) 407}.

\bibitem{Feynman:1963fq}
R.~Feynman and J.~Vernon, F.L., \emph{The theory of a general quantum system
  interacting with a linear dissipative system},
  \href{https://doi.org/10.1016/0003-4916(63)90068-X}{\emph{Annals Phys.}
  {\bfseries 24} (1963) 118}.

\bibitem{Marolf:2006nd}
D.~Marolf and S.F.~Ross, \emph{{Boundary Conditions and New Dualities: Vector
  Fields in AdS/CFT}},
  \href{https://doi.org/10.1088/1126-6708/2006/11/085}{\emph{JHEP} {\bfseries
  11} (2006) 085} [\href{https://arxiv.org/abs/hep-th/0606113}{{\ttfamily
  hep-th/0606113}}].

\bibitem{Kovtun:2003wp}
P.~Kovtun, D.T.~Son and A.O.~Starinets, \emph{{Holography and hydrodynamics:
  Diffusion on stretched horizons}},
  \href{https://doi.org/10.1088/1126-6708/2003/10/064}{\emph{JHEP} {\bfseries
  10} (2003) 064} [\href{https://arxiv.org/abs/hep-th/0309213}{{\ttfamily
  hep-th/0309213}}].

\bibitem{Grozdanov:2015kqa}
S.~Grozdanov and N.~Kaplis, \emph{{Constructing higher-order hydrodynamics: The
  third order}}, \href{https://doi.org/10.1103/PhysRevD.93.066012}{\emph{Phys.
  Rev. D} {\bfseries 93} (2016) 066012}
  [\href{https://arxiv.org/abs/1507.02461}{{\ttfamily 1507.02461}}].

\bibitem{Haehl:2015foa}
F.M.~Haehl, R.~Loganayagam and M.~Rangamani, \emph{{The Fluid Manifesto:
  Emergent symmetries, hydrodynamics, and black holes}},
  \href{https://doi.org/10.1007/JHEP01(2016)184}{\emph{JHEP} {\bfseries 01}
  (2016) 184} [\href{https://arxiv.org/abs/1510.02494}{{\ttfamily
  1510.02494}}].

\end{thebibliography}

\providecommand{\href}[2]{#2}\begingroup\raggedright\endgroup

\end{document}